\documentclass[usenatbib]{mnras} 

\usepackage{newtxtext,newtxmath}
\usepackage{color}
\usepackage{booktabs}

\usepackage[T1]{fontenc}
\usepackage{ae,aecompl}

\usepackage{graphicx}	
\usepackage{amsmath}	
\usepackage{amssymb}	
\usepackage{float} 
\usepackage{threeparttable, tablefootnote}  

\newcommand{\eg}{{\sl e.g.}, }        
\newcommand{\Rtwoh}{R_{\rm 200c}}        
\newcommand{\rtwoh}{R_{\rm 200c}}        
\newcommand{\mtwoh}{M_{\rm 200c}}        
\newcommand{\Mtwoh}{M_{\rm 200c}}

\newcommand{\Mstargal}{M_{\star}}
\newcommand{\mstargal}{M_{\star}}
\newcommand{\Mstar}{M_{\star, \rm tot}}
\newcommand{\mstar}{M_{\star, \rm tot}}
\newcommand{\MstarBCG}{M_{\star,\rm BCG}}

\newcommand{\mhalo}{M_{\rm halo}}
\newcommand{\Mhalo}{M_{\rm halo}}

\newcommand{\kms}{{\rm \, km~s}\ensuremath{^{-1}}}
\newcommand{\msol}{\ensuremath{\, {\rm M}_\odot}}    
\newcommand{\msun}{\ensuremath{\, {\rm M}_\odot}} 
\newcommand{\kpc}{\ensuremath{\, {\rm kpc}}}         
\newcommand{\mpc}{\ensuremath{\, {\rm Mpc}}}

\newcommand{\Nsat}{\ensuremath{N_{\rm sat}}}
\newcommand{\NSat}{\ensuremath{N_{\rm sat}}}
\newcommand{\Mstarsat}{\ensuremath{M_{\rm \star,sat}}}
\newcommand{\Mstartot}{\ensuremath{M_{\rm \star,tot}}}
\newcommand{\zformation}{\ensuremath{z_{\rm form}}}

\definecolor{bleudefrance}{rgb}{0.19, 0.55, 0.91}
\definecolor{purple}{RGB}{128, 0, 128}

\begin{document}
\title[Multi-Sim Massive Halo Property Statistics]{Stellar Property Statistics of Massive Halos from Cosmological Hydrodynamics Simulations: Common Kernel Shapes}

\author[Dhayaa Anbajagane]{Dhayaa Anbajagane$^{1\star}$, August E. Evrard$^{1, 2}$, Arya Farahi$^{3, 4}$, David J. Barnes$^5$, \newauthor Klaus Dolag$^{6, 7}$, Ian G. McCarthy$^8$, Dylan Nelson$^6$, Annalisa Pillepich$^{9}$
\\
\\
$^{1}$Department of Physics and Leinweber Center for Theoretical Physics, University of Michigan, Ann Arbor, MI 48109, USA\\
$^2$Department of Astronomy, University of Michigan, Ann Arbor, MI 48109, USA\\
$^3$ Michigan Institute for Data Science, University of Michigan, Ann Arbor, MI 48109, USA\\
$^4$McWilliams Center for Cosmology, Department of Physics, Carnegie
Mellon University, Pittsburgh, Pennsylvania 15312, USA\\
$^5$Department of Physics, Kavli Institute for Astrophysics and Space Research, Massachusetts Institute of Technology, Cambridge, MA 02139, USA\\
$^6$Max-Planck Institut f\"ur Astrophysik, Karl-Schwarzschild Str. 1, D-85741 Garching, Germany\\
$^7$University Observatory Munich, Scheinerstr. 1, 81679 M\"unchen, Germany\\
$^8$Astrophysics Research Institute, Liverpool John Moores University, 146 Brownlow Hill, Liverpool L3 5RF, UK\\
$^{9}$Max-Planck-Institut f\"ur Astronomie, K\"onigstuhl 17, 69117 Heidelberg, Germany
}

\date{Accepted XXX. Received YYY; in original form ZZZ}

\pubyear{2019}

\label{firstpage}
\pagerange{\pageref{firstpage}--\pageref{lastpage}}
\maketitle

\begin{abstract}
We study stellar property statistics, including satellite galaxy occupation, of massive halo populations realized by three cosmological hydrodynamics simulations:  BAHAMAS + MACSIS, TNG300 of the IllustrisTNG suite, and Magneticum Pathfinder.  The simulations incorporate independent sub-grid methods for astrophysical processes with spatial resolutions ranging from $1.5$ to $6$ kpc, and each generates samples of $1000$ or more halos with $\Mhalo > 10^{13.5} \msun$ at redshift $z=0$. Applying  localized, linear regression (LLR), we extract halo mass-conditioned statistics (normalizations, slopes, and intrinsic covariance) for a three-element stellar property vector consisting of: i) $\Nsat$, the number of satellite galaxies with stellar mass, $M_\star \! >  \! 10^{10} \msun$ within radius $\Rtwoh$ of the halo; ii) $\Mstar$, the total stellar mass within that radius, and; iii) $\MstarBCG$, the gravitationally-bound stellar mass of the central galaxy within a $100 \, \rm kpc$ radius.  Scaling parameters for the three properties with halo mass show mild differences among the simulations, in part due to numerical resolution, but there is qualitative agreement on property correlations, with halos having smaller than average central galaxies tending to also have smaller total stellar mass and a larger number of satellite galaxies.  
Marginalizing over total halo mass, we find the satellite galaxy kernel, $p(\ln\Nsat \,|\, \Mhalo, z)$ to be consistently skewed left, with skewness parameter $\gamma \! = \!  -0.91 \pm 0.02$, while that of  $\ln\Mstar$ is closer to log-normal, in all three simulations.  
The highest resolution simulations find $\gamma \! \simeq \!  -0.8$ for the $z \! = \! 0$ shape of $p(\ln\MstarBCG \,|\, \Mhalo, z)$ and also that the fractional scatter in total stellar mass is below 10 percent in halos more massive than $10^{14.3} \msol$.  We provide a Gaussian mixture fit to the low redshift $\Nsat$ kernel as well as LLR parameters tabulated for halos more massive than $10^{13.5} \msol$ in all simulations.
\end{abstract}



\section{Introduction}

Clusters of galaxies, and the underlying dark matter halos that host these systems, are important to study as their population behavior is sensitive to both the expansion history and the gravitational growth of large-scale structure in our Universe \citep{Allen2011CosmologicalClusters}.

When using clusters of galaxies for cosmology, the statistical relationship between an observable cluster property and the total mass of its host halo, which we call the mass-property relation (MPR), is a key model element \citep[\eg][]{Rozo2010SDSSclusterCosmo, Mantz2010XrayCountsCosmo, Vikhlinin2009, Zhang2011HIFLUGCS,  deHaan2016SPTCosmo, Pillepich2018DEForecastseRosita, Bocquet2019SPTCosmo, Mulroy2019LoCuSS,Costanzi2019SDSSmethods}. Current and near-future experiments will expand cluster sample sizes into the tens of thousands \citep[][]{DES2005, Laureijs2011EUCLID, Pillepich2012eRositaForecasts, Merloni2012eRositaScienceBook, Predehl2014eROSITAonSRG, Spergel2015WFIRST,SimonsObservatory}, allowing for improved understanding of the MPR for hot gas and stellar properties.  This understanding, along with careful modeling of survey selection, are crucial elements that empower studies of cosmic acceleration and of new physics using massive halos.

When the observable property is a count of galaxies above some size (luminosity, stellar mass, etc.) threshold, the MPR is analogous to the Halo Occupation Distribution \citep[HOD,][]{BerlindWeinberg2002HODmodel, Cooray2002HaloStructure, Hearin:2013, Zentner2014GalaxyAssemblyBias, Hearin:2016, Zehavi2018AssemblyBias}. The stellar property statistics of dark matter halos lie within the broad category of the "galaxy--halo connection" which was reviewed recently by \citet{WechslerTinker2018GalaxyHaloConnection}. The focus of this work is the high mass population of halos, each of which host multiple bright galaxies. 

The statistics of the stellar component properties --- galaxy occupation, central galaxy stellar mass, and total stellar mass of a halo --- across a broad range of total halo mass and redshift is a fundamental outcome of the complex astrophysical processes that drive galaxy formation \citep[see, \eg ][ for a detailed discussion]{Pillepich2018FirstGalaxies}.  

A low-accretion rate mode of supermassive black hole (SMBH) feedback within the cores of large galaxies \citep{Croton2006TheGalaxies, DeLucia2006Egals} is employed to solve the problem of excessive cooling and star formation at the centers of groups and clusters seen in early simulations \citep{KatzWhite1993, Evrard1994Two-FluidFormation}. Jet-driven, turbulent feedback from SMBH accretion appears to be an important regulator of the thermodynamic state of core gas \citep{McNamara2012AGNFeedback, Voit2015NaturePrecip} and its inclusion in cosmological simulations has significantly improved the fidelity of galaxy and hot gas properties within the population of high mass halos \citep{RagoneFigueroa2013BCGsims, Gaspari2013ChaoticColdAccretion, Hirschmann2014CosmologicalDownsizing, Khandai2015MassiveBlack, Rasia2015CoolCores, Hahn2017RhapsodyG,  McCarthy2017TheCosmology, Kaviraj2017HorizonAGN,  Nelson2018TNGcolors,  Pillepich2018FirstGalaxies}.

While hydrodynamic and magnetohydrodynamic methods have improved dramatically in terms of both numerical resolution and astrophysical treatments \citep{KravtsovBorgani2012, Vogelsberger2019SimReview}, direct inter-comparisons of different numerical solutions reveal varying degrees of inconsistency \citep[{\sl e.g.},][]{Scannapieco2012AquilaCodeComp, Elahi2016NiftyIII}. The aim of this work is to compare compressed statistical summaries of the aforementioned stellar properties for populations of massive haloes realized by independent state-of-the-art methods. We apply a local linear regression (LLR) approach first used by \citet{Farahi2018LocalizedCovariance} to describe the statistics of hot gas and stellar mass, conditioned on total halo mass, for halos realized by the BAHAMAS and MACSIS simulations.  As long as halo properties are well behaved functions of mass and redshift, the LLR method is highly effective at compressing the full range of discrete population measurements into a small number of statistical parameters. 

This work expands on \citet{Farahi2018LocalizedCovariance} by: i) using multiple stellar properties associated with a halo and; ii) performing a verification test using results from multiple simulation teams. 

We utilize three cosmological hydrodynamics simulations --- a superset of BAHAMAS and MACSIS, the IllustrisTNG 300-1 run, and the Magneticum Pathfinder 500 Mpc volume --- each of which contains $>1000$ halos with $\Mtwoh > 10^{13.5} \msun$ at $z = 0$. 
Bootstrap resampling of each discrete population is used to estimate statistical uncertainties in scaling relation parameters. As we show below, the statistical power of these large halo samples is reflected by relatively small errors in the recovered LLR parameters. The different simulations often produce results in mild statistical tension with one another, in which case the range of behavior in the quantity of interest can be considered as a first estimate of the global theoretical uncertainty in that parameter.

While such tensions exist for many of the derived LLR parameter values, we also find areas of congruence, particularly in the fundamental forms of mass-conditioned property kernels. Congruent results offer a necessary step of {\sl verification} \citep[\eg][]{Salvadori2019}, 
meaning that halo populations with consistent stellar MPRs emerge from independent solutions of the equations governing the complex, non-linear system of large-scale structure. A {\sl validation} step using observational data must be done using observable proxies for the intrinsic true properties we use here. Observational analysis with careful treatment of sample selection is emerging \citep{Mantz2016WtGScaling, Farahi2019DESY1X-Ray, Mulroy2019LoCuSS,  Bocquet2019SPTCosmo} but we do not attempt detailed comparisons to observational samples in this paper.

We employ an $\Mtwoh$ spherical overdensity mass scale convention\footnote{The radius $\rtwoh$ satisfies $ 3 M(<\rtwoh) / (4 \pi \rtwoh^3) = 200 \rho_{\rm crit}(z)$, where $\rho_{\rm crit}(z)$ is the critical density of the universe, and $\mtwoh \equiv M(<\rtwoh)$}, and define $\Nsat$ as the integer count of satellite galaxies with $\Mstarsat > 10^{10} \msun$ lying within that radius\footnote{Slight adjustments are made to normalize the mean cosmic baryon fraction, as described in $\S$ \ref{sec:properties}}. The form of the conditional likelihood, $p(\ln \Nsat \, | \, \Mhalo, z)$, a core component of HOD models, is a particular area of focus, and a key finding of our study is that all three simulations produce a consistent {\sl shape} for both this kernel and that of the total stellar mass within $\Rtwoh$.

We also find that, at fixed halo mass, satellite galaxy number is anti-correlated with the stellar mass of the central galaxy of a halo, as would be expected if central galaxies, which we refer to as the brightest central galaxy (BCG), grow primarily at late times by cannibalizing satellite galaxies \citep{TremaineRichstone1977, DeLucia:2007}.  Larger than average BCG stellar masses are also associated with magnitude gaps measured with respect to lower ranked galaxies \citep[\eg][]{GoldenMarxMiller2018}.  

The structure of this paper is as follows. In $\S$\ref{sec:Simulations} we describe the simulation samples while $\S$\ref{sec:Method} describes the localized linear regression (LLR) method applied to generate summary statistics for each simulation's halo sample. We examine satellite galaxy scaling relations in $\S$\ref{sec:LLR_and_PDF}, along with relevant aspects of scaling relations of total stellar mass and central galaxy stellar mass. Here, we also provide a two-component Gaussian mixture model (GMM) fit to $p(\Nsat \, | \, M ,z)$ that describes the consistent shape seen in all simulation ensembles at $z < 1$.  The interrelationships of BCG stellar mass with other properties is explored in $\S$\ref{sec:SecondarySelection}. We elaborate briefly on the future of such multi-simulation comparison studies in $\S$ \ref{sec:Discussion}, and summarize our findings in $\S$\ref{sec:Summary}. Appendices provide complete results for all stellar properties (\ref{Appendix:StellarLLRfits}) as well as the GMM and LLR parameter tables (\ref{Appendix:All_Params}).


\section{Simulations and Halo Populations}
\label{sec:Simulations}

\begin{table*}
	\centering
    \caption{Simulation characteristics and $z=0$ halo population sizes. Empirical sources for tuning sub-grid parameters are given in the last column and consist of the Galaxy Stellar Mass Function (GSMF), SuperMassive Black Hole scaling (SMBH), Metallicity scaling (Metals), and cluster hot gas mass fraction $<R_{500c}$ (CL $f_{\rm gas}$). All assume a flat $\Lambda {\rm CDM}$ cosmology, so $\Omega_\Lambda = 1- \Omega_m$, and their cosmic mean baryon fraction is $f_b \equiv \Omega_b/\Omega_m$. The MGTM output is actually $z = 0.03$ and B100 (used only for testing resolution) is $z = 0.12$ while the other two are exactly $z=0$. See text for references.}
	\begin{threeparttable}
		\begin{tabular}{l|c|c|c|c|c|c|c|c|l}
			Simulation\tnote{$\diamondsuit$} & $L$ [Mpc]\tnote{$\ast$} & $\Omega_m$ & $f_b$ & $\sigma_8$ & $\epsilon_{DM}^{z=0} \,\rm [kpc]$ & 
			$m_\star \, [\msol]$\tnote{$\spadesuit$}  & $\log_{10}(M_{20})$\tnote{$\heartsuit$} & 
			$N_{\rm sam}$\tnote{$\dagger$} & Calibration \\
            \hline
                BM  & $596$ & $0.3175$ & $0.154$  & $0.834$ & $5.96$ & $1.2 \times 10^9$  & $15.6$ & $9430$ & GSMF, CL $f_{\rm gas}$ \\
                B100 & $143$ & $0.2793$ & $0.166$  & $0.821$ & $2.86$ & $1.4 \times 10^{8}$  & $14.1$ & $96$ & GSMF, CL $f_{\rm gas}$\\
                MGTM & $500$ & $0.2726$ & $0.167$ & $0.809$ & $5.33$& $5.0 \times 10^7$  & $14.9$ & $4207$ & SMBH, Metals, CL $f_{\rm gas}$ \\
                TNG300 & $303$ & $0.3089$ & $0.157$ & $0.8159$ & $1.48$& $1.1 \times 10^{7}$ & $14.6$ & $1146$ & See \citet{Pillepich2018SimulatingGalaxy} \\
            \hline
            \end{tabular}
    \begin{itemize}
        \item[$^\diamondsuit$]{See text for description of acronyms.}
    	\item[$^\ast$]{Comoving simulation cube length except for MACSIS (subset of the BM data), which subsamples a $3.2$ Gpc cubic volume. }
    	\item[$^\spadesuit$]{Initial stellar particle mass. }
    	\item[$^\heartsuit$]{Upper limit of LLR regression at $z=0$, the 20${\rm th}$ most massive halo mass, in $\msol$.}
    	\item[$^\dagger$]{Number of halos with total mass, $\mtwoh > 10^{13.5} M_\odot$. The number above $10^{13.8}\msol$ for BM is $\approx 4400$.}
    \end{itemize}
	\end{threeparttable}
	
	\label{tab:samples}
\end{table*}

We analyze three different simulations --- BAHAMAS + MACSIS (BM), TNG300-1 from the IllustrisTNG project (TNG300), and the 500 Mpc high-resolution box 2b from Magneticum Pathfinder (MGTM) --- with characteristics summarized in Table~\ref{tab:samples}.  We also include a $z=0.12$ output of a higher resolution BAHAMAS run, a $143 \mpc$ volume (labelled B100), to examine resolution dependence of the derived statistics.  Collectively, the simulations produce nearly 15000 halos with total mass, $\mtwoh > 10^{13.5} \msol$, at $z=0$.

The simulations use slightly different cosmological parameters, but all assume a flat geometry with matter density parameter in the range $\Omega_m \in [0.2726, 0.3175]$.  The cosmic baryon fraction, $f_b \equiv \Omega_b / \Omega_m$, which varies from a low of $0.154$ for BM to a high of $0.167$ for MGTM, is important for setting the normalization of the stellar mass formed within halos and so we make small adjustments to the satellite galaxy stellar mass cutoff described below.  

The BAHAMAS simulations \citep{McCarthy2017TheCosmology} use a version of the Smoothed Particle Hydrodynamics (SPH) code Gadget \citep{Springel2005Gagdet2} to model a 596 Mpc periodic cube.  The MACSIS ensemble \citep{Barnes2017Macsis} comprises 390 ``zoom-in'' simulations of individual halo regions, drawn from a parent 3.2 Gpc N-body simulation, that completely resamples the most massive halos in the large parent volume. The MACSIS resimulations use the same code base and have the same numerical resolution and astrophysical treatments as BAHAMAS but 
the resimulation technique enables LLR fits extending to $\mtwoh = 4 \times 10^{15} \msol$ at $z=0$. 
The BM sample is the superset of BAHAMAS and MACSIS populations. Conversely, the 140 Mpc Bahamas higher resolution simulation (B100) contains only 96 halos with $\mtwoh > 10^{13.5} \msol$ at $z=0$, only a few dozen of which lie above $10^{14} \msol$.

\begin{figure}
	\includegraphics[width = \columnwidth]{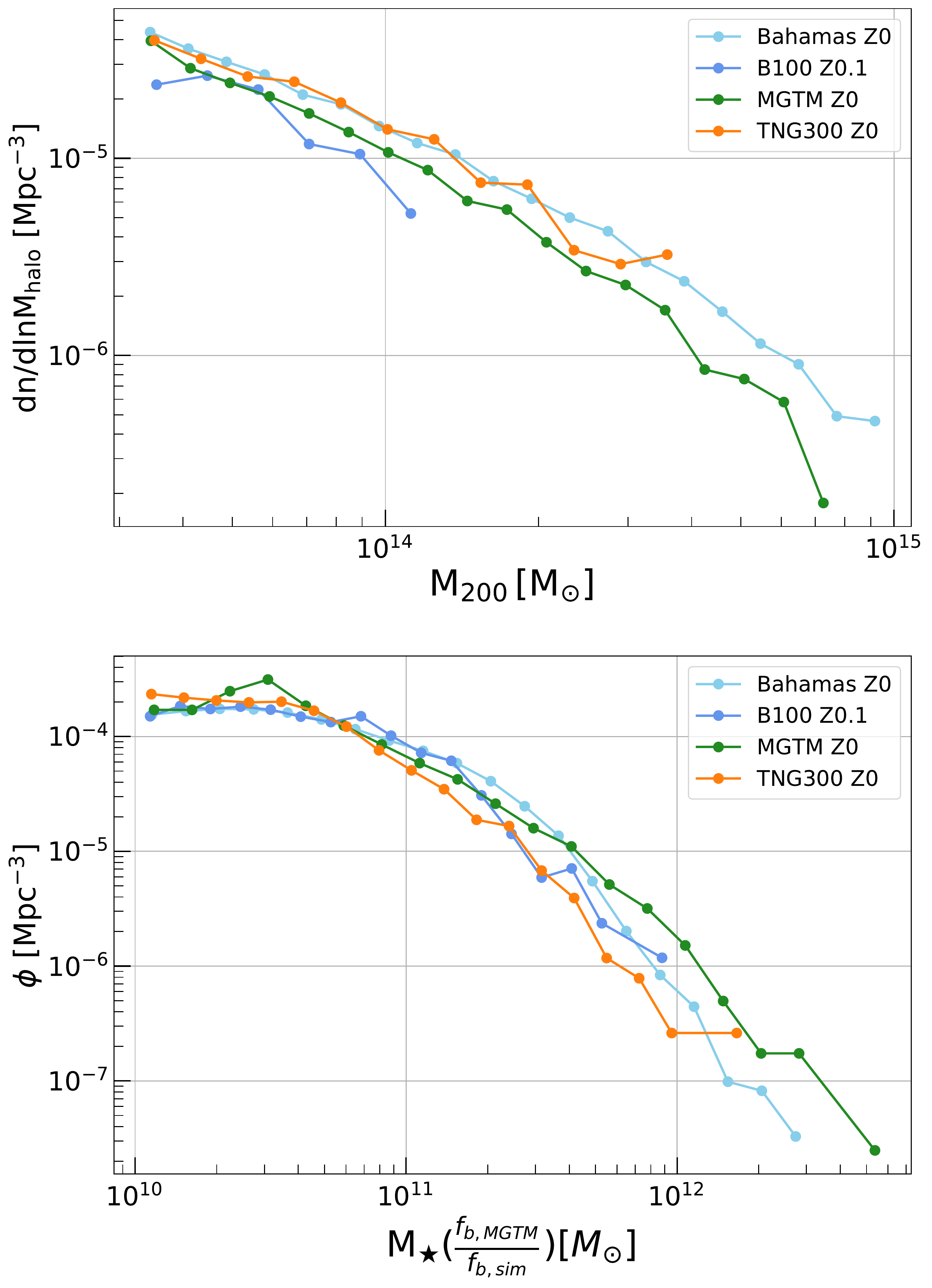}
	\caption{\textbf{Upper:} Halo Mass function of all four simulations for $\Mhalo > 10^{13.5} \msol$ at $z = 0$ (BAHAMAS, TNG300), $z = 0.03$ (MGTM), and $0.12$ (B100). \textbf{Lower:} Satellite galaxy stellar mass function within $\rtwoh$ in halos with $\mtwoh > 10^{13.5} \msol$ at $z = 0$ (BAHAMAS, TNG300), $z = 0.06$ (MGTM), and $0.12$ (B100). Stellar masses for BAHAMAS, B100, and TNG300 are adjusted to match the MGTM baryon fraction.}
\label{fig:Stellar_Mass_Function}
\end{figure}

The MGTM 500 Mpc run \citep{Hirschmann2014CosmologicalDownsizing} uses a different fork of Gadget with an improved fluid solver \citep{Donnert2013RadioHaloes} to model the co-evolution of supermassive black holes and their host galaxies. While both BM and MGTM employ variants of the same base Gadget code, the detailed treatments of star formation, black hole seeding and growth, and feedback from these compact populations were developed independently.  

IllustrisTNG \citep{Pillepich2018FirstGalaxies, Nelson2018FirstBimodality, Springel2018FirstClustering, Marinacci2018FirstFields, Naiman2018FirstEuropium,Nelson2019TNGPublicData} is a follow-up to the Illustris simulations \citep{Vogelsberger2014Illustris} based on the deformable-mesh hydroydynamics solver, \textsc{Arepo} \citep{Springel2010EMesh}. As detailed in \citet{Pillepich2018FirstGalaxies}, the stellar mass functions and stellar mass to halo mass relations are in reasonable agreement with observational and empirical constraints. At the same time, \citet{Springel2018FirstClustering} show that the observed large-scale spatial clustering of galaxies, and its dependence on mass and color, is also reproduced. In this paper, we use the TNG300 run, the simulation of the TNG suite with the largest volume.

The last four columns of Table~\ref{tab:samples} highlight important similarities and differences among the simulations. The gravitational softening length used for dark matter and stars is similar in BM and MGTM ($5-6 \kpc$) but both are a factor $\sim 3$ larger than that of TNG300. TNG300 and MGTM employ star particles that are significantly less massive than that of BM. Because BM models a $10^{10} \msol$ stellar mass galaxy with as few as nine star particles, we find below that the variance in satellite galaxy count is enhanced with respect to both B100 and the other two simulations.

Resolution convergence is a non-trivial issue for all hydrodynamical simulations of galaxy formation. For example, \citet{Pillepich2018SimulatingGalaxy} and \citet{Pillepich2018FirstGalaxies} analyze the rate of convergence in stellar mass contents of halos in the TNG model, finding shifts of $\sim$40\% in galaxy stellar mass between the two highest levels of resolution of TNG at the high-mass end. The three simulations span a range of length, mass, and time resolution, so the differences in stellar mass statistics represents a convolution of numerical (\eg resolution) and physical model (\eg forms of feedback) differences.

Given the modest range of cosmologies explored by the simulations in the ensemble, one expects similar dark matter halo mass functions at $z=0$. The top panel of Figure~\ref{fig:Stellar_Mass_Function} shows that this is the case, with the high $\Omega_m$ cosmologies of BM and TNG300 shifted above MGTM, which has comparatively lower values of both $\Omega_m$ and $\sigma_8$.

The satellite galaxy stellar mass function (S-GSMF) for groups and clusters, shown in the bottom panel of Figure~\ref{fig:Stellar_Mass_Function}, is derived from the full collection of satellite galaxies within $\rtwoh$ of our target halo population with $\mtwoh > 10^{13.5} \msol$. The S-GSMF space density is computed here using the full simulation volume (not the volume occupied by halos), and stellar masses have been normalized to the baryon fraction of MGTM.

There is fairly good agreement in the S-GSMF among the simulations for galaxies with $10^{10} < M_\star /M_\odot < 10^{11}$ but the populations diverge somewhat at high masses. Note that the BAHAMAS simulation, despite the higher stellar particle mass, shows no sign of incompleteness down to the $10^{10} \msol$ stellar mass limit.  At $10^{12} \msol$, the MGTM simulation produces nearly seven times as many galaxies per unit volume as the TNG300 solution.  Given that MGMT has the lowest space density of high mass halos (top panel of Figure~\ref{fig:Stellar_Mass_Function}), we anticipate that the HOD normalization of MGTM is higher than those of the other two simulations.  The differences in the outcome of the $z\sim0$ GSMF are a consequence of different underlying choices and implementations of stellar and black hole feedback leading to different effective outcomes in the regulation and quenching of star formation.

\subsection{Halo Finding and Catalogs}\label{sec:haloProperties}

Our analysis methods employ catalog-level products derived independently by each simulation team. The methods operate to identify common halo and stellar properties listed in Table~\ref{tab:common_definitions}.

\begin{table}
	\centering
    \caption{Property definitions.}	\label{tab:common_definitions}
	\begin{threeparttable}
	\begin{tabular}{l|l}
	    Quantity & Definition  \\
        \hline
            Halo center & Minimum gravitational potential\tnote{$\ast$} \\
            Halo total mass, $\mtwoh$ & All species within $\rtwoh$ sphere \tnote{$\heartsuit$} \\
            Halo stellar mass, $\Mstar$ & All stellar particles within $\rtwoh$ \\
            Galaxy center & Most bound particle\tnote{$\dagger$} \\
            Galaxy stellar mass, $\Mstargal$ & Gravitationally-bound stellar particles  \\
         \hline
    \end{tabular}
    \begin{itemize}
        \item[$^\ast$]{FOF links of 0.2 (BM, TNG) and 0.16 (MGTM) mean separation.}
        \item[$^\heartsuit$]{TNG halo masses use only FOF particle set.}
        \item[$^\dagger$]{Of any species (BM, TNG) or collisionless only (MGTM).}
    \end{itemize}
\end{threeparttable}
\end{table}

The identification of the parent population of halos is done in a similar fashion across the four simulations, with a percolation (friends-of-friends) step followed by an identification of locally-bound sub-structures using the SUBFIND algorithm \citep{Springel2001Subfind,Dolag2009Subfind}. Halo centers are identified as the location of the most-bound particle and $\Mtwoh$ enclosed masses are derived using all particle species.

Stellar properties of galaxies within sub-halos are then derived. The common use of SUBFIND means that the galaxy stellar masses are well aligned across the simulations, employing common definitions of the center as well as the local binding energy condition.

Our study is based on halo catalogs defined by a total mass threshold, $ \mtwoh > 10^{13.5} \msun$, and we examine snapshots at $z = 0$, $0.5$, $1$, $1.5$ and $2$. Due to their larger volumes, BM and MGTM offer samples of more than 4000 halos at $z = 0$ while the higher resolution TNG300 simulation yields 1130.

\subsection{Stellar property vector}\label{sec:properties} 

We use a vector of stellar properties listed in Table~\ref{tab:stellar_props}. The satellite galaxy count, $\Nsat$, is the number of galaxies within $\rtwoh$ having individual stellar mass $\Mstargal > \left( \frac{f_{\rm\,b,\,Sim}}{f_{\rm\, b, \, MGTM}} \right) \times 10^{10} \msol$. We apply the linear correction to account for global baryon fraction differences across the simulations (see Table~\ref{tab:samples}) and arbitrarily normalize to the MGTM value. The mass limits for BM and TNG300 are thus $\sim 8\%$ smaller than the MGTM value.

This stellar mass threshold is chosen to balance resolution considerations with discreteness in $\Nsat$ counts.  At $10^{10} \msol$, we are working with galaxies resolved by 9 (BM), 73 (B100), 200 (MGTM), and 910 star particles (TNG300).  Lowering the threshold would decrease the minimum number of stellar particles in our selected galaxies leading to unacceptably small values for BM. Raising the threshold instead leaves us with more halos exhibiting either no satellites or just 1 or 2 of them, which complicates our calculations using $\ln \Nsat$.  Our stellar mass cutoff of $\sim10^{10} \msol$
lies significantly below the $\sim \! 10^{10.8} \msol$ knee of the observed stellar mass function \citep{Moustakas:2013}.

The rest of the stellar property vector consists of $\mstar$, the total stellar mass within $\Rtwoh$ as well as $\MstarBCG$, the central galaxy's bound stellar mass within a fixed sphere of 100 kpc physical radius. 

\begin{table}
\caption{Property components of LLR regression vector, $\mathbf{S}$.}\label{tab:stellar_props}
\begin{threeparttable}
	\begin{tabular}{l|l}
    Symbol & Quantity  \\
    \hline
    $\NSat$ & Count of non-central galaxies within $\rtwoh$\tnote{$\dagger$} \\
    $\Mstar$ & Total stellar mass within $\rtwoh$ ($\msol$).\\
    $\MstarBCG $  & Central galaxy stellar mass within 100 kpc ($\msol$). \\
    \hline
    \end{tabular}
    $^\dagger$ Stellar mass-limited, $\mstargal > \left( \frac{f_{\rm \, b,\,Sim}}{f_{\rm \,b, \, MGTM}} \right) \times 10^{10} \msol$.
    \end{threeparttable}
\end{table}

\section{Local Linear Regression Method}
\label{sec:Method}

From dimensional arguments one can infer that integral properties of massive halos, such as aggregate stellar mass or global X-ray temperature measured at some redshift will, in the mean, scale as power-laws with total system mass \citep{Kaiser:1986, BryanNorman:1998}. Due to variations in formation history and dynamical state, any individual halo will be offset from the population mean, and this intrinsic dispersion is often assumed to be log-normal in form \citep{Evrard2014Statistics}.  

Linear regression of the simple least-squares variety has been a canonical method used to characterize cluster scaling laws, but its utility is limited by the fact that it reduces full population statistics for a given property down to three numbers: a slope, normalization and variance/standard deviation. With large halo samples extracted from cosmological simulations we can perform a more sensitive analysis using {\sl localized} linear regression \citep{Farahi2018LocalizedCovariance}. The LLR method generates mass-conditioned estimates of the slope, normalization, and property covariance, where the term 'mass-conditioned' implies that we are determining these parameters given a certain halo mass and redshift. 

\subsection{Mass-conditioned parameters and normalized residuals}\label{sec:scaling}

Using natural logarithms of the properties, $\mathbf{s} = \ln\mathbf{S}$, the population mean of the log of property, $S_a$, at a fixed redshift scales with halo mass $M$ as 
 \begin{equation} \label{eq:scalingmodel}
 \langle s_a \, | \, M , z \rangle \ = \ \pi_a(M_c,z) + \alpha_a(M_c,z) \ln(M/M_c) , 
\end{equation}
where $M_c$ is a mass scale of interest, and the log-linear relation has slope, $\alpha_a(M_c,z)$, and intercept, $\pi_a(M_c,z)$, that, in general, depend on both redshift and the chosen halo mass scale. The subscript $a$ denotes the property under consideration and, at mass $M_c$, the modal and median value of $S_a$ is $e^{\pi_a(M,z)}$. 

The mass dependence of the fit parameters is found by applying a mass-dependent weight factor centered on the chosen mass scale.  Letting $M_c \rightarrow M$ for simplicity, we minimize the weighted square error,
\begin{equation}\label{eq:squareError}
\epsilon^2_a(M, z) = \sum\limits_{i=1}^{n} \ w_i^2 \ (s_{a,i}- \alpha_a(M,z) \mu_i - \pi_a(M,z) )^2,
\end{equation}
where $\mu_i \equiv \ln (M_{{\rm halo},i}/M)$, the sum $i$ is over all halos, and $w_i$ is the mass-dependent Gaussian weight factor, 
\begin{equation} \label{eq:weights}
 w_i \ = \ \frac{1}{\sqrt{2 \pi} \sigma_{\rm LLR} } \exp\left\{-\frac{\mu_i^2}{2\sigma_{\rm LLR}^2} \right\},
\end{equation}
with $\sigma_{\rm LLR}$ the width of the log-mass filter. For ideal mass localization we want $\sigma_{\rm LLR}$ to be very small, but the finite sample sizes from the simulations prevent us from using too small a value.  We use $\sigma_{\rm LLR} = 0.46$, equivalent to $0.2$ dex in halo mass. Our results are relatively insensitive to this choice, but choosing too small a value leads to noisy features at high mass where the population density is sparse.  

As the central filter mass scale, $M$, is varied in fixed logarithmic steps, we estimate the local slope, $\widetilde{\alpha}_a(M,z)$, intercept, $\widetilde{\pi}_a(M,z)$, and scatter, $\widetilde{\sigma}_a(M,z)$, parameters by minimizing the locally weighted square error, equation~(\ref{eq:squareError}).  The residual deviation in property $a$ for a specific halo of mass $M_i$ is then defined as 
\begin{equation}\label{eq:residual}
    \delta s_{a,i} \equiv s_{a,i} -  \widetilde{\pi}_a(M_i,z), 
\end{equation}
where the second term is determined by linear interpolation of the values sampled uniformly in the log of halo mass.

These residuals are combined to form the halo mass-conditioned property covariance, 
\begin{equation} \label{eq:r-estimator}
{\rm COV(s_a,s_b)} = A \sum\limits_{i=1}^{n}w_{i} ~ \delta s_{a,i} ~ \delta s_{b,i},
\end{equation}
with normalizing pre-factor 
\begin{equation} \label{eq:A}
A = {\sum\limits_{i=1}^{n}w_{i}} \ /\ \left[ {\left(\sum\limits _{i=1}^{n}w_{i}\right)^{2}-\sum\limits _{i=1}^{n}w_{i}^{2}}  \right].
\end{equation}
The corresponding property pair correlation coefficient is 
\begin{equation} \label{eq:rab}
r_{a, b} = \frac{{\rm COV(s_a,s_b)} }{\sigma_a \sigma_b}, 
\end{equation}
where $\sigma_a = \sqrt{ {\rm COV(s_a,s_a)}}$ is the intrinsic scatter in property $a$ at fixed halo mass, and similarly for property $b$.

Finally, we focus attention below on the {\sl normalized residuals} in logarithmic properties, defined as 
\begin{equation}\label{eq:normedResidual}
    \widehat{\delta} s_{a,i} \equiv \frac{\delta s_{a,i}}{\widetilde{\sigma}_a }   
    = \frac{ s_{a,i} - \widetilde{\pi}_a(M_i,z) } {\widetilde{\sigma}_a }  
\end{equation}


\begin{figure*}

    \includegraphics[width = 0.89\columnwidth]{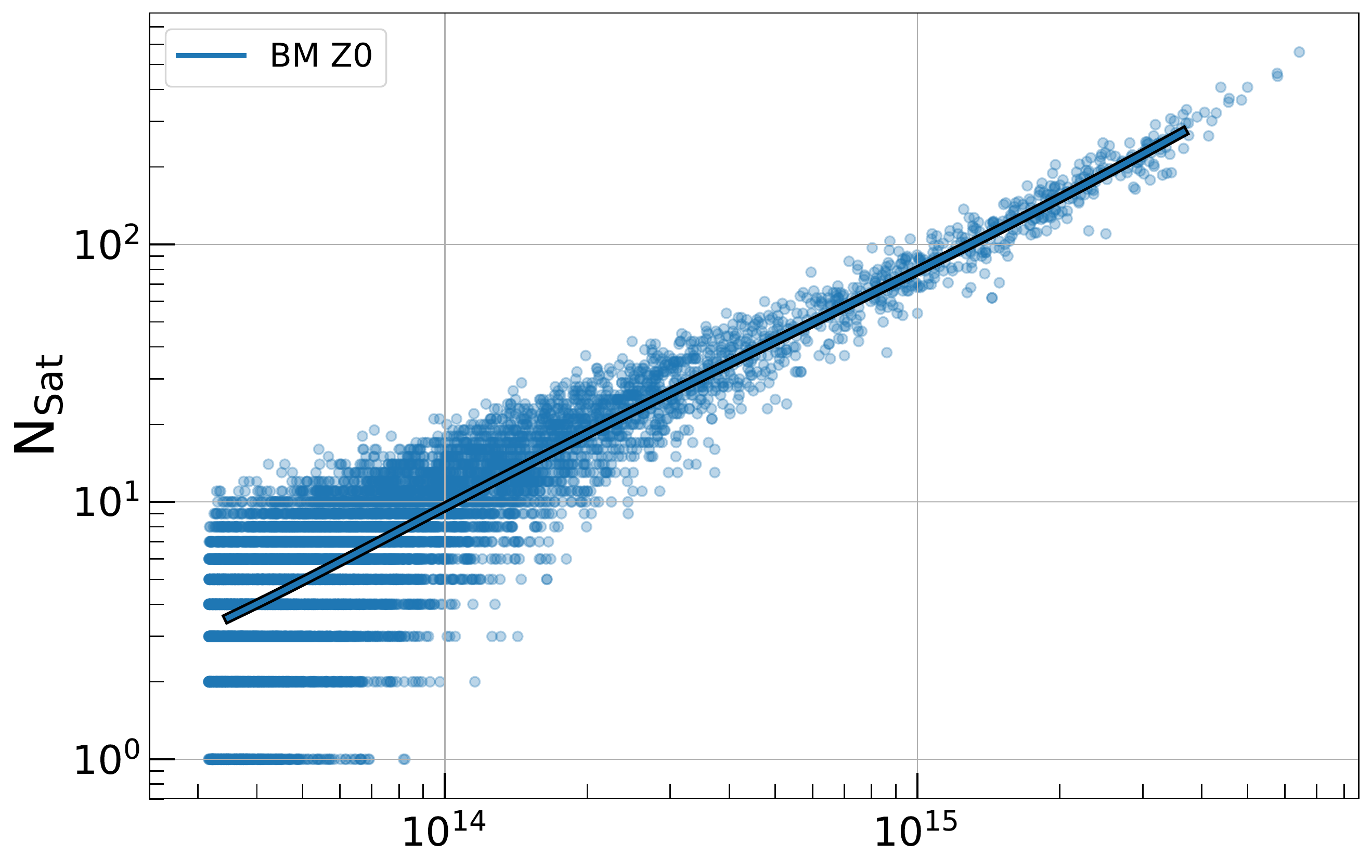}
    \includegraphics[width = 0.85\columnwidth]{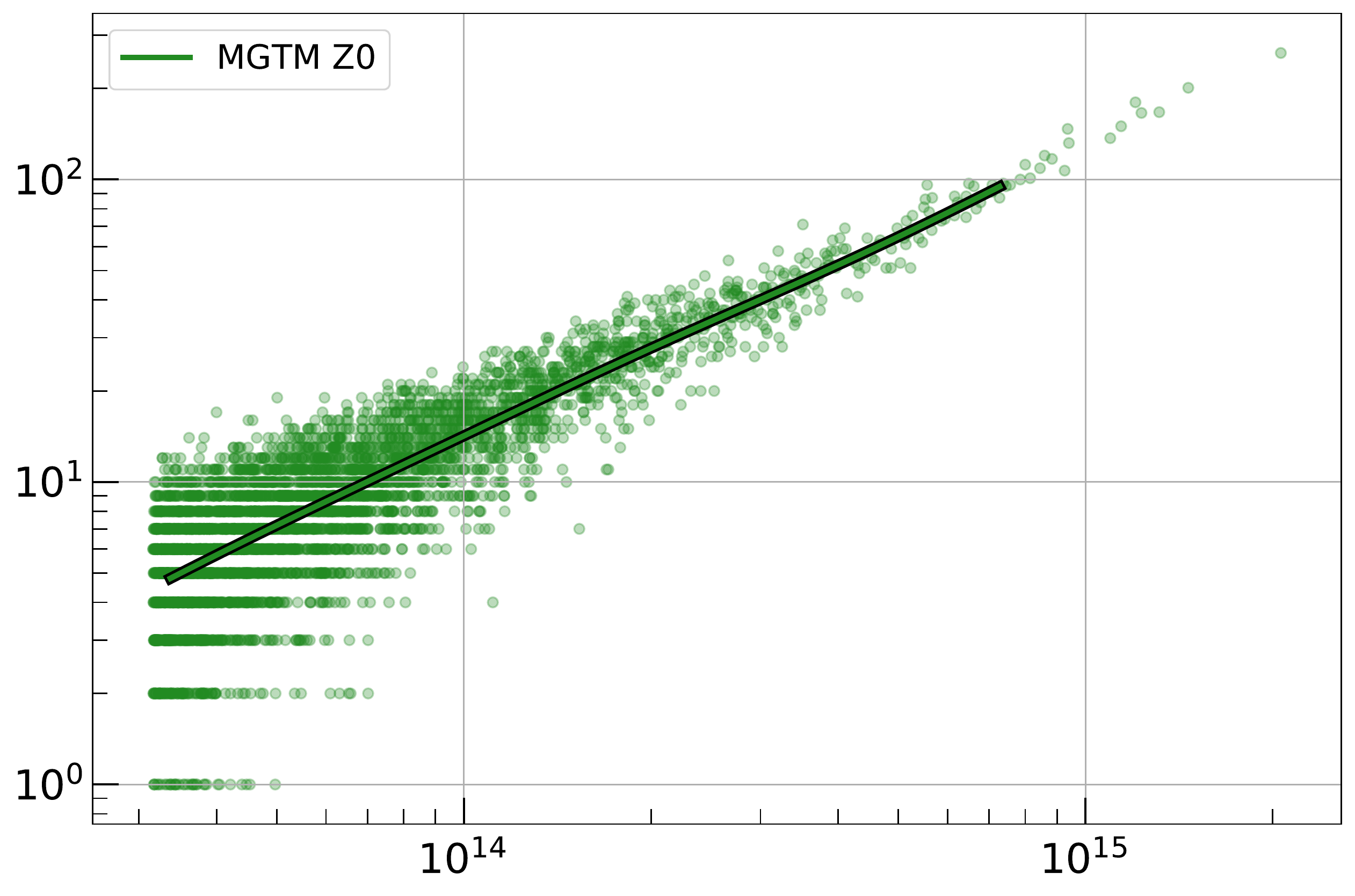}
    \includegraphics[width = 0.89\columnwidth]{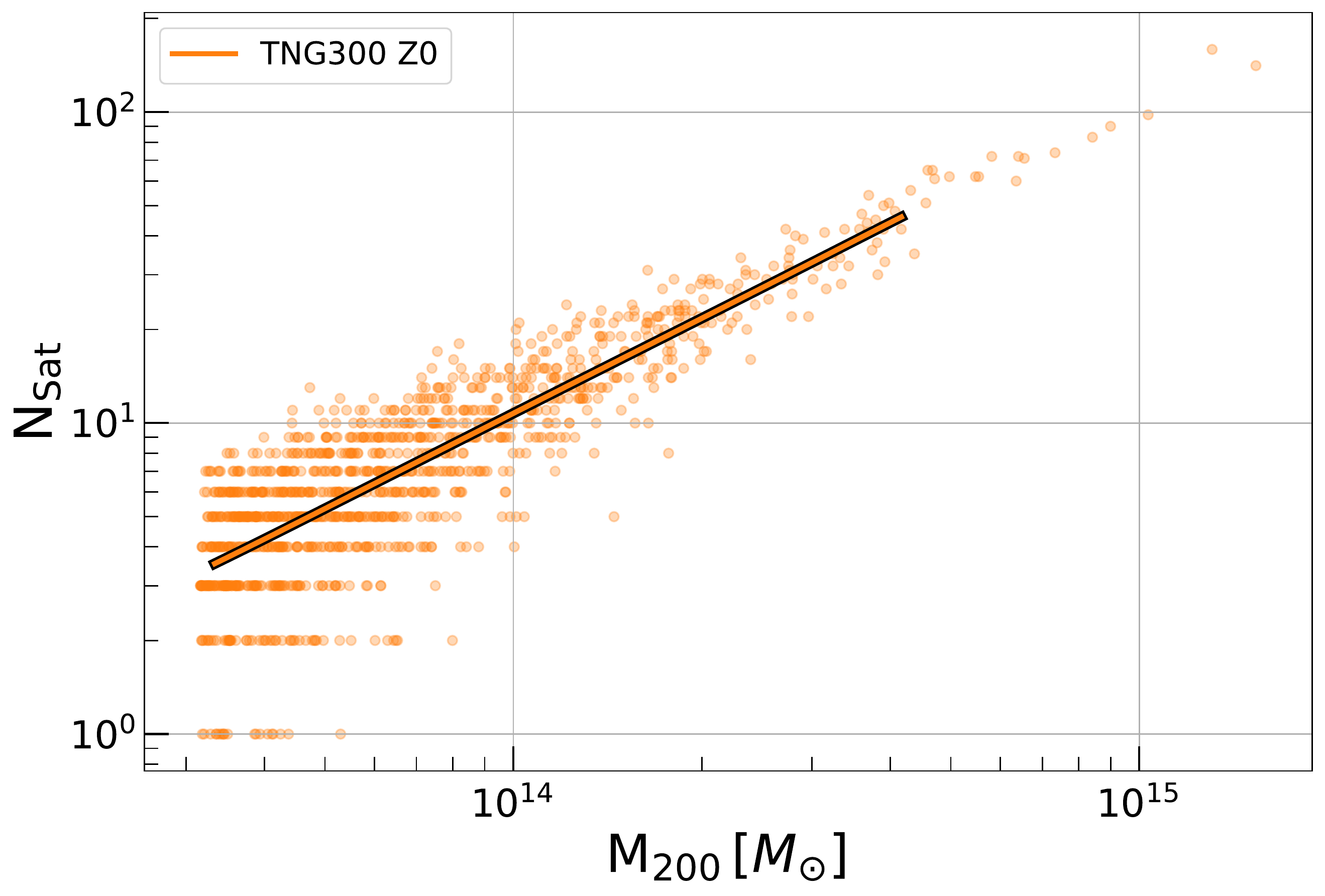}
    \includegraphics[width = 0.85\columnwidth]{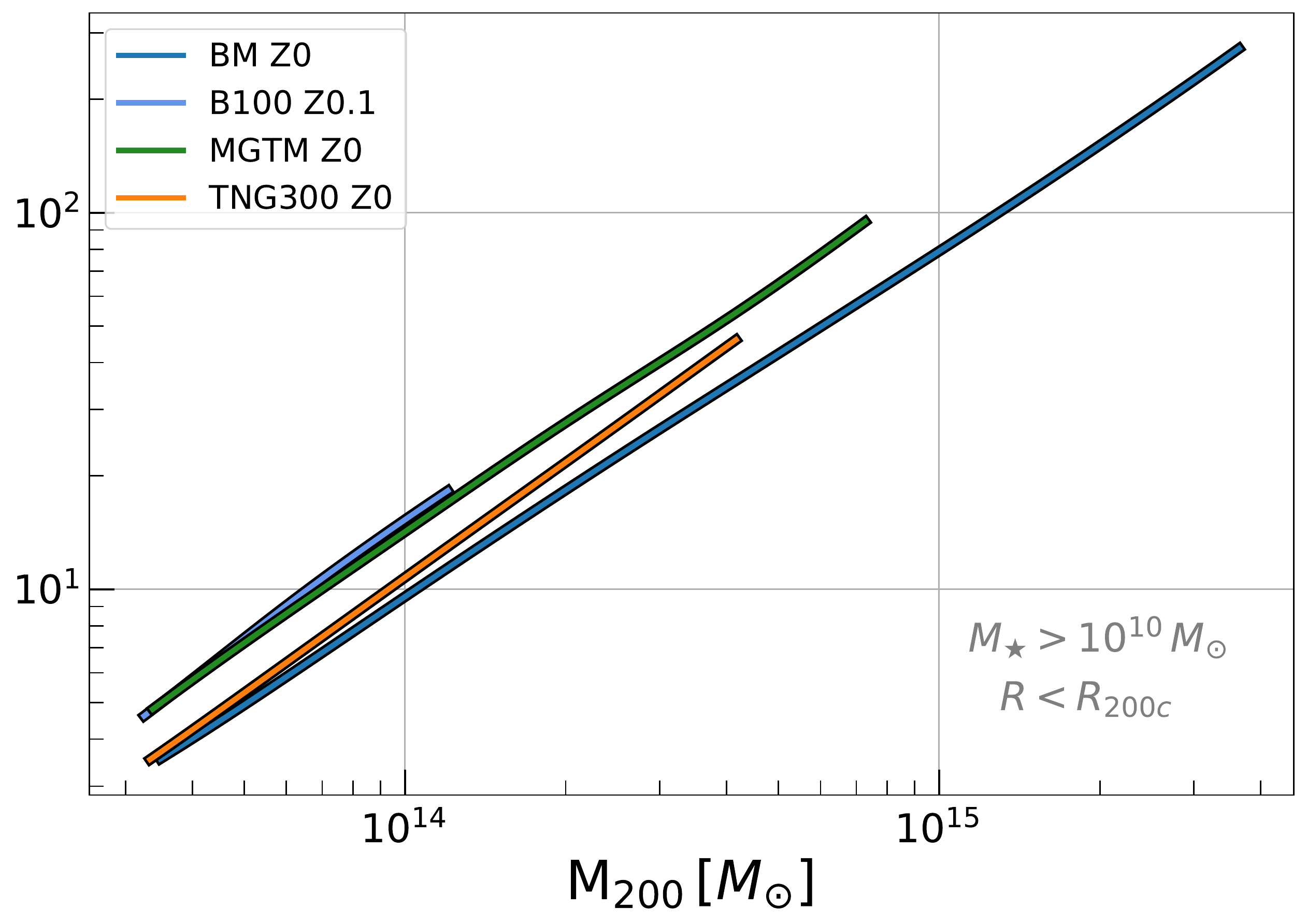}
    
    \caption{Number of stellar-mass limited satellite galaxies, $\Nsat$, as a function of total halo mass in the halo populations of each simulation at $z \! = \! 0$. Lines show mean LLR fits from the lower halo mass limit of $10^{13.5} \msun$ to an upper limit determined by the 20th most massive halo of each sample. The lower right panel compares mean behaviors and includes results from B100, the higher resolution run of the BAHAMAS simulation from the BM sample.
    }
    
    \label{fig:LLR Fits}
\end{figure*}

\section{LLR Scaling and Kernel Shapes of Stellar Properties}
\label{sec:LLR_and_PDF}

We begin by presenting $z = 0$ scaling behaviors of the three stellar properties with $\Mhalo$, finding qualitative agreement in many respects but also discrepancies in the details. Marginalizing over halo mass, we obtain estimates of the kernel shapes and find consistent support for $p(\ln \Nsat | M,z)$ to be negatively skewed while the kernel of total stellar mass is much closer to log-normal. We then examine off-diagonal elements of the mass-conditioned property covariance as motivation for exploring secondary selection by $\MstarBCG$, presented in section \ref{sec:SecondarySelection}.  

We provide tables of LLR fit parameters for these properties in Appendix~\ref{Appendix:StellarLLRfits}.

\subsection{Present-epoch Stellar Property Scaling Relations }
\label{sec:Scaling-Relations}

\begin{figure}
    \includegraphics[width = \columnwidth]{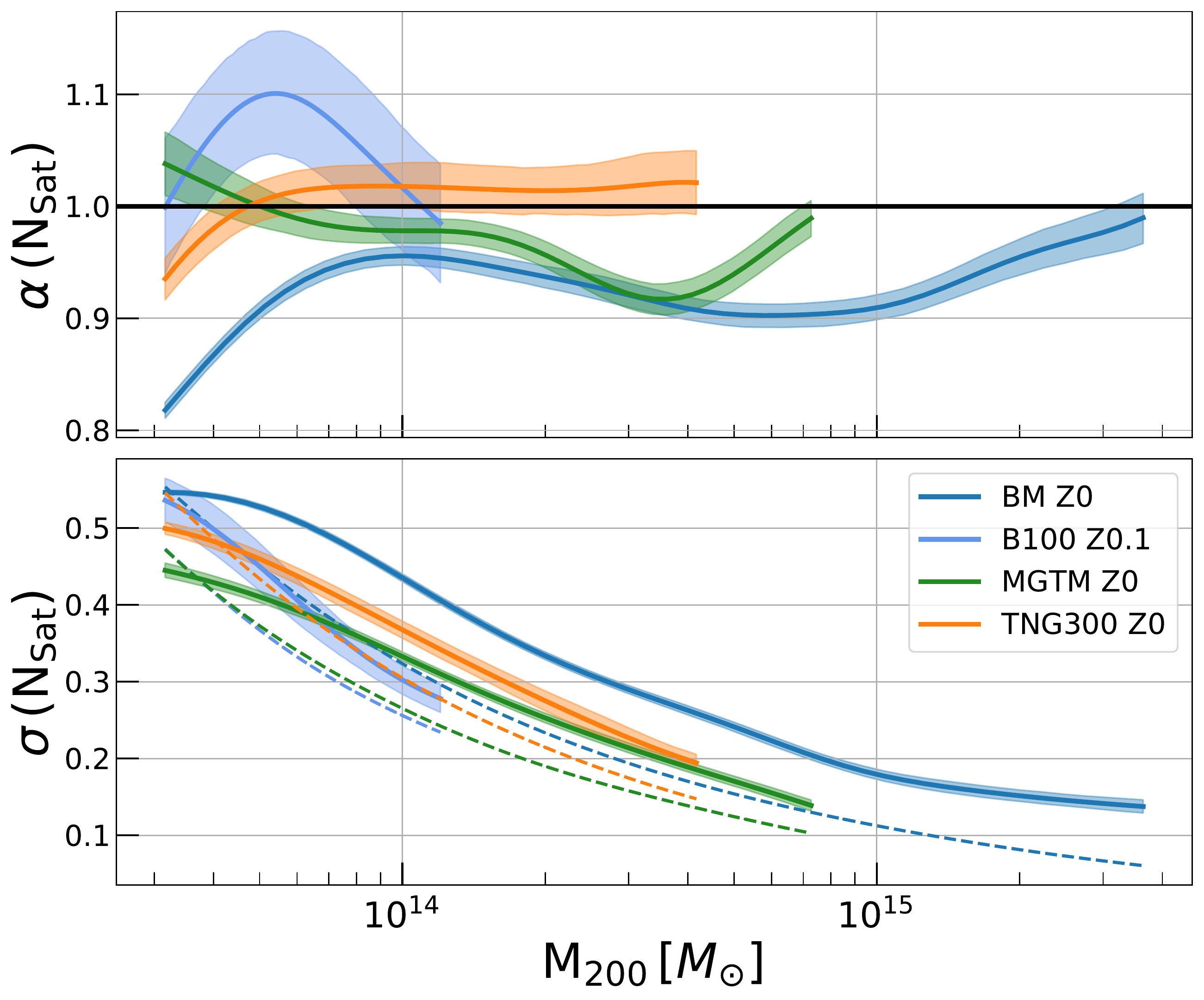}
    \vspace{-0.4 cm}
    \caption{LLR slope, $\alpha$ (top), and natural logarithmic scatter, $\sigma$ (bottom), of the $\Nsat$ scaling with halo mass for the $z = 0$ populations shown in Figure~\ref{fig:LLR Fits}. Dashed lines show the Poisson-expected fractional scatter, $\langle  \Nsat \, | \, M,z  \rangle^{-1/2}$ obtained from the mean satellite number as a function of halo mass. B100, the high-resolution 140 Mpc BAHAMAS run, exhibits smaller scatter over the mass range accessible within that volume, reducing the tension with the other simulation outcomes. }
    \label{fig:Params}
\end{figure}

Figure~\ref{fig:LLR Fits} shows the $z=0$ scalings of satellite galaxy counts, $\Nsat$, with halo mass for the four simulation populations.  Solid lines show the LLR mean behaviors, which are inter-compared in the bottom right panel.  As anticipated from the space densities of Figure~\ref{fig:Stellar_Mass_Function}, the MGTM simulation has a higher normalization compared to the other two simulations. Numerical resolution is an important factor; the B100 simulation has $58\%$ more galaxies per halo compared to the BM model, whose dark matter particle mass is $\sim9$ times larger than that of B100. A shift of similar order of magnitude is found for the TNG300 suite. TNG300-1, the highest resolution run, has $86\%$ more galaxies per halo than TNG300-2, whose dark matter particle mass is $8$ times that of the TNG300-1 run. The quoted values are the mean shifts for halos with $\Mhalo > 10^{13.8}\msun$.

The local slope and scatter for $\Nsat$ as a function of halo mass are presented in Figure~\ref{fig:Params}, with shaded regions showing $68\%$ confidence intervals from bootstrap resampling. The local slopes lie close to the simplest self-similar expectation of unity, with values in the range $0.90 - 1.05$.  

Above a halo mass of $\sim 10^{14} \msol$, where the mean occupation numbers are ten or larger, the scatter in $\ln \Nsat$ declines with mass in a manner that roughly follows Poisson expectations, $\langle  \Nsat \, | \, M,z  \rangle^{-1/2}$, shown as dashed lines. The scatter in the BM model is significantly larger than this, but the low stellar mass resolution of this simulation appears to be adding extra variance. Evidence for this is given by the B100 simulation result. The scatter in $\ln \Nsat$ near $10^{14} \msol$ drops from $44 \%$ (BM) to $30\%$ (B100), a factor of 2 decline in variance that brings B100 much closer to the scatter values seen in TNG300 and MGTM populations. We confirm that a similar increase in variance occurs for the lower resolution TNG300-2 simulation.  At $10^{14}\msol$, the scatter in $\ln \Nsat$ is $48\%$ in the TNG300-2 run, compared to $37 \%$ for TNG300-1.

\begin{figure*}
    \includegraphics[width =  \columnwidth]{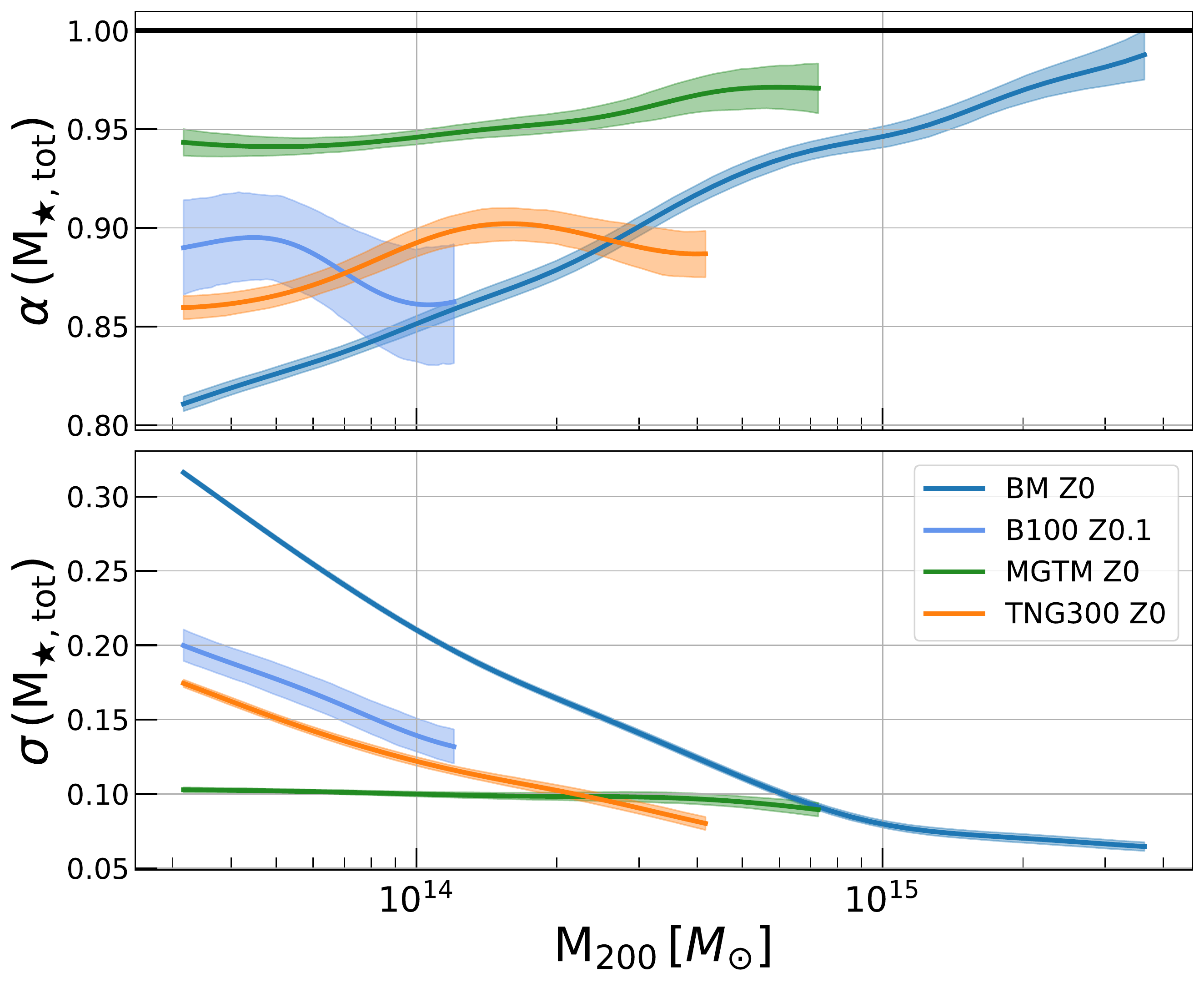}
    \includegraphics[width = \columnwidth]{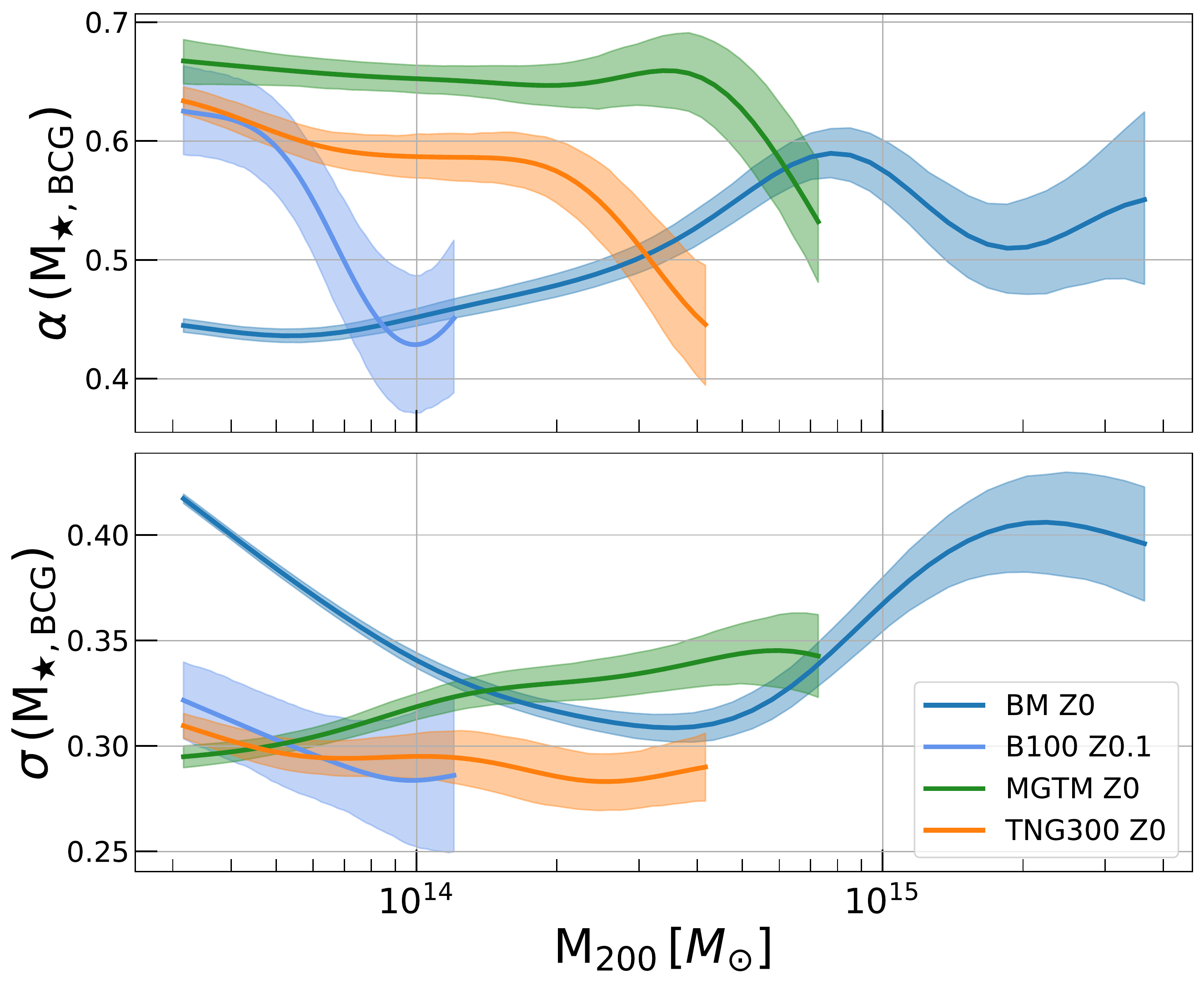}
    \caption{LLR parameters (slope, $\alpha$, and natural log scatter $\sigma$) for the $\Mstar-\Mhalo$ (left) and $\MstarBCG-\Mhalo$ (right) scaling relations for all simulations at $z = 0$.  The underlying data for $\Mstar$ and $\MstarBCG$ in each simulation are shown in Figure~\ref{fig:MStar_Z0_allsims} and Figure~\ref{fig:LLR_MStar_BCG100}, respectively. }
    \label{fig:LLR_Stellar_params}
\end{figure*}

For the sake of economy, we show the raw data for the scalings of central and total stellar mass measures ($\MstarBCG$ and $\Mstar$) with halo mass at $z=0$ in Appendix~\ref{Appendix:StellarLLRfits}.  The mean stellar mass fraction within $\Rtwoh$ of a $10^{14} \msol$ halo ranges from a high of $0.025$ in MGTM to a low of $0.013$ in TNG300, with Bahamas intermediate. These population values lie within the range of individual cluster stellar mass fractions inferred from small observational samples (see Figure~3 from \citet{Tremmel2019Romulusc}, and also work from \citet{Pillepich2018FirstGalaxies} for additional comparisons). The stellar mass fractions are fairly stable with redshift in all models (see tables in Appendix \ref{Appendix:All_Params}) but its absolute value is again sensitive to resolution. Mean stellar masses in the higher resolution B100 run are elevated by $\sim 25\%$ relative to those of BM. 

The slope and scatter of $\Mstar$ and $\MstarBCG$ derived from the simulations are compared in Figure~\ref{fig:LLR_Stellar_params}. For the $\Mstar-\Mhalo$ relation, all simulations show sub-linear scaling, but approach self-similarity ($\alpha = 1$) at high halo mass, a result already published for the BM simulation \citep{Farahi2018LocalizedCovariance}.

An observational study of 21 nearby clusters using an $R_{500c}$ scale finds a somewhat shallower slope for the total stellar mass scaling, $\alpha = 0.6 \pm 0.1$ \citep{Kravtsov:2018}. The tension in slope may reflect physical or numerical deficiencies in the simulations or it may reflect systematic differences in the two quantities being compared --- true quantities in simulations versus those inferred from multi-band photometry and other observations.  Future work using synthetic observations to analyze the simulation expectations directly in the space of survey observables, including intracluster light, is needed to explore this discrepancy in more detail.  

For the central galaxy scaling, all simulations display slopes that are sub-linear, lying in the range $[0.45, \, 0.65]$, a range that encompasses values derived from the UniverseMachine semi-analytic models \citep{Bradshaw2019}.

Empirical studies of the BCG stellar mass slope find contradictory results. Some are in agreement with our results \citep{GoldenMarxMiller2018} while some are shallower \citep{Zhang2016, Kravtsov:2018, Mulroy2019LoCuSS}. The seemingly inconsistent results can be mitigated by redshift evolution of the slope \citep{GoldenMarxMiller2019}. 
We note, as quantified by \citet{Pillepich2018FirstGalaxies}, that both the slope and scatter of the stellar mass-halo mass (SMHM) relations depend sensitively on the operational definition of stellar mass. For example, different choices of aperture for the $\MstarBCG$ calculation results in slopes varying from $\sim0.5$ to $0.75$ in TNG300, and also in TNG100, the $100 \,\rm Mpc$ box simulation from the TNG suite \citep[see table 4 of][]{Pillepich2018FirstGalaxies}.

The scatter in the BCG stellar mass relation lies in the range $0.3 - 0.4$ for all simulations over the entire mass range. This range encompasses the empirical value of $0.39 \pm 0.07$ found by \citet{Kravtsov:2018} but it is lower than the value of $0.5$ derived from the UniverseMachine analysis of \citet{Bradshaw2019}. 

Consistency with \citet{Bradshaw2019} is found for the scatter in total stellar mass. The simulations also find that the scatter in $\Mstar$ is below $10 \%$ for halos above $\sim 10^{14.3} \msol$. This supports the observational finding that total $K$- band luminosity ($\propto \Mstar$) is a tight mass proxy for high mass clusters \citep{Chiu2016, Mulroy2014LkMwl, Mulroy2019LoCuSS} selected by X-ray flux.

We note that the MGTM simulation finds no trend of scatter with halo mass while the other two simulations, along with the UniverseMachine study, find that the scatter in total stellar mass monotonically decreases with increasing halo mass. In MGTM, the constant scatter in total stellar mass arises from compensating effects of central and satellite galaxy contributions. At low halo masses, the scatter in the central galaxy stellar mass is lower than at high masses, but the scatter in satellite count is higher, leaving the scatter in total stellar mass nearly constant.  

\subsection{Kernel shapes of normalized residuals}
\label{sec:LLR_PDFs}

\begin{figure}
    \includegraphics[width = \columnwidth]{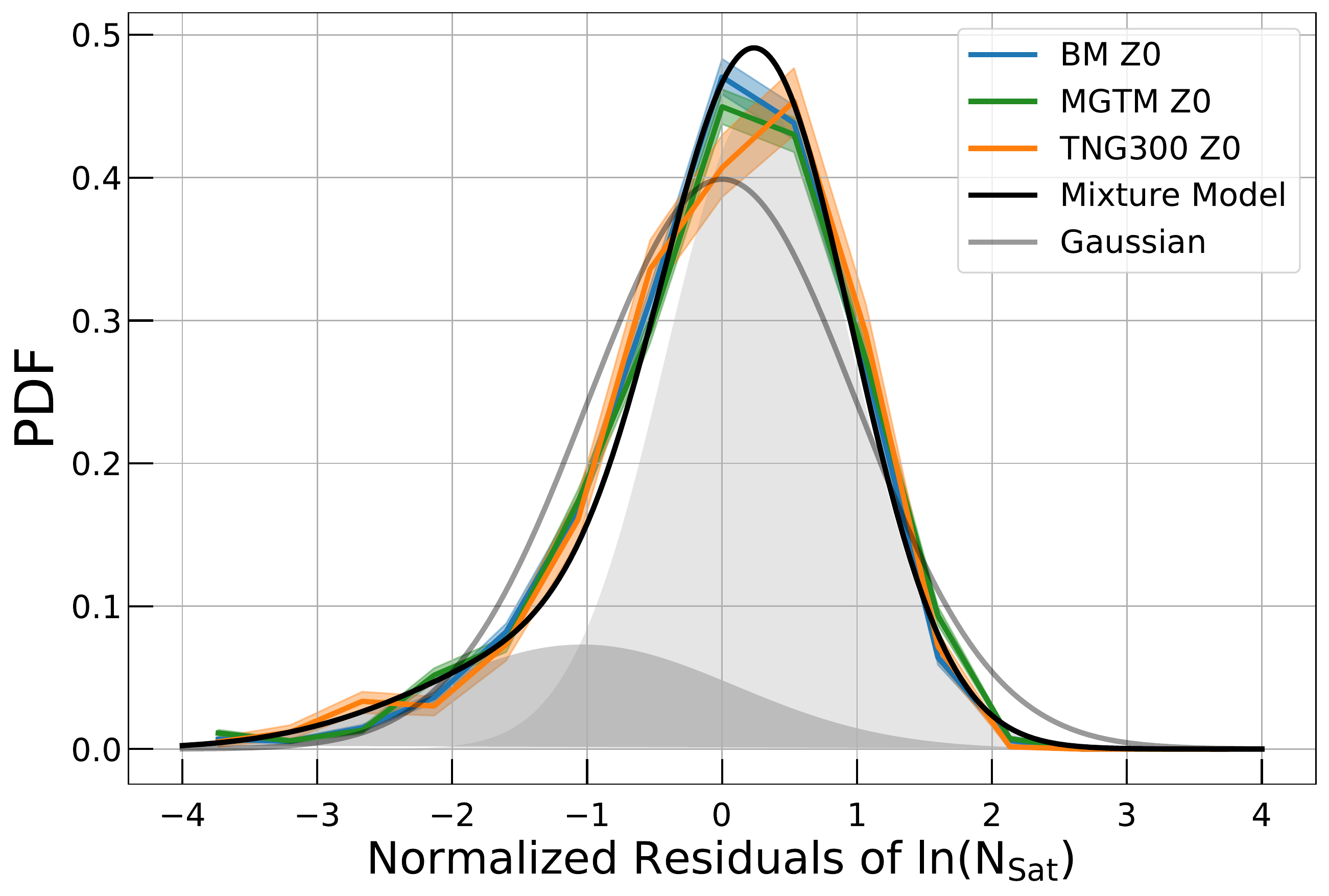}
    \caption{Mean mass-conditioned kernel --- PDF of the normalized residuals, equation~(\ref{eq:normedResidual}) --- for $\ln \Nsat$ at $z=0$ using a lower halo mass limit of $10^{13.8} \msun$ for BM (to avoid discreteness features from the lowest $\Nsat$ halos, see text) and $10^{13.5} \msun$ for the other two simulations.
    A Gaussian mixture model (bold black line, with two components shaded) provides a much better fit than a single Gaussian (grey line). GMM fit parameters for the combined $z \le 1$ samples across all simulations are given in Table~\ref{table:Uber_Sample_values} and individual sample values in Table~\ref{table:Skew_and_GMM_Params}.  
    }
\label{fig:LLR_PDF_NSat}
\end{figure}

Forward modeling counts of massive halos as a function of an observable property, $s$, is sensitive to the assumed shape of the conditional kernel, $p (s \, | \, M,z) $ \citep[\eg][]{Shaw2010Non-GaussianRelations,Erickson2011ProjectionModel, Costanzi2019SDSSmethods}. Moving beyond the mild tensions in stellar property variance seen above, we seek here to test whether consistent kernel shapes emerge from the different simulation treatments.  

While the complex coupling of physical and numerical factors makes it difficult to prove that kernel shapes should be preserved under transformations in the simulation control space, we take a purely empirical approach and simply ask whether consistent forms emerge.  While lack of consistency in kernel shapes would be troubling, we caution that consistency is a necessary, but not sufficient, condition that the simulations have converged to the true form.  

\subsubsection{Satellite galaxy count kernel}
\label{sec:NsatPDF}

At halo masses near our cutoff value of $10^{13.5} \msol$, the mean occupation number is less than ten and the fractional scatter is larger than $\sim 0.4$. The odds of encountering a halo with $\Nsat=0$ is therefore non-negligible. The BM simulation has 72 such satellite-empty halos, a higher frequency than MGTM and TNG300, both of which have 5. For BM, we therefore introduce a cutoff of $10^{13.8} \msol$ when we extract the normalized residual kernel.  Above this modified mass, we find only two halos with zero satellites. The skewness and other statistics are not strongly affected by this choice.\footnote{An alternative approach where we smooth the occupation count by adding random deviates in the range $\pm 0.5$ to $\Nsat$ produces a similar effect of removing the discreteness feature.} In all simulations, the small number of empty halos are removed before producing LLR statistics.

The PDF of the normalized, mass-conditioned residuals, equation~(\ref{eq:normedResidual}), in $\ln \Nsat$, defines the kernel shown for the $z=0$ halo populations in Figure \ref{fig:LLR_PDF_NSat}.  
Confidence bands are constructed from $1000$ bootstrap realizations of the samples, through computing $1\sigma$ confidence intervals for the probability in each normalized residual bin.

The three simulation populations exhibit very similar kernel shapes evident in their overlapping contours. The standard normal, $G(x,1)$\footnote{Using notation $G(x-\mu, \sigma) \, dx = \frac{1}{\sqrt{2 \pi} \sigma} \exp (-\frac{(x-\mu)^2}{2\sigma^2}) \, dx $.}, shown as the grey line, is not a good description of this left-leaning distribution, which has a normalized skewness value of $\gamma \equiv {\rm E} [\left(\frac{x-\mu}{\sigma}\right)^3] = -0.9$.

A natural extension, one that enables efficient calculation with the analytical framework of \citet{Evrard2014Statistics}, is a two-component Gaussian mixture, 
\begin{equation} \label{eqn:GMM}
    \Pr (x) = f_1\, G(x - \mu_1, \sigma_1) + (1 - f_1)\, G(x -\mu_2, \sigma_2) ,
\end{equation}
where $f_1$ is the weight of a Gaussian with mean, $\mu_1,$ and standard deviation, $\sigma_1$, and $1-f_1$ the weight of a second Gaussian with mean, $\mu_2$, and standard deviation, $\sigma_2$.

The GMM result, shown as the bold black line with individual components as gray, shaded regions, provides a good fit to the reduced population statistics of all three simulations. Performing a Bayesian Information Criterion (BIC) test confirms that a two component model is the most optimal at replicating this distribution. Increasing the number of components in the GMM is not supported by the BIC criteria and adds no significant improvement in fitting the residuals. We considered an alternative fit using an Edgeworth expansion \citep{Shaw2010Non-GaussianRelations} but the mixture model is preferred because, unlike the Edgeworth expansion, it guarantees a positive-definite probability distribution. 

The GMM fit parameters and skewness measures are presented in section \ref{sec:redshift}, where we explore the dependence of this shape on redshift. The dominant component of the GMM, representing 80\% of the halo population, is centered near 0.3 and has normalized variance of $(0.68)^2$, smaller than the complete population by a factor of two. The remaining one-fifth of the population is centered near $-1$ with variance $(1.13)^2$. Given the important role of variance in cluster cosmological applications \citep[\eg][]{Allen2011CosmologicalClusters}, empirical methods to separate these populations could yield significant benefits.

\subsubsection{Total and central stellar mass kernels} \label{sec:MstarPDFs} 

\begin{figure}
  \includegraphics[width = \columnwidth]{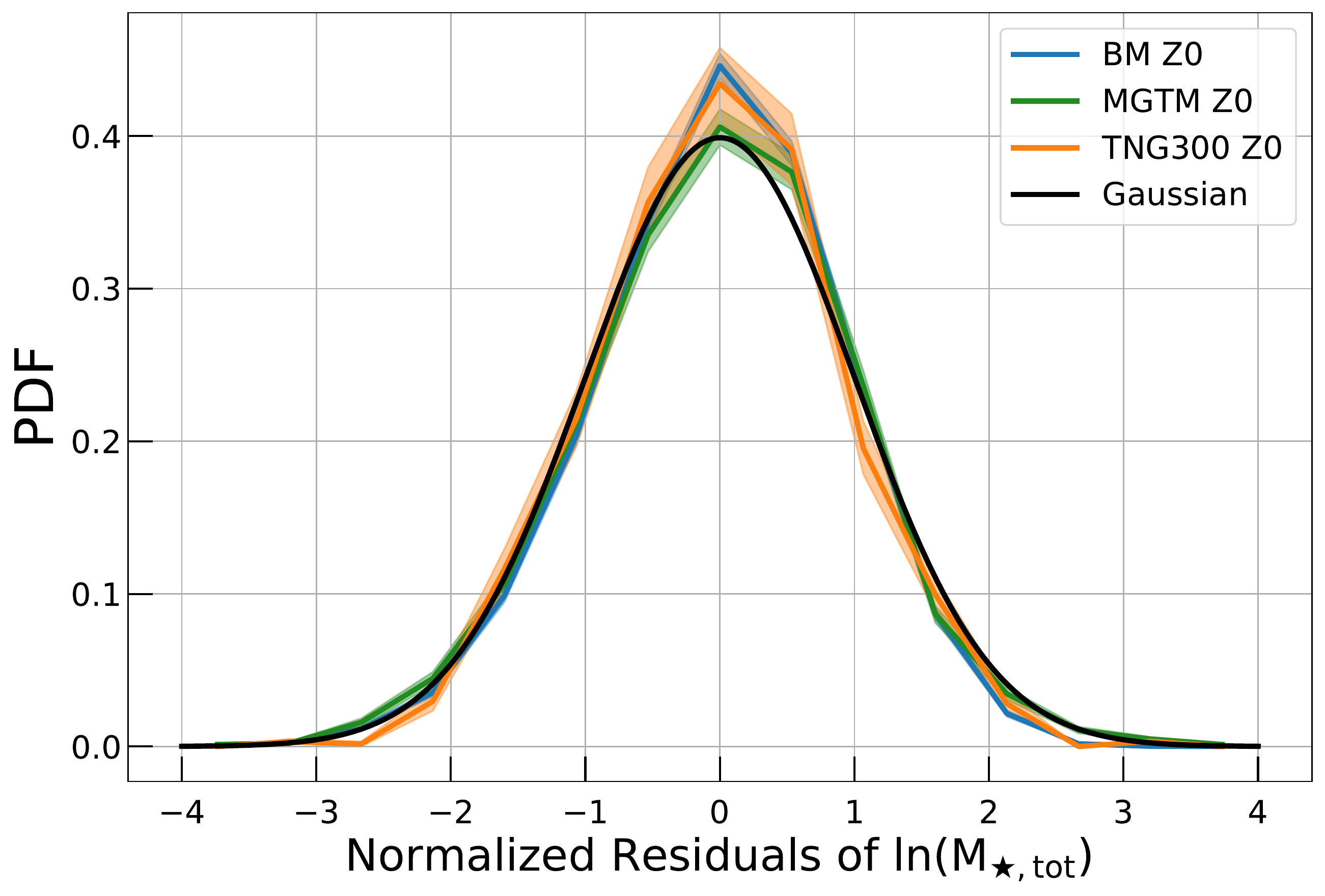}
  \includegraphics[width = \columnwidth]{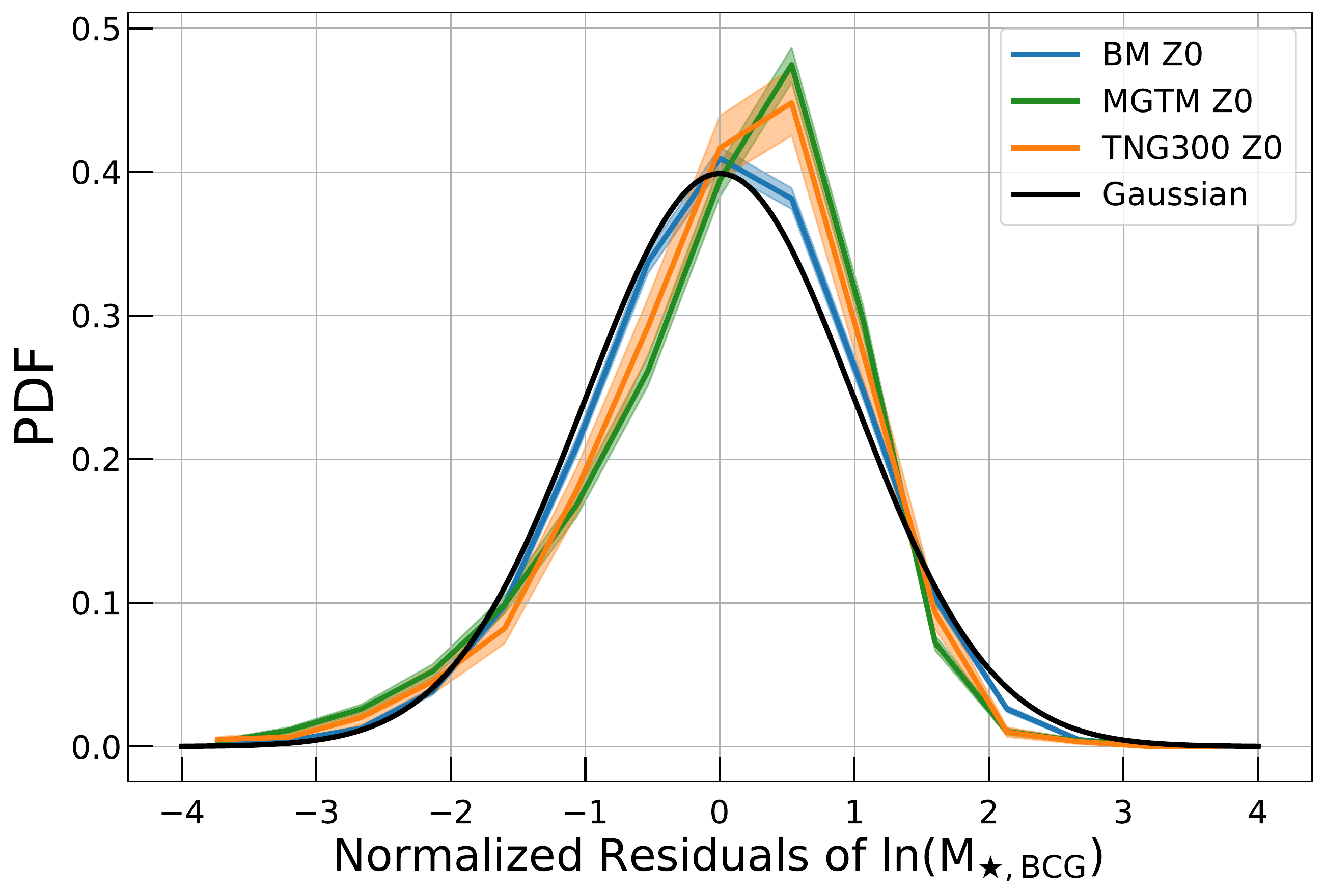}
  \caption{Mean mass-conditioned kernels for $\ln \mstar$ (top) and $\ln\MstarBCG$ (bottom) at $z=0$ for halos with mass above $10^{13.5} \msun$ in all simulations. The solid black line shows the standard normal distribution.
  }
\label{fig:LLR_PDF_MStar}
\end{figure}

Unlike the skewed residuals in satellite galaxy count, the residuals in total stellar mass within $\Rtwoh$, shown in the top panel of Figure~\ref{fig:LLR_PDF_MStar}, are much closer to Gaussian in shape.  The skewness is consistent with zero for TNG300 and MGTM, and the BM value of $\gamma = -0.24 \pm 0.02$ at $z=0$ value is much smaller than the $-0.9$ value displayed by the satellite galaxy counts. The Gaussianity of the residuals in $\ln \Mstar$ for TNG300 and MGTM confirm the same result found previously for the BM sample only by  \citet{Farahi2018LocalizedCovariance}. 

When a system is subject to many random multiplicative factors, the central limit theorem argues for a kernel shape that is log-normal, or Gaussian in log-space \citep{Adams1996AClouds}. We postulate that the formation of individual stellar particles in cosmological simulations is dictated by such multiplicative factors, while the aggregated effort to form many individual star particles into a single galaxy entails fewer effective degrees of freedom and so can deviate more from log-normality.

The kernel shape for central galaxy stellar mass, shown in the bottom panel of Figure~\ref{fig:LLR_PDF_MStar}, is also negatively skewed. The shapes for TNG300 and MGTM show good agreement, with $\gamma \simeq -0.8$, while BM tends closer to log-normality, with $\gamma \simeq -0.3$.  Unfortunately, the shape of the kernel from the higher resolution B100 run is not well defined because of the much smaller sample of halos available in that simulation. Given the low stellar particle resolution of BM, we tentatively promote the $-0.8$ value as more likely, but defer more careful analysis to future simulations with higher resolution.  


\subsection{Mass-conditioned Correlations in Stellar Properties} \label{sec:Correlations} 

\begin{figure}
    \includegraphics[width = \columnwidth]{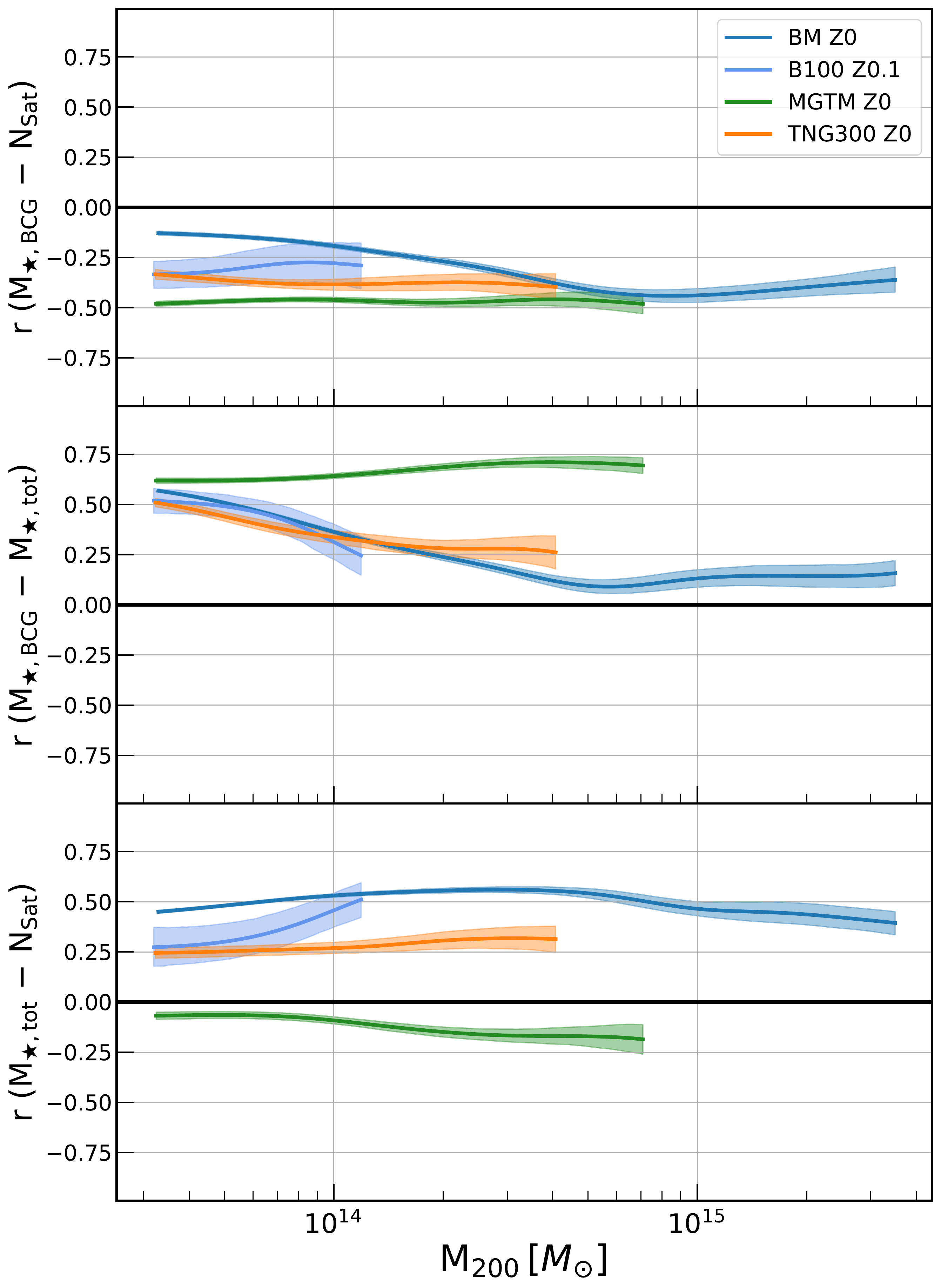}
    \caption{Correlation coefficients, at fixed $\mhalo$, of $\Nsat$ and $\MstarBCG$ (top),  $\Mstar$ and $\MstarBCG$ (middle), and $\Mstar$ and $\Nsat$ (bottom).}
    \label{fig:Correlations}
\end{figure}

There are physical reasons to anticipate correlations between elements of the stellar property vector conditioned on total halo mass. The tidal disruption and accretion of satellite galaxy stellar material onto a halo's central galaxy is a well-known mechanism for producing an anti-correlation between $\Nsat$ and $\MstarBCG$.

It has long been postulated that central galaxies in groups and clusters increase their stellar mass via mergers \citep[\eg][]{TremaineRichstone1977}. Recent high resolution hydrodynamical simulations \citep{Cui2014diffuselightsims, Hydrangea2017, ROMULUSC2019, Bahe2019Satellites} and semi-analytic models \citep{Croton2006TheGalaxies, DeLucia:2007, Bradshaw2019} confirm the growth of centrals at the expense of satellites.

The Hydrangea suite of simulations \citep{Bahe2019Satellites} find that, across the mass range we study here, the majority of galaxies accreted at $z <2$ do not survive to the present. While some caution is required, as even the highest resolution studies may not yet be converged \citep[][but see appendix of \citet{Bahe2019Satellites}]{vandenBosch2018}, observational studies also support the growth of central galaxies and intracluster light over time \citep{Zhang2016, Tang2018diffuselightsims, Zhang2019ICL}.  In the Illustris and TNG300 simulations, galaxies with stellar mass greater than a few $10^{11} \msol$ are mostly made of ex-situ stars accumulated through the merging and accretion of material from other galaxies \citep[see][and references therein]{Rodriguez2016StellarMassAssemblyIllustris, Pillepich2018FirstGalaxies}.

As a result of this dynamical processing, all simulations display mildly anti-correlated behavior between $\Nsat$ and $\MstarBCG$ at $z=0$.  The top panel of Figure~\ref{fig:Correlations} shows that correlation coefficient, $r \simeq -0.4$, is nearly independent of halo mass in all realizations.  The smaller values of the BM correlation at low halo masses are driven by the larger variance in  $\ln \Nsat$ seen in that run.  The high resolution B100 model, with lower scatter in $\Nsat$ compared to BM, yields a larger correlation coefficient consistent with the values seen in the TNG300 and MGTM.

The correlation coefficients of the remaining pairs are sensitive to the central galaxy stellar mass statistics, particularly the normalization and scatter, and the MGTM central galaxy population is extreme in both measures.  As a result, the mass-conditioned correlation coefficients of $\Mstar$ and $\MstarBCG$, as well as $\Nsat$ and $\Mstar$ (middle and lower panels of Figure~\ref{fig:Correlations}, respectively) show behaviors for which MGTM differs from the others.

For the pairing of $\Mstar$ and $\MstarBCG$ at fixed halo mass, stronger correlation is seen in MGTM because that simulation has a very low scatter in total stellar mass, making the role of central galaxy variations more prominent. The central galaxies in MGTM also contribute the largest fraction of total stellar mass. At $10^{14} \msun$, the stellar mass fraction of the central galaxy, $\frac{\MstarBCG}{\Mstar}$, in the simulations are 0.42 (MGTM), 0.40 (TNG300), 0.35 (B100), and 0.27 (BM).

One would reasonably anticipate a positive correlation between $\Nsat$ and $\Mstar$ at fixed $\Mhalo$, as halos with more satellite galaxies should also have a larger total stellar mass. The bottom panel of Figure~\ref{fig:Correlations} shows that the BM, B100, and TNG300 simulations follow that expectation, albeit with somewhat different magnitudes between $0.25$ and $0.5$. The MGTM simulation, however, exhibits a weak {\sl anti-correlation} between these two properties.  This counterintuitive result is explained by the non-Gaussian scatter in the full space of residuals that we examine next.

\subsubsection{Non-Gaussian features in residual space} \label{sec:nonGaussianity}

\begin{figure*}
    \includegraphics[width = 0.68\columnwidth]{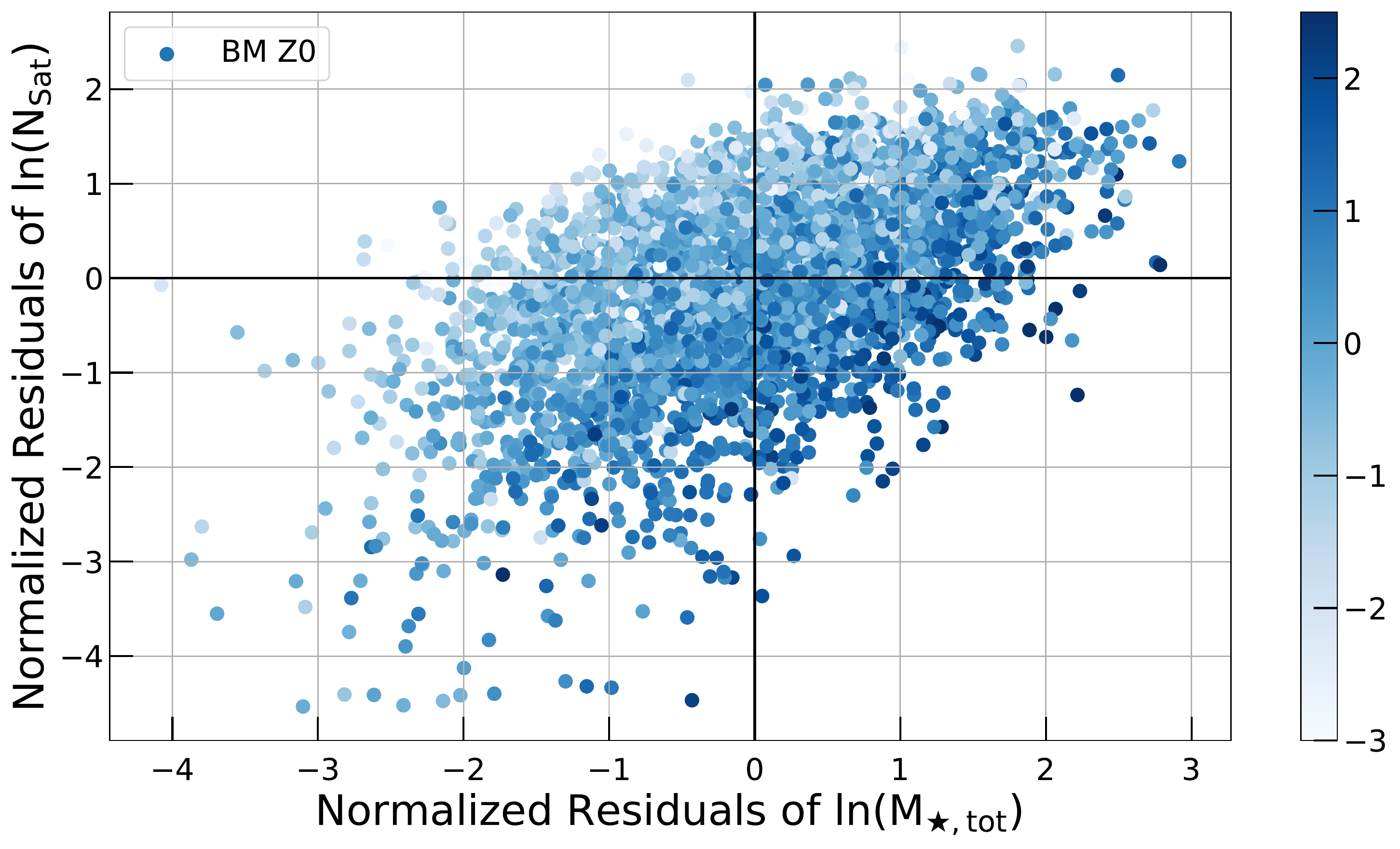}
    \includegraphics[width = 0.65\columnwidth]{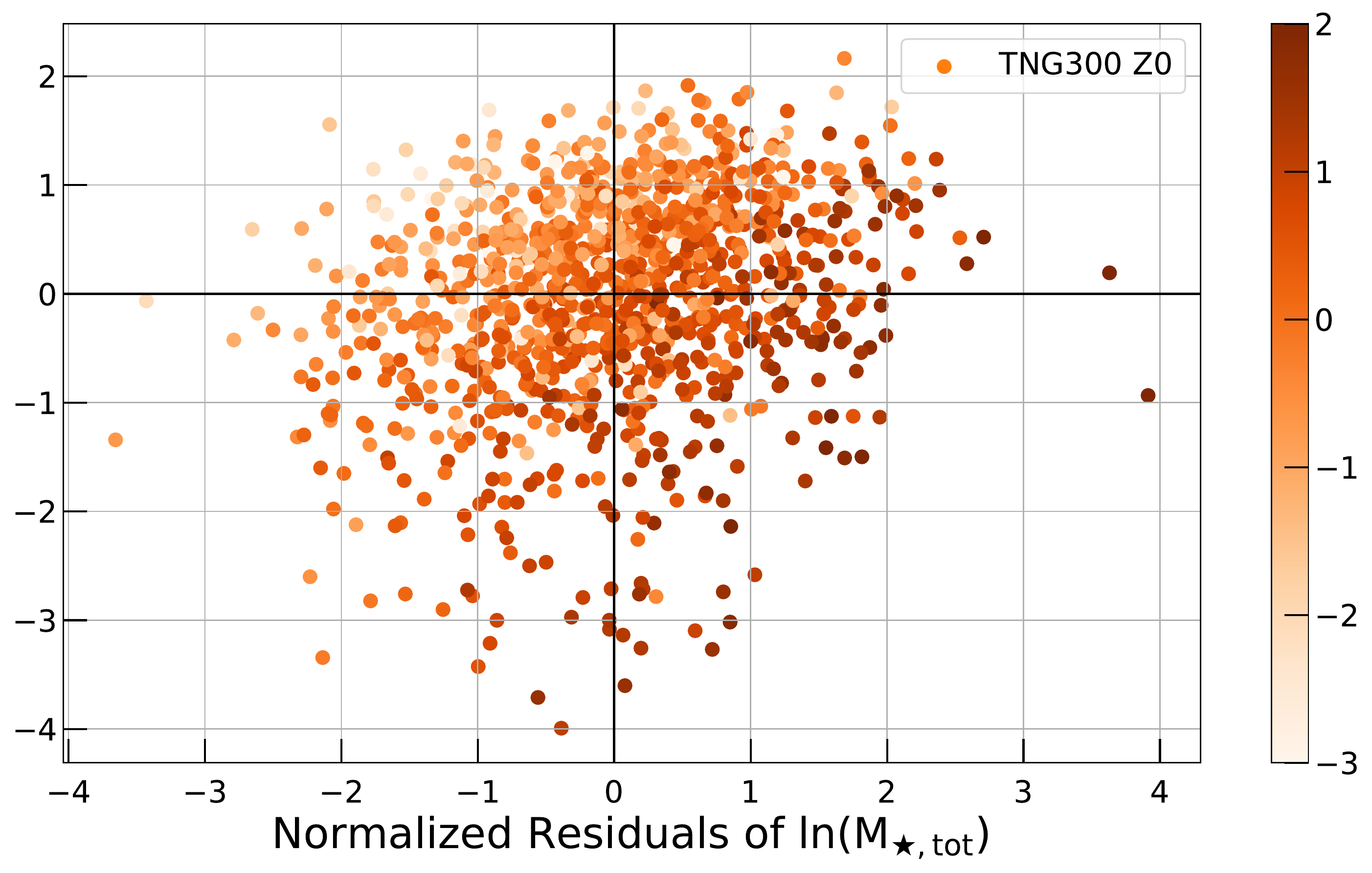}
    \includegraphics[width = 0.68\columnwidth]{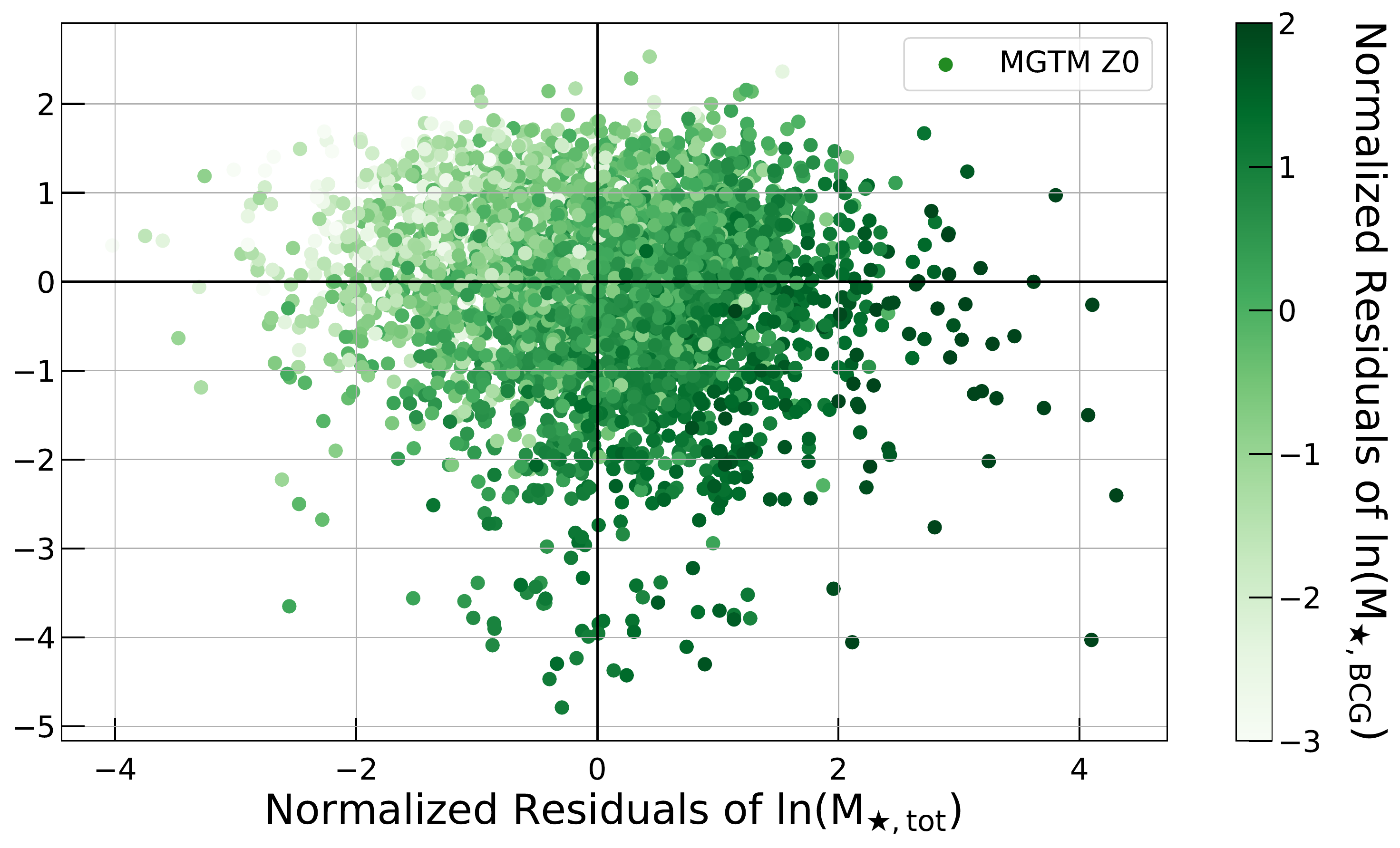}
    \caption{Normalized residuals of $\Nsat$ and $\Mstar$ colored by the $\MstarBCG$ residuals for all halos at $z \! = \! 0$. The BM sample uses a halo mass threshold of $10^{13.8}\msun$ while the others use $10^{13.5}\msun$.}
    \label{fig:2DResiduals_Mstar_Nsat}
\end{figure*}

Except for total stellar mass, the one-dimensional kernels in Figures~\ref{fig:LLR_PDF_NSat} and \ref{fig:LLR_PDF_MStar} display non-Gaussian features.  To expand the view into the full 3-D space of residuals, Figure \ref{fig:2DResiduals_Mstar_Nsat} shows normalized residuals in $\Nsat$ and $\Mstar$ for each halo colored by its $\MstarBCG$ residuals.  

It is evident by eye that the three simulations exhibit somewhat different forms. The BM and TNG300 residuals show positive correlation in $\Nsat$ and $\Mstar$ with shapes that are approximately elliptical.  In particular, all halos with extreme low satellite galaxy counts (given a halo mass) also have low total stellar mass.  The lower right quadrant, corresponding to halos with low numbers of satellites but high total stellar mass, is relatively vacant. 

The MGTM residuals, in contrast, include a few outlying points in this lower-right quandrant, and it is these systems that drive the weak anti-correlation between $\Nsat$ and $\Mstar$ seen for this simulation in the lower panel of  Figure~\ref{fig:Correlations}. 

Note that the applied point colors change in the same manner in all the simulations, with low to high $\MstarBCG$ residuals running from the top-left to the bottom-right. This pattern reflects the rough agreement of the correlations involving $\MstarBCG$ shown in the top two panels of Figure~\ref{fig:Correlations}.

\subsection{Redshift behavior and low-$z$ GMM $\Nsat$ fit}
\label{sec:redshift}

\begin{figure}
    \centering
    \includegraphics[width = \columnwidth]{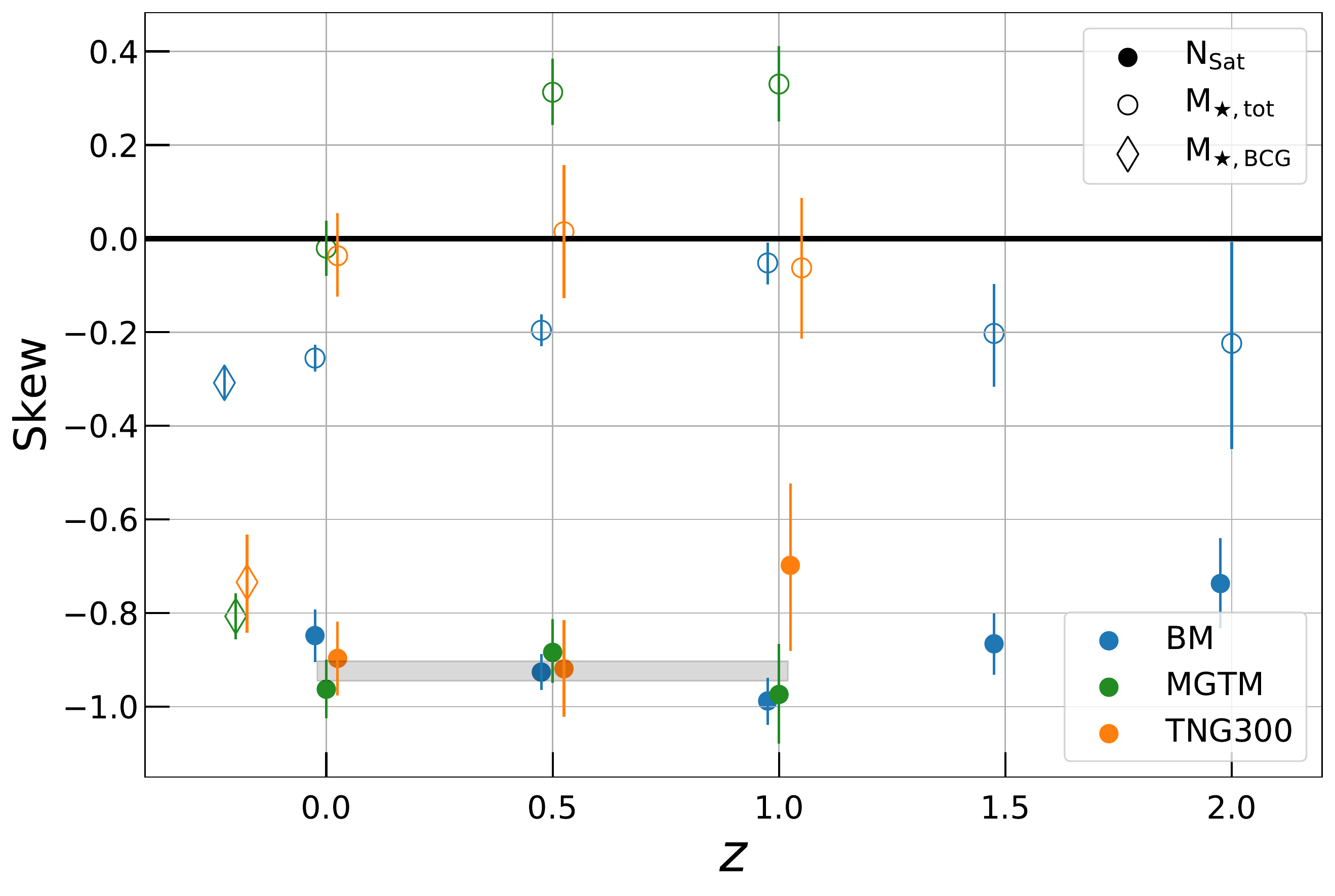}
    \caption{Skewness of the normalized, mass-conditioned residuals in $\Nsat$ (solid circles) and $\Mstar$ (open circles) for samples with at least 300 halos above $10^{13.5} \msol$ at each redshift.  The $\MstarBCG$ skewness (open diamond) is shown only at $z = 0$, offset by $-0.05$ in redshift to improve readability. The gray band is a $68\%$ interval for the $\Nsat$ skewness obtained using the combined halo populations of all three simulations at $z = 0$, $0.5$, and $1$.}
    \label{fig:Sample_Skewness}.
\end{figure}

Returning to the issue of kernel shapes, we find that the skewness in the $\Nsat$ and $\Mstar$ kernel shapes varies little with redshifts $z \le 1$. Figure~\ref{fig:Sample_Skewness} shows these values at discrete redshifts for simulation samples with at least 300 halos above $10^{13.5} \msol$. 

As noted previously, the BCG stellar mass (shown at $z=0$ only) is skew negative with $\gamma \simeq -0.8$ in both MGTM and TNG300, while the value for BM is much smaller, $-0.3$. For B100 we find $\gamma = -0.71 \pm 0.26$, where the large error reflects the small sample size of $< 100$ halos. 

The skewness in total stellar mass varies across the simulations, but is limited to the range $[-0.3,3]$.  The TNG300 results are consistent with zero at all redshifts while zero values are found by MGTM and BM at some redshifts. 

In contrast, the skewness in $\ln \Nsat$ is both consistent across simulations and persistent in redshift. Having verified robustness of the $\Nsat$ kernel shape, we combine the $z=0$, $0.5$, and $1$ redshift samples from all three simulations into a superset ensemble of more than 26,000 halos. Parameters from this superset, given in Table~\ref{table:Uber_Sample_values}, are precisely constrained by this large halo ensemble, with statistical uncertainties of a few percent in most parameters.  

\begin{figure}
    \centering
    \includegraphics[width =\columnwidth]{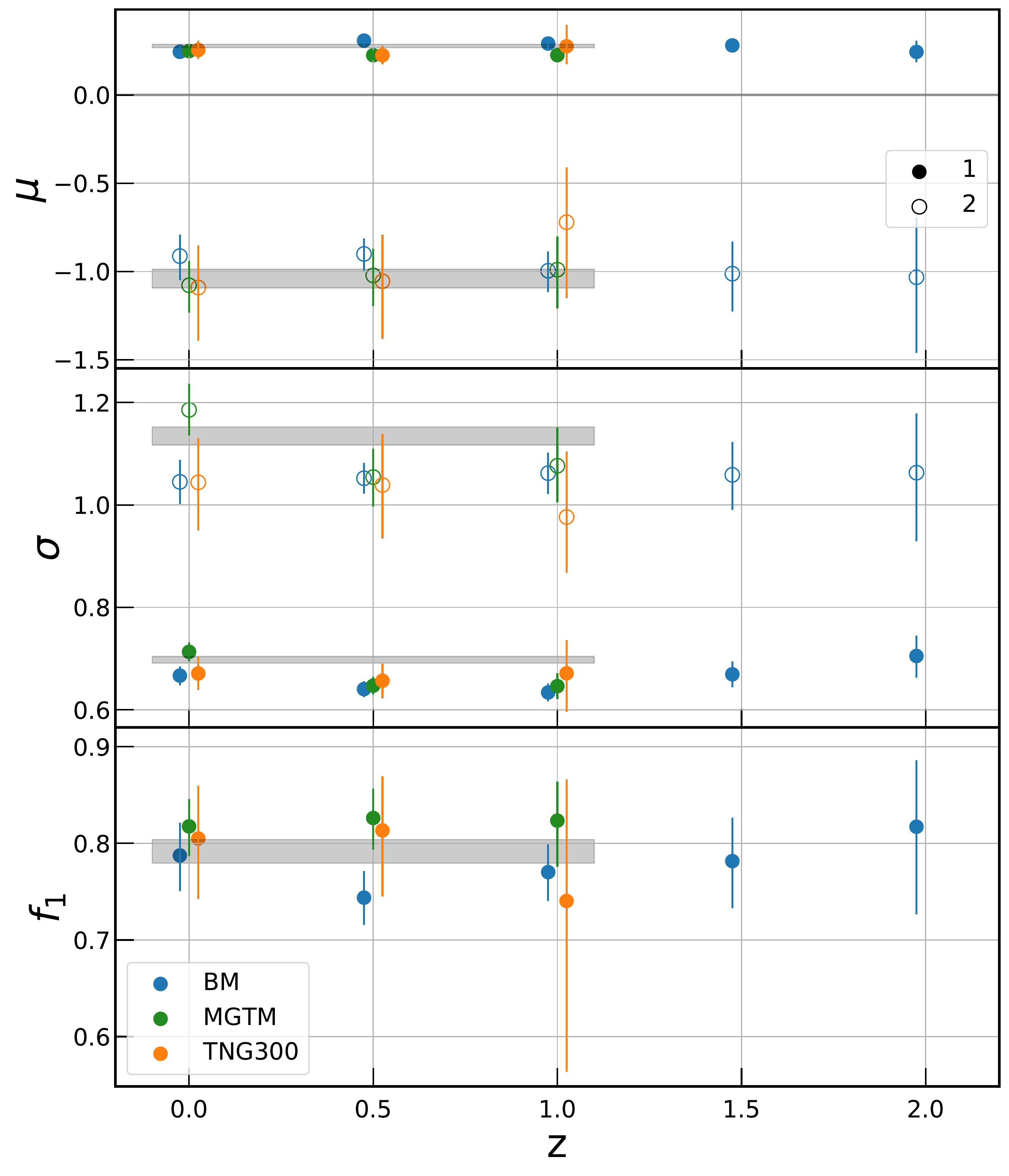}
    \caption{Gaussian mixture model parameters for the $\Nsat$ kernel, equation~(\ref{eqn:GMM}), with solid and open circles giving the first and second components, respectively. Values for each simulation are given in Table~\ref{table:Skew_and_GMM_Params}. Gray bands are $68\%$ confidence bands derived from the superset of all simulation populations at $z = 0$, $0.5$ and $1$ (see Table~\ref{table:Uber_Sample_values}).}
    \label{fig:GMM_Params}
\end{figure}

Beyond skewness, there is also good agreement in the GMM parameters of the $\Nsat$ kernel, shown for the three different simulations as a function of redshift in Figure~\ref{fig:GMM_Params}. The superset sample values of Table~\ref{table:Uber_Sample_values}, shown as the grey bands in Figure~\ref{fig:GMM_Params}, indicate that the halo population consists of an 80\% majority with mean $0.28 \pm 0.01$ and dispersion $0.68 \pm 0.01$ along with a wider, left-leaning minority having mean $-1.04 \pm 0.05$ and scatter $1.13 \pm 0.02$.  In the next section, we use importance sampling to trace how these components map to different distributions in central galaxy stellar mass.  

\begin{table}
 \begin{center} 
 	\caption{Satellite galaxy kernel skewness, $\gamma$, and GMM fit parameters for the combined $z = 0, 0.5, 1$ halo populations of all three simulations above a halo mass of $10^{13.5} \msol$ ($10^{13.8} \msol$ for BM at $z = 0$ only), with uncertainties from bootstrap (skewness) and MCMC posterior sampling (GMM parameters).  The superset contains a total of $26,\!332$ halos. }
    \label{table:Uber_Sample_values}
    \begin{tabular}{c | c}
            Parameter & Value\\
    	    \hline 
            $\gamma$    &   $-0.91 \pm 0.02$\\
            $f_1$       &   $0.79 \pm 0.01$\\
            $\mu_1$     &   $0.28 \pm 0.01$\\
            $\mu_2$     &   $-1.04 \pm 0.05$\\
            $\sigma_1$  &   $0.68 \pm 0.01$ \\
            $\sigma_2$  &   $1.13 \pm 0.02$ \\
    \hline 
  		\end{tabular}
    \end{center}
 \end{table}

\begin{figure}
    \centering
    \includegraphics[width = \columnwidth]{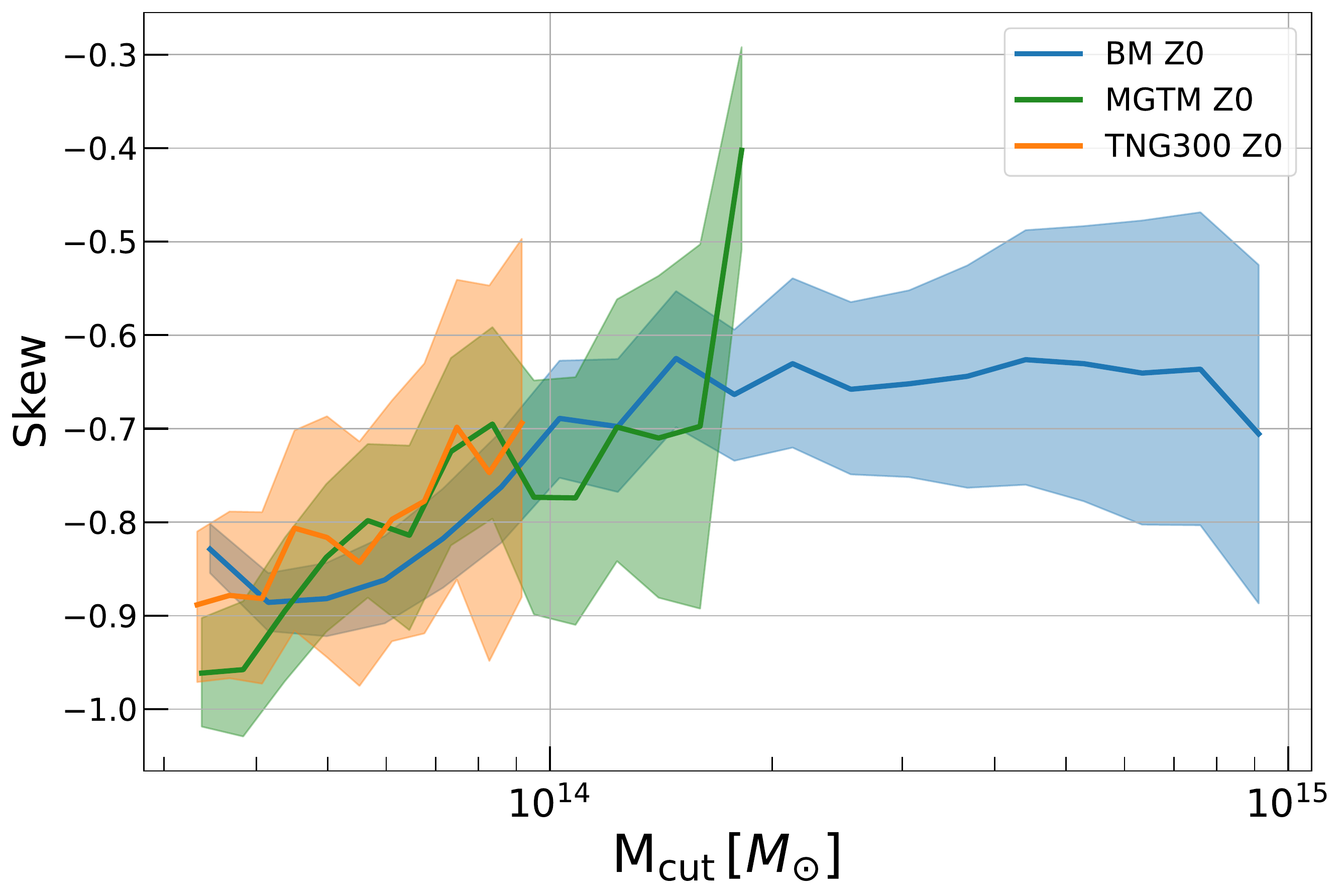}
    \caption{Skew in the $\ln \Nsat$ kernel as a function of lower mass limit, $M_{\rm cut}$. The maximum $M_{\rm cut}$ value is set by a requirement that samples contain at least $300$ halos.
    }
    \label{fig:Skew_CutoffFunction}
\end{figure}

Due to the steepness of the cosmic mass function, the shape of the $\Nsat$ kernel is heavily weighted by halos near the cutoff mass scale of $10^{13.5} \msol$. In Figure~\ref{fig:Skew_CutoffFunction}, we show how the skewness runs with applied cutoff mass, up to a limit for each sample at which the number of halos falls below 300.  While somewhat arbitrary, the 68\% bootstrap uncertainties for smaller samples become large and the results uninformative.

At $10^{14} \msol$, all simulations show that the shape is somewhat less skewed, with $\gamma \sim -0.75$.  At higher masses, we rely solely on the massive MACSIS sample, which displays asymptotic behavior to $\sim -0.65 \pm 0.15$. 
We leave it to future work with larger simulation ensembles to address this question in more detail.


\section{Toward survey validation: Secondary selection effects } \label{sec:SecondarySelection}

\begin{figure}
      \centering
      \includegraphics[width = \columnwidth]{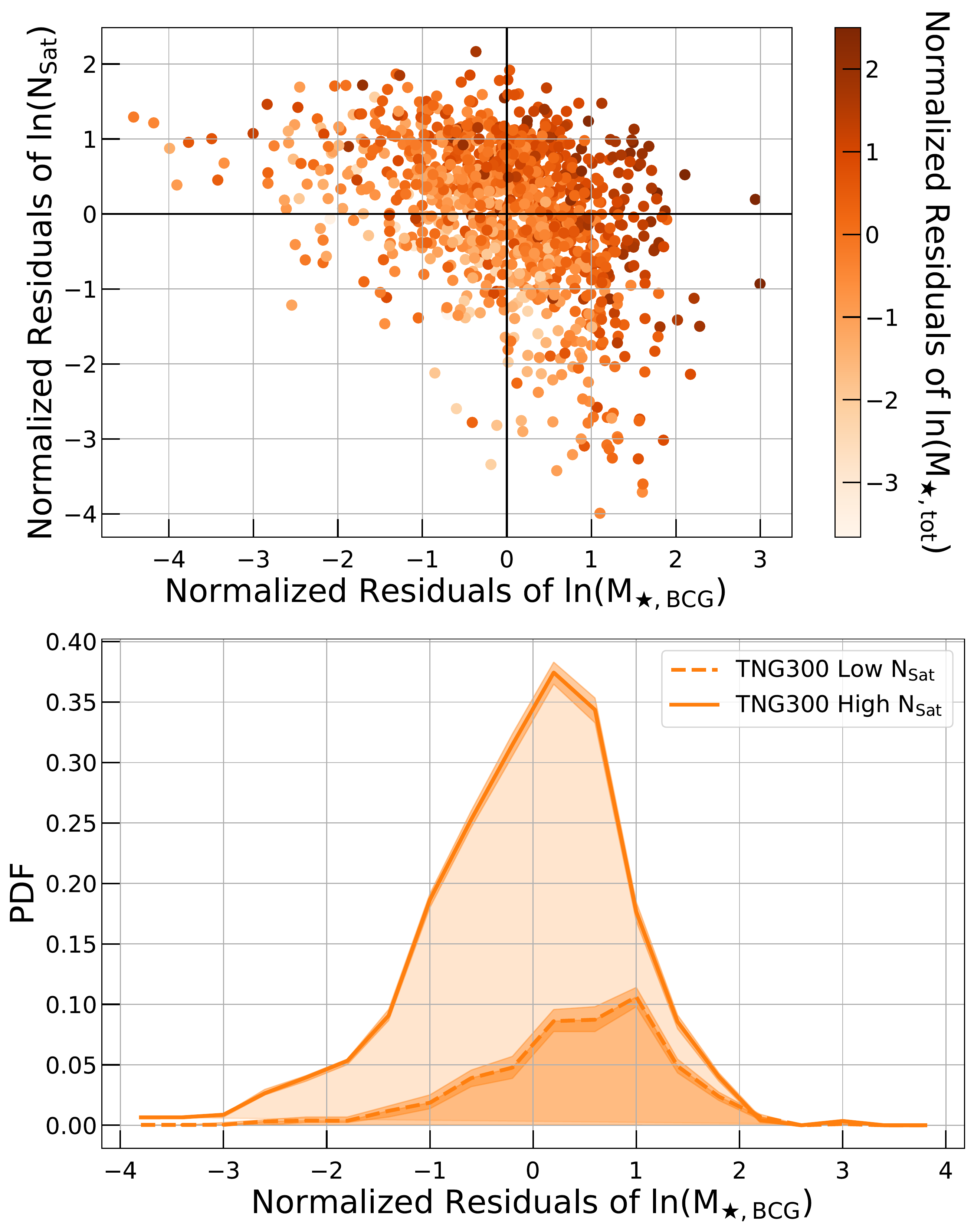}
      \caption{Mixture model demonstration for the TNG300 halo population at $z=0$. {\bf Upper:} Distribution of normalized residuals in the $\ln \MstarBCG- \ln \Nsat$ plane, colored by the residual in $\ln \Mstar$.  {\bf Lower:}
      Importance-sampling of the two GMM components in satellite galaxy count produces offset kernels in $\ln \MstarBCG$. The darker shading at high $\MstarBCG$ corresponds to the darker shaded GMM component at low $\Nsat$ in Figure~\ref{fig:LLR_PDF_NSat}, and vice-versa for the light shading, as anticipated by the residuals above.}
\label{fig:covarImportanceSample}
\end{figure}

In this section we explore how secondary selection in BCG stellar mass affects the statistical properties of satellite count and total stellar mass. For the TNG300 simulation we also explore secondary selection in halo formation epoch. 

Application of secondary selection to cluster surveys requires a statistical model relating cluster properties, especially those involved in selection, to our primary selection variable. The consistent patterns exhibited by these simulations are testable with current surveys when selection and projection effects are properly included.  A preparatory step toward sample modeling could be to use synthetic Chandra and XMM observations of the simulated halo ensembles \citep{Biffi2012PHOX, Lebrun2014,  Koulouridis2018XXLSurvey,  ZuHone2018ClusterMergerCatalog} to explore expectations for cluster samples selected by core-excised X-ray flux \citep{Mantz2018Center-excised}. 

The ultimate aim is to {\sl validate} these expectations in observed cluster samples with high quality, uniform optical properties, such as SDSS \citep{York2000SDSS}, DES \citep{DES2005} and, in the future, LSST \citep{Ivezic2019LSST} and Euclid \citep{Laureijs2011EUCLID, Racca2016EuclidDesign}. Such validation will require an observable mass proxy, such as weak lensing mass or hot gas mass, that itself is likely to correlate with the stellar properties under consideration \citep{Wu:2015, Farahi2018LocalizedCovariance, Farahi:2019anticorrelation}.

\subsection{Secondary selection on $\MstarBCG$}
\label{sec:selectBCG}

The correlation structure in the top two panels of Figure~\ref{fig:Correlations} provides a lever arm for secondary selection in BCG stellar mass.  We first explore this structure using the mixture model in satellite galaxy count, $\Nsat$.  

The top panel of Figure~\ref{fig:covarImportanceSample} displays the residual correlations of $\MstarBCG$ and $\Nsat$ for the $z=0$ TNG300 halo population (results are similar for the other two simulations).  This panel is merely a rotated version of the middle panel in Figure~\ref{fig:2DResiduals_Mstar_Nsat}.  An anti-correlation is apparent, with non-Gaussian tails in both directions.  As shown below, the tail to low central galaxy stellar mass is associated with late-forming systems.  

The lower panel of Figure~\ref{fig:covarImportanceSample} illustrates the utility of the Gaussian mixture model for $\Nsat$ to stratify the halo population in a property correlated with it.  The two shaded regions shown in the lower panel of Figure~\ref{fig:covarImportanceSample} are built from importance sampling the two components of the mixture whose parameters are given in Table~\ref{table:Uber_Sample_values}.  
Halos associated with the minority component, the broad tail displaced to lower $\Nsat$, possess central galaxies with higher $\MstarBCG$ values (mean of $0.4$ and width of $0.7$) shown by the darker shaded component (consistent with the darker shaded component of Figure~\ref{fig:LLR_PDF_NSat}). This region overlaps with the dominant, lighter-shaded component that has a mean of $-0.09$ and standard deviation $0.99$.  Both components are skewed negative with similar values of $\gamma = -0.75$, reflecting the non-Gaussian structure of the residuals in both components.

This structure implies that sub-samples of halos with lower than average (given their halo mass) central galaxy stellar masses, those below roughly $-2 \sigma$, are comprised almost exclusively of the dominant component in $\Nsat$.  Halos with higher than average central galaxy stellar masses, in contrast, consist of an $\Nsat$ mixture in which the minority component is enhanced but not dominant.

\begin{figure}
    \centering
    \includegraphics[width = \columnwidth]{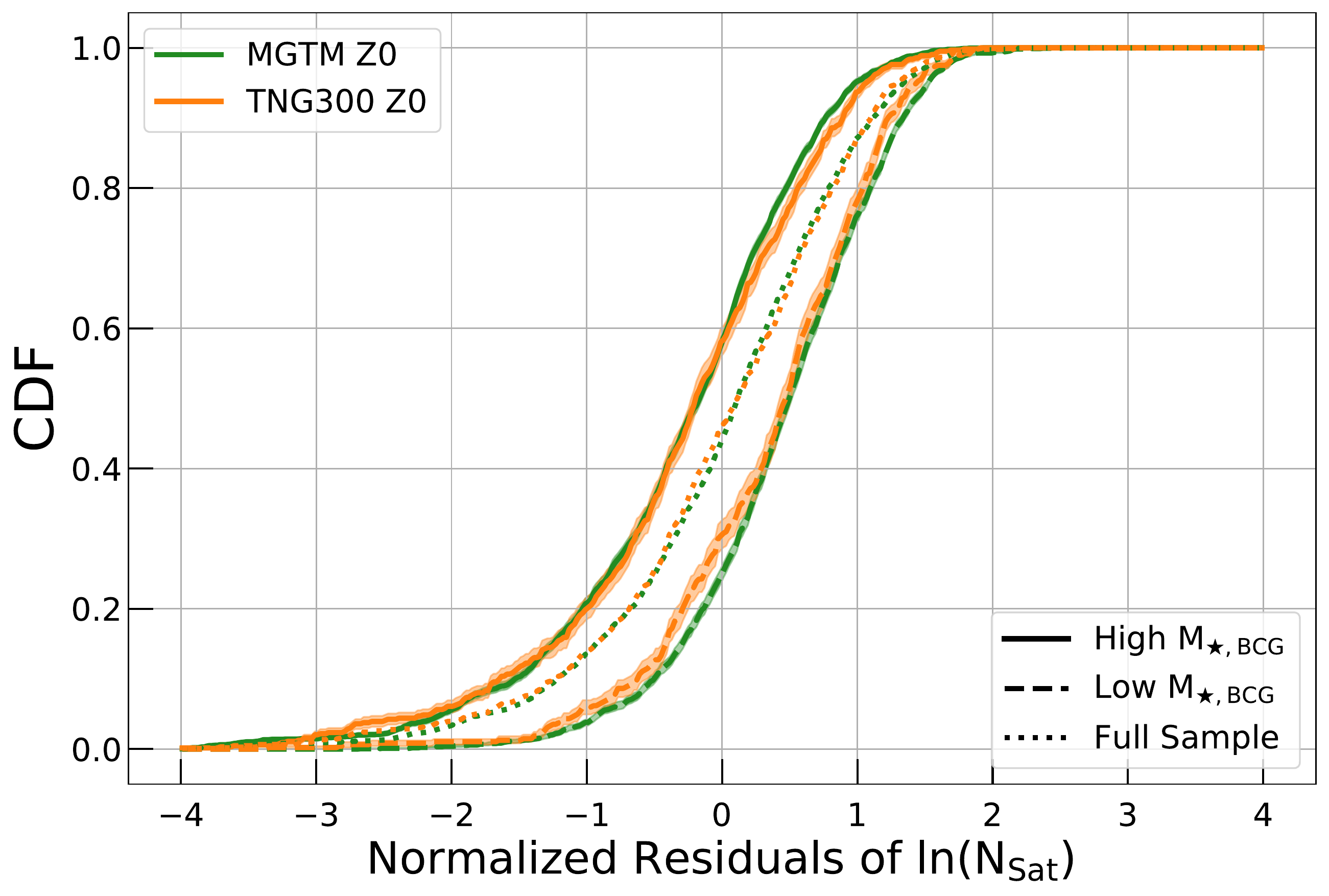}
    \includegraphics[width = \columnwidth]{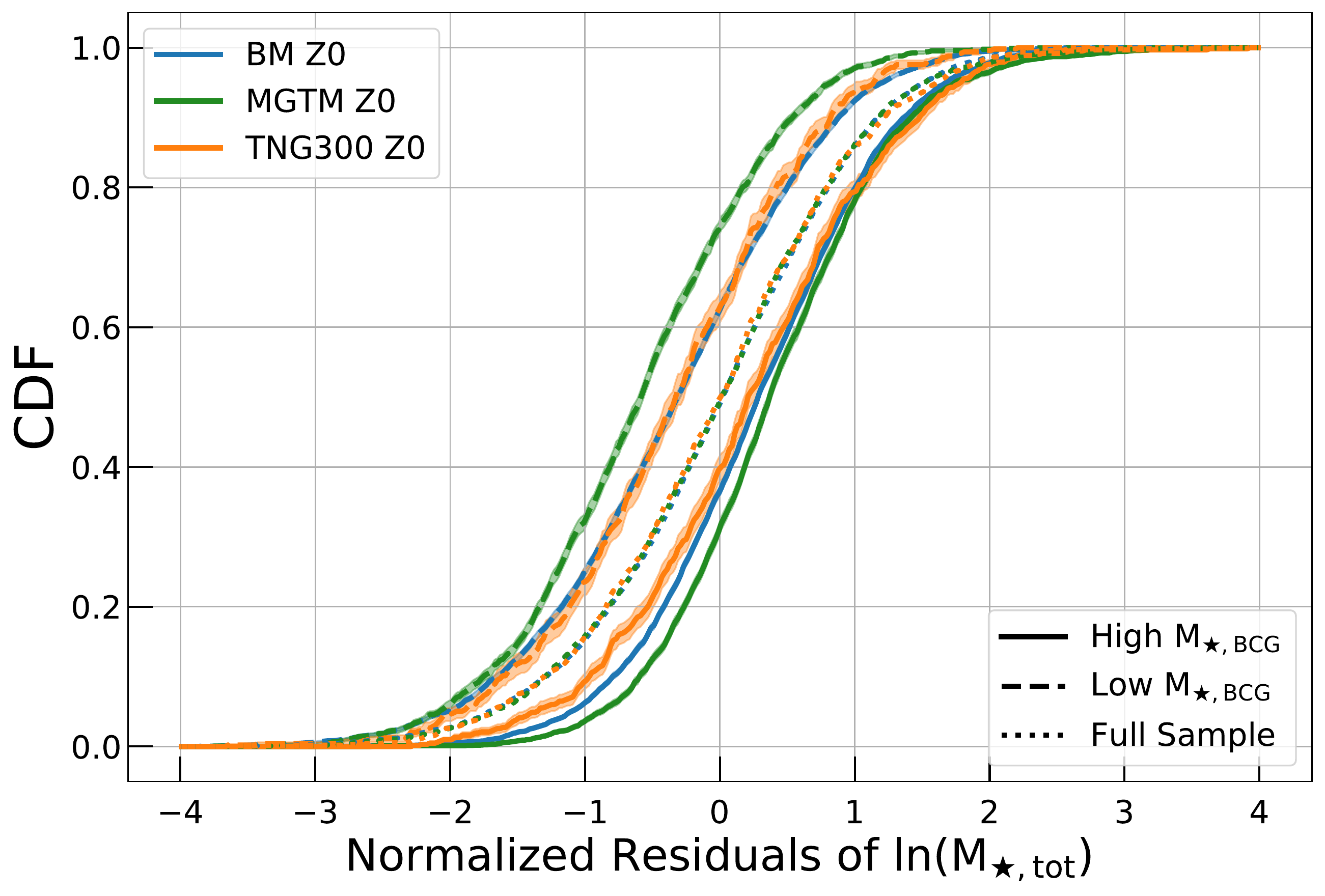}
    \caption{Cumulative distribution functions for residuals in $\ln\Nsat$ (upper) and $\ln\Mstar$ (lower) conditioned on relative $\MstarBCG$ and total halo mass. Solid lines show halos with higher than LLR-averaged $\MstarBCG$ while dashed lines show the opposite. Dotted lines show the behavior of the full population, unconditioned on $\MstarBCG$. The BM sample is omitted from the top plot (see text). 
    }
    \label{fig:CDF_BCG_Residual_Split}.
\end{figure}

We can expand on this result by treating $\MstarBCG$ as a secondary selection variable (total halo mass being the primary selection). 
We divide each simulation sample into two subsets based on whether a halo's central galaxy stellar mass lies above or below the LLR mean expectation at that halo mass, meaning we are selecting secondarily on the sign of the residual, equation~(\ref{eq:residual}). Note that, due to the non-Gaussian shape of the $\MstarBCG$ kernel, this does not split the population into sub-samples of equal size.

Figure~\ref{fig:CDF_BCG_Residual_Split} shows cumulative distribution functions (CDFs) for $\Nsat$ (top) and $\Mstar$ (bottom) for the full population (dotted lines) as well as the high and low $\MstarBCG$ halo subsets. Because the scatter in $\Nsat$ for the BM simulations is spuriously enhanced by its coarse stellar mass resolution (see Figures~\ref{fig:Params} and  \ref{fig:Correlations}), we do not show that model in the top panel.

Secondary selection by $\MstarBCG$ generates fairly dramatic shifts in the CDFs of both $\ln \Nsat$ and $\ln \Mstar$.  Halos with lower than average BCG stellar masses tend to have both higher numbers of satellite galaxies as well as lower total stellar masses. Table~\ref{table:CDF_metrics} lists root-mean-square and maximum values of the CDF offsets, $|x_2 - x_1|$, in normalized $\Nsat$ or $\Mstar$ deviation, where $x$ is the cumulant location at which the integrated probability takes some fixed value, ${\rm CDF}(x_2) = {\rm CDF}(x_1) = {\rm constant}$. To minimize discreteness effects in the rare event tails of these distributions, values in the table are limited to CDF values in the range $(0.1, 0.9)$. 

The rms values, in Table~\ref{table:CDF_metrics}, give us the same information as the correlations, as evidenced by the simulation ordering of the cumulant shifts reflecting the ordering of the $\MstarBCG - \Nsat$ and $\MstarBCG - \Mstar$ correlations in Figure~\ref{fig:Correlations}. For residual CDFs in both $\ln \NSat$ and $\ln \Mstar$, MGTM shows the largest rms offset, and its correlations also have the largest magnitude. For the residual CDFs of $\ln \Mstar$, BM and TNG300 show similar deviations, since their correlations are in agreement for a large part of the halo mass range.

\begin{table}
 \begin{center} 
 	\caption{Kernel offsets, the shifts in normalized cumulants shown as the dashed and solid lines in Figures~\ref{fig:CDF_BCG_Residual_Split} and Figure~\ref{fig:CDF_z_form_Split}, for sub-samples split by $\MstarBCG$ and (for TNG300 only) $\zformation$.   }
    \begin{tabular}{lcccccc}
        	\label{table:CDF_metrics}
        	 & \multicolumn{2}{c}{\textbf{$\Nsat$}} & \multicolumn{2}{c}{\textbf{$\Mstar$}} & \multicolumn{2}{c}{\textbf{$\MstarBCG$}} \\
            Sample (selection) & rms & max & rms & max & rms & max \\
        	\hline
         	 BM ($\MstarBCG$) & $(0.30)$ & $(0.39)$ & $0.63$ & $0.86$ & -- & -- \\
             MGTM ($\MstarBCG$) & $0.72$ & $1.03$  & $0.98$ & $1.16$
             & -- & -- \\
         	 TNG300 ($\MstarBCG$) & $0.63$ & $1.02$  & $0.57$ & $0.65$ & -- & -- \\
         	 TNG300 ($\zformation$)  &$0.41$ & $0.59$ & $0.18$ & $0.39$  & $0.79$ & $1.22$ 
  		\end{tabular}
    \end{center}
 \end{table}

\subsection{Secondary selection on Formation Epoch, $\zformation$}

The mass-conditioned covariance among stellar properties and non-Gaussian kernel shapes in $\Nsat$ and $\MstarBCG$ are related to the formation histories sampled by these discrete halo populations.  For example, \citet{Bradshaw2019} use the semi-analytic UniverseMachine model \citep{Behroozi2019UniverseMachine} to demonstrate that relative BCG stellar mass is correlated with the age of a halo, while total stellar mass is nearly independent of age.

We examine this behavior for TNG300, the simulation for which data to derive formation time estimates are publicly available. Analyzing the merger tree of each $z=0$  halo, we define the formation redshift, $\zformation$, as the epoch at which the total mass of a halo falls to half of its final value.  After LLR fitting $\zformation$ versus halo mass, we condition the residuals of $\ln \Nsat$, $\ln \Mstar$, and $\ln \MstarBCG$ on the sign of the $\zformation$ residuals.

\begin{figure}
    \centering
    \includegraphics[width = \columnwidth]{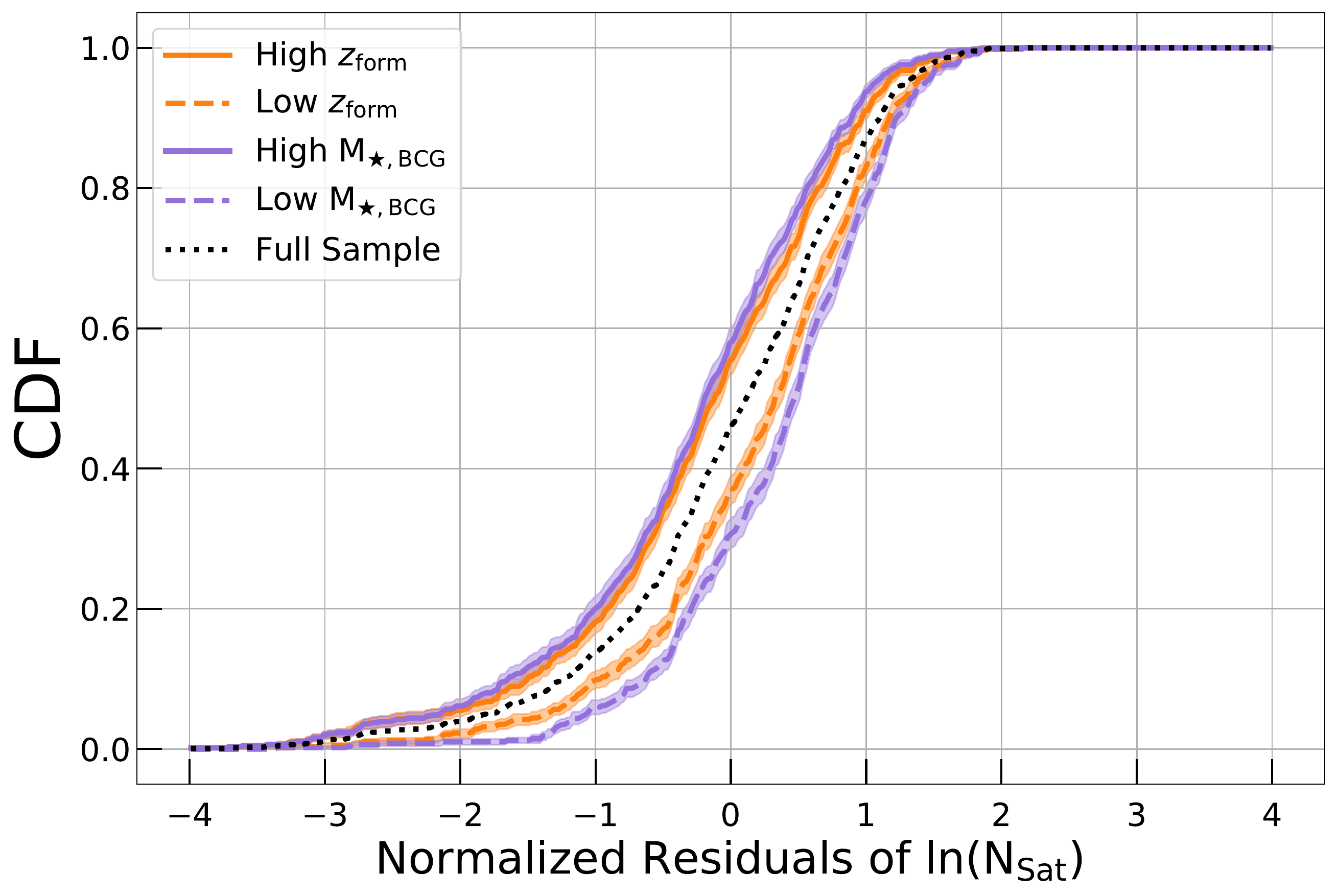}
    \includegraphics[width = \columnwidth]{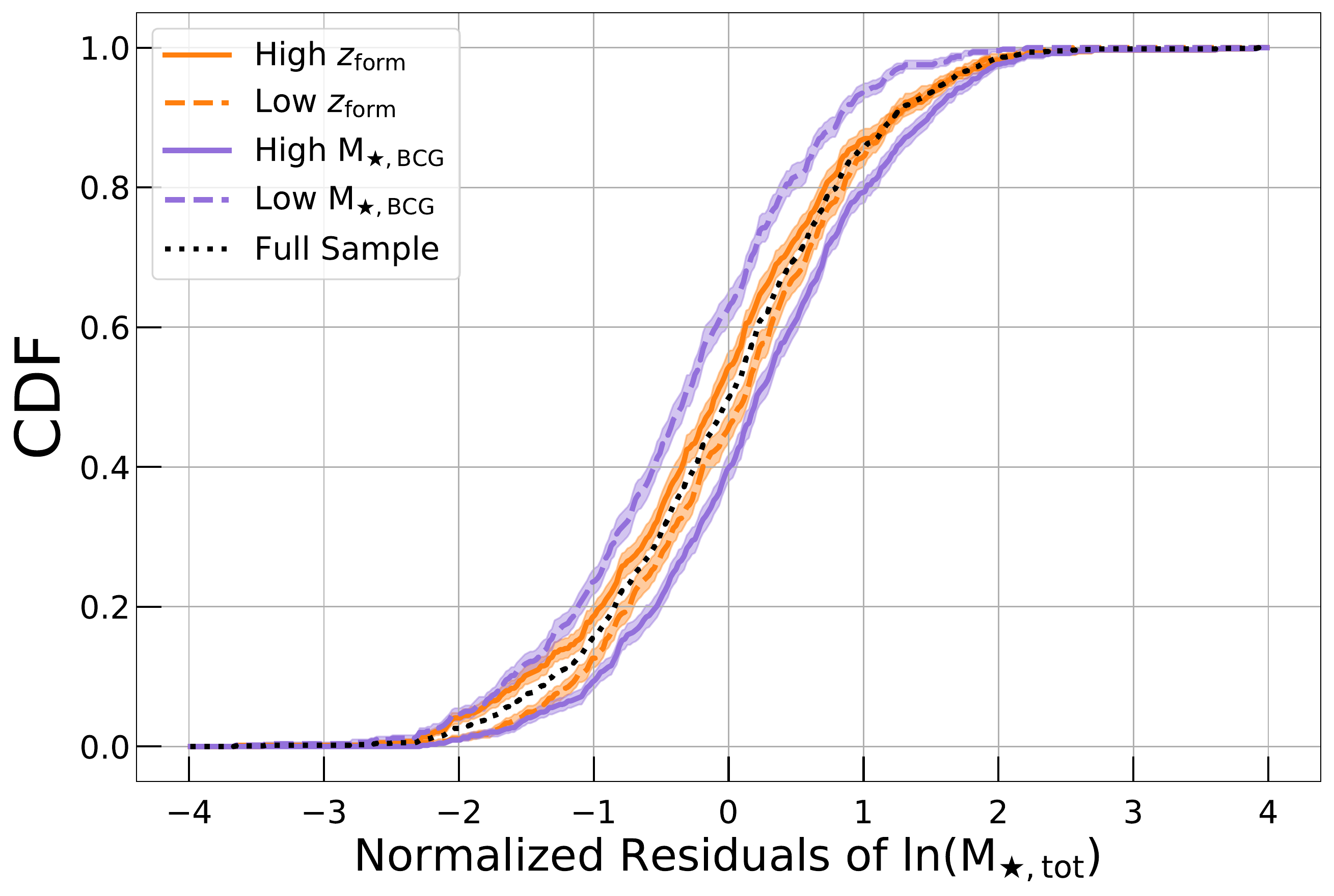}
    \includegraphics[width = \columnwidth]{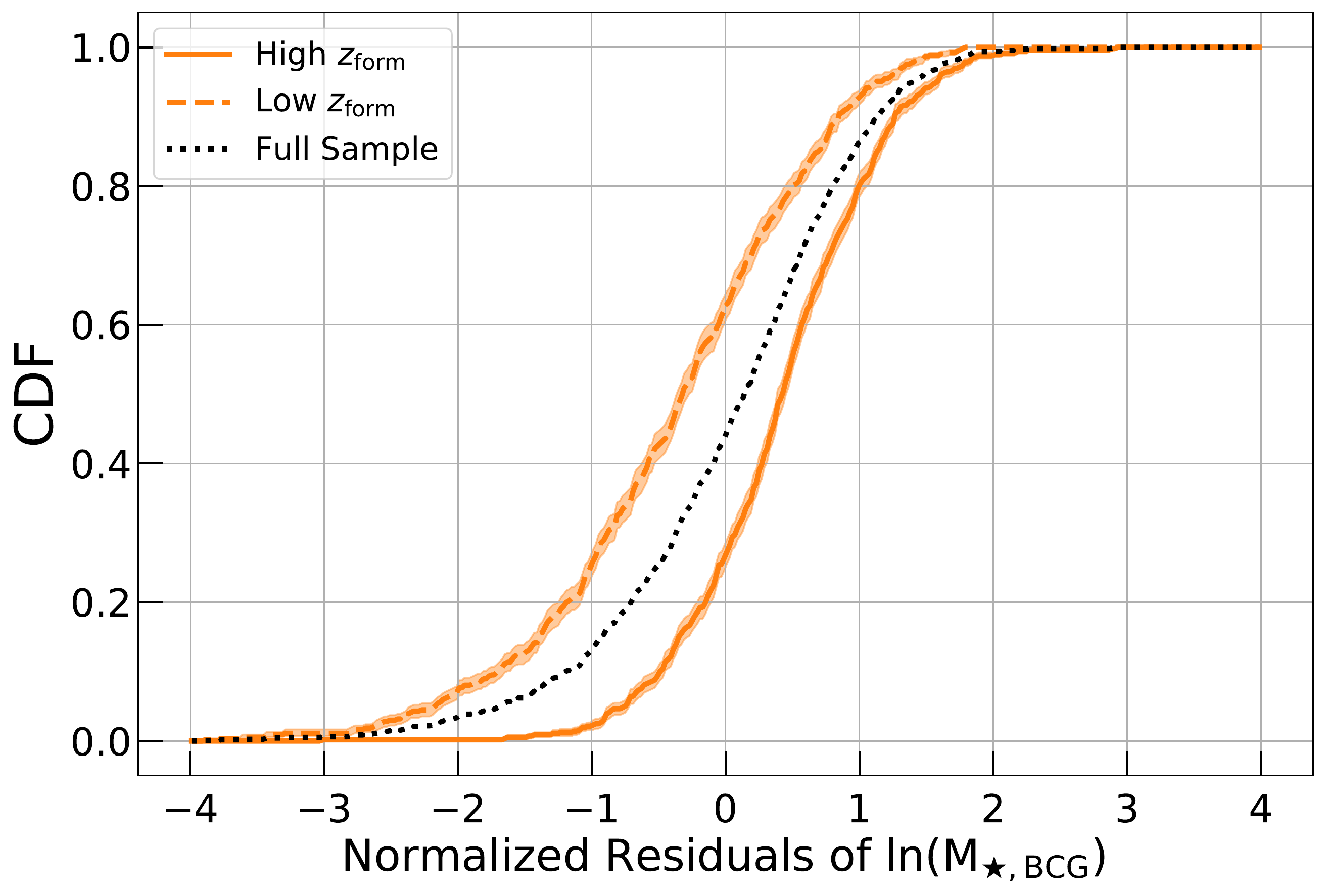}
    \caption{For the TNG300 $z=0$ sample, cumulative distribution functions for satellite galaxy counts (top) and total stellar mass (middle) conditioned on halo formation time, $\zformation$ (orange) and $\MstarBCG$ (purple). The bottom panel shows the residual CDFs of the central galaxy stellar mass conditioned on $\zformation$.
    }
    \label{fig:CDF_z_form_Split}.
\end{figure}

In Figure~\ref{fig:CDF_z_form_Split}, we present the residual CDFs for the two sub-populations in all three stellar properties. For the upper two panels, showing $\Nsat$ and $\mstar$, we compare to secondary conditioning using $\MstarBCG$, presented above. 

The $\Nsat$ CDF (top panel) shows similar deviations when conditioned on either $\MstarBCG$ or $\zformation$. Namely, halos of a younger age (i.e. with lower than average $\zformation$) and with lower central galaxy stellar mass are surrounded by a larger number of satellite galaxies. Our results agree with those of \cite{Bose2019Revealingthegalaxy}, who find a split in the $\Nsat$ scaling relations when conditioning on $\zformation$ for a TNG sample spanning a wider range in halo mass.

In contrast, the $\Mstar$ CDF (middle panel) shows differences between the two secondary selection variables. The positive correlation with $\MstarBCG$ produces a shift of $0.57$ in the $\Mstar$ CDF split by central galaxy stellar mass. A much weaker correlation with $\zformation$ yields a smaller shift of $0.16$ in the $\Mstar$ CDF. 
The latter finding is in qualitative agreement with \citet{Bradshaw2019}, who find no offset in the stellar mass-halo mass (SMHM) relation when conditioning on $\zformation$.

The central galaxy stellar mass (bottom panel of Figure~\ref{fig:CDF_z_form_Split}) is most sensitive to formation history, with an $0.8 \sigma$ rms shift in the CDF. The tail below $-1 \sigma$ in $\MstarBCG$ is almost exclusively late forming halos. These shifts are again in qualitative agreement with \citet{Bradshaw2019}, who find a difference of 0.2 dex in the $\MstarBCG-\Mhalo$ relation between the top and bottom 20\% ranked halos in $\zformation$.   

Figure~\ref{fig:CDF_z_form_Split} provides a view of how formation redshift maps onto the space of residuals in total and BCG stellar masses in the TNG300 population. Loci of constant formation time are oriented roughly along the diagonal, with early forming systems tending to have brighter than average BCGs overall but with total stellar masses that span the full range above and below the mean.  

\begin{figure}
    \centering
    \includegraphics[width = 0.9 \columnwidth]{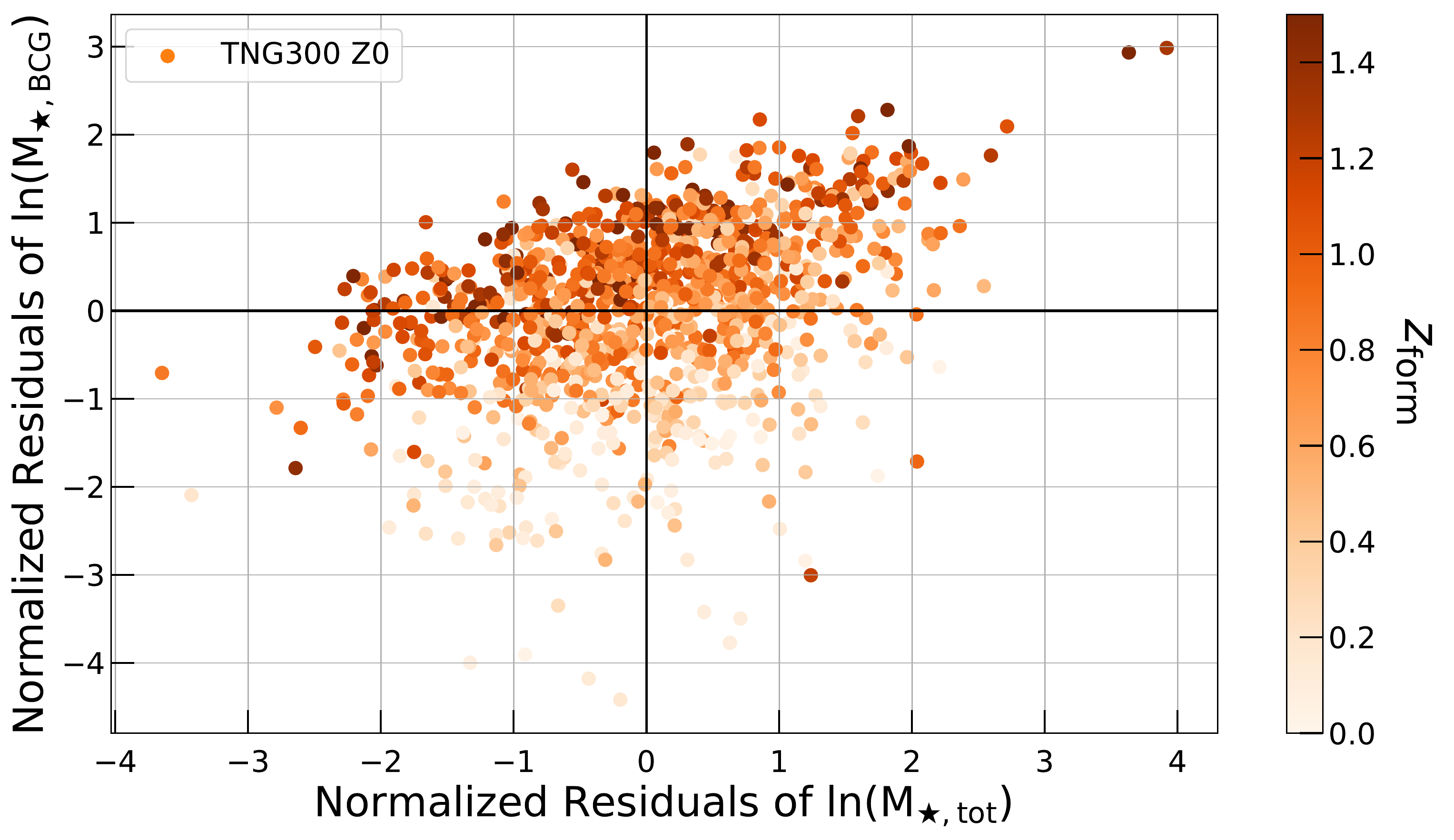}
    \caption{Normalized residuals in the $\Mstar-\MstarBCG$ plane colored by $\zformation$ for the TNG300 $z=0$ halo sample.
    }
    \label{fig:Mstar_MstarBCG_zform}.
\end{figure}

Other hydrodynamic simulation studies have also studied the satellite galaxy HOD conditioned on $\zformation$, finding that younger haloes have preferentially more satellites \citep{Bose2019Revealingthegalaxy, Artale2018TheImpactof}. This feature they explain by older halos losing more satellites to mergers with central galaxies, an interpretation  supported by \citet{Matthee2017TheOriginaOfScatter} and \citet{Bradshaw2019}.

\citet{Bradshaw2019} also find that the ex-situ stellar mass, material obtained through mergers with satellites, correlates strongly with $\zformation$, with central galaxies in older halos containing more ex-situ stellar mass. Conversely, the in-situ stellar mass shows no $\zformation$ dependence.


\section{Discussion}
\label{sec:Discussion}

\subsection{Property extensions and scale dependence of halo population statistics}
\label{methodology} 

The capacity of cosmological hydrodynamical simulations has expanded to the point where multiple simulation methods can produce halo populations containing many millions of objects across the full range of resolved scales, from individual galaxies to rich clusters of galaxies.  Future work can investigate the degree of self-similarity of our findings by considering a wider range in halo mass.  

In general, integrated physical properties connect to halo mass and redshift in a way that combines secular scaling of mean behavior with stochastic variations for individual objects. We intend to expand our study to include more features, such as hot gas masses, X-ray temperatures and luminosities, and galaxy velocities.  For example, \citet{Farahi2018LocalizedCovariance} find that the joint kernel of hot gas mass and total stellar mass is well described by a simple two-dimensional Gaussian with moderate anti-correlation between these mass fractions. The anti-correlation reflects the fact that, compared to the $10^{12} \msol$ halos hosting individual bright galaxies that lose considerable fraction of their baryons \citep{Bregman2018}, the cosmic mix of baryons is more nearly retained within the deep gravitational potential wells of clusters.  Verifying this and other multi-phase signatures in TNG and MGTM solutions remains to be done.

Our study is focused on verifying stellar property statistics for massive halo populations derived from simulations that employ multiple cosmological hydrodynamics methods.  However, other approaches, particularly semi-analytic methods (SAMs) that assign galaxy properties to N-body halos and sub-halos in a manner constrained by empirical data, are also capable of producing population-level expectations.  The qualitative agreement we find with \citet{Bradshaw2019} on property correlations, and other features such as running of $\Nsat$ scatter with mass, should be investigated more carefully to gain insights into the robustness of our findings and the limitations of both SAM and full hydro methods.  

An open question is whether SAM methods produce kernel shapes for satellite galaxy counts and central galaxy stellar mass that are negatively skewed to the degree seen in the three cosmological simulations studied here.  It is worth noting that, compared to cosmological hydrodynamics simulations, SAM populations require significantly less computational time and energy to produce, meaning that sensitive kernel shape measurements may be possible using large populations even at mass scales above $10^{15} \msol$.  Direct comparison of SAM and hydro population statistics would be a preliminary step in this direction. 

\subsection{Implications for optical cluster survey likelihoods }\label{surveys} 

The intrinsic satellite galaxy count is a core ingredient of models that use optical cluster counts in richness and redshift to constrain cosmology \citep{Rozo2010SDSSclusterCosmo, Chiu2019,  Costanzi2019SDSSmethods}.  However, due primarily to line-of-sight projection, the optical richness measured for a cluster is a noisy measure of the $1+\Nsat$ occupation intrinsic to halos \citep[\eg][]{Cohn2007}, an expectation confirmed by spectroscopic follow-up studies of SDSS redMaPPer clusters \citep{Sohn2018HectospecRedMapper}. 

Cluster richness can be modeled as a sum of partial contributions from halos \citep{Farahi2016SDSSspec}, so future work may lead to likelihoods employing a mixture of mixtures, one mixture representing contributions of multiple halos to the richness $\lambda$, the other mixture expressing intrinsic richness at a given halo mass (the one we study here). 

We note that secondary selection to identify the narrow $\Nsat$ component is a potential boon to cosmological studies. Shifts in mean halo mass associated with property selection scale as the variance \citep{Evrard2014Statistics}, so the factor two smaller variance in satellite galaxy count at fixed halo mass for the narrow component (Table~\ref{table:Uber_Sample_values}) could potentially be exploited to more accurately calibrate mean masses via stacked weak lensing analysis \citep[\eg][]{McClintock2019DESRMlensingmass}. 

\subsection{Cyberinfrastructure improvements }\label{cyberinfrastructure} 

Our study has been greatly enabled by the availability of full data releases of the IllustrisTNG simulations \citep{Nelson2019TNGPublicData}, and a partial public release of Magneticum Pathfinder \citep{Ragagnin2017WebPortal}.  The democratization of the data, through the availability of uniform, catalog-level simulation products to the public, is key to permitting more in-depth analyses of halo population statistics derived from multiple cosmological simulations.

Looking even further, reproducible computational science benefits from having open community access to the specific simulation methods, both production and analysis code bases, used in simulation studies \citep[\eg][]{Stodden:2016}. The scale of simulation data volumes makes it difficult to move high-resolution data to a central location, but a future in which distributed, containerized analysis environments  \citep{Raddick2019SciServerCompute} operate using improved discoverability standards \citep{Languignon2017IVOA} could greatly simplify and empower verification studies of the type we perform here.


\section{Summary}
\label{sec:Summary}

Motivated by the need to verify the highly detailed solutions produced by complex cosmological simulations, we perform a statistical study of the stellar and satellite galaxy contents within halo populations produced by three large cosmological hydrodynamics simulations. These include simulated samples from BAHAMAS and MACSIS, a volume from the Magneticum suite, and TNG300 from the IllustrisTNG Project. We focus here on features that describe the galaxy--halo connection --- the stellar-mass limited satellite galaxy occupation, $\Nsat$,  the stellar mass of the central galaxy, $\MstarBCG$, and the total stellar mass within the halo, $\Mstar$ --- in halos with total mass exceeding $10^{13.5} \msol$.

By applying a local linear regression method, we summarize the form of the mass-conditioned kernel, $\Pr (\ln \mathbf{S} \, | \,\Mhalo, z)$, where $\mathbf{S}$ is the set of aforementioned stellar properties. We provide 
local regression fit parameters for these properties --- normalizations, slopes, and covariance --- sampled over roughly two orders of magnitude in halo mass at $z=0$, $0.5$, $1$ and $2$. 
The $z=0$ halo samples contain from $1000$ to $9000$ objects, and this large scale allows us, by marginalizing over halo mass, to analyze the mean shape of the conditional kernel for each stellar property.
Our analysis effectively compresses a large amount of raw output data into a highly compact analytic form useful for modeling statistical likelihoods. 

Our main results are summarized in the following points. 
\begin{itemize}
    \item We verify common kernel shapes for the satellite count, $\Nsat$, and total stellar mass, $\Mstar$, where the former is skewed negatively, with skewness parameter $\gamma = -0.91 \pm 0.02$, and the latter being much closer to Gaussian ($|\gamma| < 0.3$).  For the population of halos above $10^{13.5}$ at $z \le 1$, we provide a two-component Gaussian mixture fit to the $\Nsat$ kernel (Table~\ref{table:Uber_Sample_values}), but note the shape is mildly sensitive to this halo mass threshold. The $z=0$ kernel shape of $\MstarBCG$ is also negatively skewed, with MGTM and TNG300 finding skewness $-0.8$.
    
    \item While the values of halo mass-conditioned regression parameters (slope, normalization, covariance) are often in mild tension among the three simulations, we find areas of qualitative agreement: i) the  scatter in $\ln \Nsat$ depends on halo mass and is slightly super-Poissonion (Figure~\ref{fig:Params}); ii) the scaling of total stellar mass is consistently sub-linear and the fractional scatter in $\Mstar$ is below $10\%$ for halos above $2 \times 10^{14} \msol$ (Figure~\ref{fig:LLR_Stellar_params}); iii) the mass-conditioned residuals in $\Nsat$ and $\MstarBCG$ are anti-correlated while those in $\MstarBCG$ and $\Mstar$ are positively correlated (Figure~\ref{fig:Correlations}).  
    
    \item All simulations find that halos with bigger than average central galaxies have both fewer than average satellite galaxies and larger than average total stellar mass. The former is consistent with a picture in which earlier-forming systems cannibalize satellite galaxies to the benefit of the BCG. 
    
    \item For TNG300, the mass-conditioned formation redshift correlates strongly with $\MstarBCG$ and $\Nsat$ but weakly with $\Mstartot$.  

    \item The structure of the mass-conditioned stellar property residuals is more similar in BM and TNG300 than MGTM (Figure~\ref{fig:2DResiduals_Mstar_Nsat}).
    
\end{itemize}

The low-order statistical measures of our study are empirically testable through careful analysis of scaling behavior in large cluster samples.  Sample selection, mis-centering, projection, and the evolution of galaxy colors are all confounding factors to be addressed in such studies. 

With nearly one million halos above $10^{14} \msol$ anticipated on the full sky  \citep{Allen2011CosmologicalClusters}, the demand for improved statistical representations linking cluster observable properties to those of the underlying halo population will only grow in the era of deep all-sky surveys across millimeter, optical-IR, and X-ray wavelengths. 

\bibliographystyle{mnras}
\bibliography{MoreRefs.bib}

\begin{thebibliography}{}
\makeatletter
\relax
\def\mn@urlcharsother{\let\do\@makeother \do\$\do\&\do\#\do\^\do\_\do\%\do\~}
\def\mn@doi{\begingroup\mn@urlcharsother \@ifnextchar [ {\mn@doi@}
  {\mn@doi@[]}}
\def\mn@doi@[#1]#2{\def\@tempa{#1}\ifx\@tempa\@empty \href
  {http://dx.doi.org/#2} {doi:#2}\else \href {http://dx.doi.org/#2} {#1}\fi
  \endgroup}
\def\mn@eprint#1#2{\mn@eprint@#1:#2::\@nil}
\def\mn@eprint@arXiv#1{\href {http://arxiv.org/abs/#1} {{\tt arXiv:#1}}}
\def\mn@eprint@dblp#1{\href {http://dblp.uni-trier.de/rec/bibtex/#1.xml}
  {dblp:#1}}
\def\mn@eprint@#1:#2:#3:#4\@nil{\def\@tempa {#1}\def\@tempb {#2}\def\@tempc
  {#3}\ifx \@tempc \@empty \let \@tempc \@tempb \let \@tempb \@tempa \fi \ifx
  \@tempb \@empty \def\@tempb {arXiv}\fi \@ifundefined
  {mn@eprint@\@tempb}{\@tempb:\@tempc}{\expandafter \expandafter \csname
  mn@eprint@\@tempb\endcsname \expandafter{\@tempc}}}

\bibitem[\protect\citeauthoryear{{Adams} \& {Fatuzzo}}{{Adams} \&
  {Fatuzzo}}{1996}]{Adams1996AClouds}
{Adams} F.~C.,  {Fatuzzo} M.,  1996, \mn@doi [\apj] {10.1086/177318}, \href
  {https://ui.adsabs.harvard.edu/abs/1996ApJ...464..256A} {464, 256}

\bibitem[\protect\citeauthoryear{{Ade} et~al.,}{{Ade}
  et~al.}{2019}]{SimonsObservatory}
{Ade} P.,  et~al., 2019, \mn@doi [\jcap] {10.1088/1475-7516/2019/02/056}, \href
  {https://ui.adsabs.harvard.edu/abs/2019JCAP...02..056A} {2019, 056}

\bibitem[\protect\citeauthoryear{{Allen}, {Evrard}  \& {Mantz}}{{Allen}
  et~al.}{2011}]{Allen2011CosmologicalClusters}
{Allen} S.~W.,  {Evrard} A.~E.,   {Mantz} A.~B.,  2011, \mn@doi [\araa]
  {10.1146/annurev-astro-081710-102514}, \href
  {https://ui.adsabs.harvard.edu/abs/2011ARA&A..49..409A} {49, 409}

\bibitem[\protect\citeauthoryear{{Artale}, {Zehavi}, {Contreras}  \&
  {Norberg}}{{Artale} et~al.}{2018}]{Artale2018TheImpactof}
{Artale} M.~C.,  {Zehavi} I.,  {Contreras} S.,   {Norberg} P.,  2018, \mn@doi
  [\mnras] {10.1093/mnras/sty2110}, \href
  {https://ui.adsabs.harvard.edu/abs/2018MNRAS.480.3978A} {480, 3978}

\bibitem[\protect\citeauthoryear{{Bah{\'e}} et~al.,}{{Bah{\'e}}
  et~al.}{2017}]{Hydrangea2017}
{Bah{\'e}} Y.~M.,  et~al., 2017, \mn@doi [\mnras] {10.1093/mnras/stx1403},
  \href {https://ui.adsabs.harvard.edu/abs/2017MNRAS.470.4186B} {470, 4186}

\bibitem[\protect\citeauthoryear{{Bah{\'e}} et~al.,}{{Bah{\'e}}
  et~al.}{2019}]{Bahe2019Satellites}
{Bah{\'e}} Y.~M.,  et~al., 2019, \mn@doi [\mnras] {10.1093/mnras/stz361}, \href
  {https://ui.adsabs.harvard.edu/abs/2019MNRAS.485.2287B} {485, 2287}

\bibitem[\protect\citeauthoryear{{Barnes}, {Kay}, {Henson}, {McCarthy},
  {Schaye}  \& {Jenkins}}{{Barnes} et~al.}{2017}]{Barnes2017Macsis}
{Barnes} D.~J.,  {Kay} S.~T.,  {Henson} M.~A.,  {McCarthy} I.~G.,  {Schaye} J.,
    {Jenkins} A.,  2017, \mn@doi [\mnras] {10.1093/mnras/stw2722}, \href
  {http://adsabs.harvard.edu/abs/2017MNRAS.465..213B} {465, 213}

\bibitem[\protect\citeauthoryear{{Behroozi}, {Wechsler}, {Hearin}  \&
  {Conroy}}{{Behroozi} et~al.}{2019}]{Behroozi2019UniverseMachine}
{Behroozi} P.,  {Wechsler} R.~H.,  {Hearin} A.~P.,   {Conroy} C.,  2019,
  \mn@doi [\mnras] {10.1093/mnras/stz1182}, \href
  {https://ui.adsabs.harvard.edu/abs/2019MNRAS.488.3143B} {488, 3143}

\bibitem[\protect\citeauthoryear{{Berlind} \& {Weinberg}}{{Berlind} \&
  {Weinberg}}{2002}]{BerlindWeinberg2002HODmodel}
{Berlind} A.~A.,  {Weinberg} D.~H.,  2002, \mn@doi [\apj] {10.1086/341469},
  \href {https://ui.adsabs.harvard.edu/abs/2002ApJ...575..587B} {575, 587}

\bibitem[\protect\citeauthoryear{{Biffi}, {Dolag}, {B{\"o}hringer}  \&
  {Lemson}}{{Biffi} et~al.}{2012}]{Biffi2012PHOX}
{Biffi} V.,  {Dolag} K.,  {B{\"o}hringer} H.,   {Lemson} G.,  2012, \mn@doi
  [\mnras] {10.1111/j.1365-2966.2011.20278.x}, \href
  {https://ui.adsabs.harvard.edu/abs/2012MNRAS.420.3545B} {420, 3545}

\bibitem[\protect\citeauthoryear{{Bocquet} et~al.,}{{Bocquet}
  et~al.}{2019}]{Bocquet2019SPTCosmo}
{Bocquet} S.,  et~al., 2019, \mn@doi [\apj] {10.3847/1538-4357/ab1f10}, \href
  {https://ui.adsabs.harvard.edu/abs/2019ApJ...878...55B} {878, 55}

\bibitem[\protect\citeauthoryear{{Bose}, {Eisenstein}, {Hernquist},
  {Pillepich}, {Nelson}, {Marinacci}, {Springel}  \& {Vogelsberger}}{{Bose}
  et~al.}{2019}]{Bose2019Revealingthegalaxy}
{Bose} S.,  {Eisenstein} D.~J.,  {Hernquist} L.,  {Pillepich} A.,  {Nelson} D.,
   {Marinacci} F.,  {Springel} V.,   {Vogelsberger} M.,  2019, \mn@doi [\mnras]
  {10.1093/mnras/stz2546}, \href
  {https://ui.adsabs.harvard.edu/abs/2019MNRAS.490.5693B} {490, 5693}

\bibitem[\protect\citeauthoryear{{Bradshaw}, {Leauthaud}, {Hearin}, {Huang}  \&
  {Behroozi}}{{Bradshaw} et~al.}{2019}]{Bradshaw2019}
{Bradshaw} C.,  {Leauthaud} A.,  {Hearin} A.,  {Huang} S.,   {Behroozi} P.,
  2019, arXiv e-prints, \href
  {https://ui.adsabs.harvard.edu/abs/2019arXiv190509353B} {p. arXiv:1905.09353}

\bibitem[\protect\citeauthoryear{{Bregman}, {Anderson}, {Miller},
  {Hodges-Kluck}, {Dai}, {Li}, {Li}  \& {Qu}}{{Bregman}
  et~al.}{2018}]{Bregman2018}
{Bregman} J.~N.,  {Anderson} M.~E.,  {Miller} M.~J.,  {Hodges-Kluck} E.,  {Dai}
  X.,  {Li} J.-T.,  {Li} Y.,   {Qu} Z.,  2018, \mn@doi [\apj]
  {10.3847/1538-4357/aacafe}, \href
  {https://ui.adsabs.harvard.edu/abs/2018ApJ...862....3B} {862, 3}

\bibitem[\protect\citeauthoryear{{Bryan} \& {Norman}}{{Bryan} \&
  {Norman}}{1998}]{BryanNorman:1998}
{Bryan} G.~L.,  {Norman} M.~L.,  1998, \mn@doi [\apj] {10.1086/305262}, \href
  {http://adsabs.harvard.edu/abs/1998ApJ...495...80B} {495, 80}

\bibitem[\protect\citeauthoryear{{Chiu} et~al.,}{{Chiu}
  et~al.}{2016}]{Chiu2016}
{Chiu} I.,  et~al., 2016, \mn@doi [\mnras] {10.1093/mnras/stw292}, \href
  {https://ui.adsabs.harvard.edu/abs/2016MNRAS.458..379C} {458, 379}

\bibitem[\protect\citeauthoryear{{Chiu}, {Umetsu}, {Murata}, {Medezinski}  \&
  {Oguri}}{{Chiu} et~al.}{2019}]{Chiu2019}
{Chiu} I.-N.,  {Umetsu} K.,  {Murata} R.,  {Medezinski} E.,   {Oguri} M.,
  2019, arXiv e-prints, \href
  {https://ui.adsabs.harvard.edu/abs/2019arXiv190902042C} {p. arXiv:1909.02042}

\bibitem[\protect\citeauthoryear{{Cohn}, {Evrard}, {White}, {Croton}  \&
  {Ellingson}}{{Cohn} et~al.}{2007}]{Cohn2007}
{Cohn} J.~D.,  {Evrard} A.~E.,  {White} M.,  {Croton} D.,   {Ellingson} E.,
  2007, \mn@doi [\mnras] {10.1111/j.1365-2966.2007.12479.x}, \href
  {https://ui.adsabs.harvard.edu/abs/2007MNRAS.382.1738C} {382, 1738}

\bibitem[\protect\citeauthoryear{{Cooray} \& {Sheth}}{{Cooray} \&
  {Sheth}}{2002}]{Cooray2002HaloStructure}
{Cooray} A.,  {Sheth} R.,  2002, \mn@doi [\physrep]
  {10.1016/S0370-1573(02)00276-4}, \href
  {https://ui.adsabs.harvard.edu/abs/2002PhR...372....1C} {372, 1}

\bibitem[\protect\citeauthoryear{{Costanzi} et~al.,}{{Costanzi}
  et~al.}{2019}]{Costanzi2019SDSSmethods}
{Costanzi} M.,  et~al., 2019, \mn@doi [\mnras] {10.1093/mnras/stz1949}, \href
  {https://ui.adsabs.harvard.edu/abs/2019MNRAS.488.4779C} {488, 4779}

\bibitem[\protect\citeauthoryear{{Croton} et~al.,}{{Croton}
  et~al.}{2006}]{Croton2006TheGalaxies}
{Croton} D.~J.,  et~al., 2006, \mn@doi [\mnras]
  {10.1111/j.1365-2966.2005.09675.x}, \href
  {https://ui.adsabs.harvard.edu/abs/2006MNRAS.365...11C} {365, 11}

\bibitem[\protect\citeauthoryear{{Cui} et~al.,}{{Cui}
  et~al.}{2014}]{Cui2014diffuselightsims}
{Cui} W.,  et~al., 2014, \mn@doi [\mnras] {10.1093/mnras/stt1940}, \href
  {https://ui.adsabs.harvard.edu/abs/2014MNRAS.437..816C} {437, 816}

\bibitem[\protect\citeauthoryear{{De Lucia} \& {Blaizot}}{{De Lucia} \&
  {Blaizot}}{2007}]{DeLucia:2007}
{De Lucia} G.,  {Blaizot} J.,  2007, \mn@doi [\mnras]
  {10.1111/j.1365-2966.2006.11287.x}, \href
  {https://ui.adsabs.harvard.edu/abs/2007MNRAS.375....2D} {375, 2}

\bibitem[\protect\citeauthoryear{{De Lucia}, {Springel}, {White}, {Croton}  \&
  {Kauffmann}}{{De Lucia} et~al.}{2006}]{DeLucia2006Egals}
{De Lucia} G.,  {Springel} V.,  {White} S. D.~M.,  {Croton} D.,   {Kauffmann}
  G.,  2006, \mn@doi [\mnras] {10.1111/j.1365-2966.2005.09879.x}, \href
  {https://ui.adsabs.harvard.edu/abs/2006MNRAS.366..499D} {366, 499}

\bibitem[\protect\citeauthoryear{{Dolag}, {Borgani}, {Murante}  \&
  {Springel}}{{Dolag} et~al.}{2009}]{Dolag2009Subfind}
{Dolag} K.,  {Borgani} S.,  {Murante} G.,   {Springel} V.,  2009, \mn@doi
  [\mnras] {10.1111/j.1365-2966.2009.15034.x}, \href
  {https://ui.adsabs.harvard.edu/abs/2009MNRAS.399..497D} {399, 497}

\bibitem[\protect\citeauthoryear{{Donnert}, {Dolag}, {Brunetti}  \&
  {Cassano}}{{Donnert} et~al.}{2013}]{Donnert2013RadioHaloes}
{Donnert} J.,  {Dolag} K.,  {Brunetti} G.,   {Cassano} R.,  2013, \mn@doi
  [\mnras] {10.1093/mnras/sts628}, \href
  {https://ui.adsabs.harvard.edu/abs/2013MNRAS.429.3564D} {429, 3564}

\bibitem[\protect\citeauthoryear{{Elahi} et~al.,}{{Elahi}
  et~al.}{2016}]{Elahi2016NiftyIII}
{Elahi} P.~J.,  et~al., 2016, \mn@doi [\mnras] {10.1093/mnras/stw338}, \href
  {https://ui.adsabs.harvard.edu/abs/2016MNRAS.458.1096E} {458, 1096}

\bibitem[\protect\citeauthoryear{{Erickson}, {Cunha}  \& {Evrard}}{{Erickson}
  et~al.}{2011}]{Erickson2011ProjectionModel}
{Erickson} B. M.~S.,  {Cunha} C.~E.,   {Evrard} A.~E.,  2011, \mn@doi [\prd]
  {10.1103/PhysRevD.84.103506}, \href
  {https://ui.adsabs.harvard.edu/abs/2011PhRvD..84j3506E} {84, 103506}

\bibitem[\protect\citeauthoryear{{Evrard}, {Summers}  \& {Davis}}{{Evrard}
  et~al.}{1994}]{Evrard1994Two-FluidFormation}
{Evrard} A.~E.,  {Summers} F.~J.,   {Davis} M.,  1994, \mn@doi [\apj]
  {10.1086/173700}, \href
  {https://ui.adsabs.harvard.edu/abs/1994ApJ...422...11E} {422, 11}

\bibitem[\protect\citeauthoryear{{Evrard}, {Arnault}, {Huterer}  \&
  {Farahi}}{{Evrard} et~al.}{2014}]{Evrard2014Statistics}
{Evrard} A.~E.,  {Arnault} P.,  {Huterer} D.,   {Farahi} A.,  2014, \mn@doi
  [\mnras] {10.1093/mnras/stu784}, \href
  {https://ui.adsabs.harvard.edu/abs/2014MNRAS.441.3562E} {441, 3562}

\bibitem[\protect\citeauthoryear{{Farahi}, {Evrard}, {Rozo}, {Rykoff}  \&
  {Wechsler}}{{Farahi} et~al.}{2016}]{Farahi2016SDSSspec}
{Farahi} A.,  {Evrard} A.~E.,  {Rozo} E.,  {Rykoff} E.~S.,   {Wechsler} R.~H.,
  2016, \mn@doi [\mnras] {10.1093/mnras/stw1143}, \href
  {https://ui.adsabs.harvard.edu/abs/2016MNRAS.460.3900F} {460, 3900}

\bibitem[\protect\citeauthoryear{{Farahi}, {Evrard}, {McCarthy}, {Barnes}  \&
  {Kay}}{{Farahi} et~al.}{2018}]{Farahi2018LocalizedCovariance}
{Farahi} A.,  {Evrard} A.~E.,  {McCarthy} I.,  {Barnes} D.~J.,   {Kay} S.~T.,
  2018, \mn@doi [\mnras] {10.1093/mnras/sty1179}, \href
  {https://ui.adsabs.harvard.edu/abs/2018MNRAS.478.2618F} {478, 2618}

\bibitem[\protect\citeauthoryear{{Farahi} et~al.,}{{Farahi}
  et~al.}{2019a}]{Farahi2019DESY1X-Ray}
{Farahi} A.,  et~al., 2019a, \mn@doi [\mnras] {10.1093/mnras/stz2689}, \href
  {https://ui.adsabs.harvard.edu/abs/2019MNRAS.tmp.2299F} {p.~2299}

\bibitem[\protect\citeauthoryear{{Farahi} et~al.,}{{Farahi}
  et~al.}{2019b}]{Farahi:2019anticorrelation}
{Farahi} A.,  et~al., 2019b, Nature Communications, \href
  {https://ui.adsabs.harvard.edu/abs/2019arXiv190702502F} {10}

\bibitem[\protect\citeauthoryear{{Gaspari}, {Ruszkowski}  \& {Oh}}{{Gaspari}
  et~al.}{2013}]{Gaspari2013ChaoticColdAccretion}
{Gaspari} M.,  {Ruszkowski} M.,   {Oh} S.~P.,  2013, \mn@doi [\mnras]
  {10.1093/mnras/stt692}, \href
  {https://ui.adsabs.harvard.edu/abs/2013MNRAS.432.3401G} {432, 3401}

\bibitem[\protect\citeauthoryear{{Golden-Marx} \& {Miller}}{{Golden-Marx} \&
  {Miller}}{2018}]{GoldenMarxMiller2018}
{Golden-Marx} J.~B.,  {Miller} C.~J.,  2018, \mn@doi [\apj]
  {10.3847/1538-4357/aac2bd}, \href
  {https://ui.adsabs.harvard.edu/abs/2018ApJ...860....2G} {860, 2}

\bibitem[\protect\citeauthoryear{{Golden-Marx} \& {Miller}}{{Golden-Marx} \&
  {Miller}}{2019}]{GoldenMarxMiller2019}
{Golden-Marx} J.~B.,  {Miller} C.~J.,  2019, \mn@doi [\apj]
  {10.3847/1538-4357/ab1d55}, \href
  {https://ui.adsabs.harvard.edu/abs/2019ApJ...878...14G} {878, 14}

\bibitem[\protect\citeauthoryear{{Hahn}, {Martizzi}, {Wu}, {Evrard}, {Teyssier}
   \& {Wechsler}}{{Hahn} et~al.}{2017}]{Hahn2017RhapsodyG}
{Hahn} O.,  {Martizzi} D.,  {Wu} H.-Y.,  {Evrard} A.~E.,  {Teyssier} R.,
  {Wechsler} R.~H.,  2017, \mn@doi [\mnras] {10.1093/mnras/stx001}, \href
  {https://ui.adsabs.harvard.edu/abs/2017MNRAS.470..166H} {470, 166}

\bibitem[\protect\citeauthoryear{{Hearin}, {Zentner}, {Berlind}  \&
  {Newman}}{{Hearin} et~al.}{2013}]{Hearin:2013}
{Hearin} A.~P.,  {Zentner} A.~R.,  {Berlind} A.~A.,   {Newman} J.~A.,  2013,
  \mn@doi [\mnras] {10.1093/mnras/stt755}, \href
  {https://ui.adsabs.harvard.edu/abs/2013MNRAS.433..659H} {433, 659}

\bibitem[\protect\citeauthoryear{{Hearin}, {Zentner}, {van den Bosch},
  {Campbell}  \& {Tollerud}}{{Hearin} et~al.}{2016}]{Hearin:2016}
{Hearin} A.~P.,  {Zentner} A.~R.,  {van den Bosch} F.~C.,  {Campbell} D.,
  {Tollerud} E.,  2016, \mn@doi [\mnras] {10.1093/mnras/stw840}, \href
  {https://ui.adsabs.harvard.edu/abs/2016MNRAS.460.2552H} {460, 2552}

\bibitem[\protect\citeauthoryear{{Hirschmann}, {Dolag}, {Saro}, {Bachmann},
  {Borgani}  \& {Burkert}}{{Hirschmann}
  et~al.}{2014}]{Hirschmann2014CosmologicalDownsizing}
{Hirschmann} M.,  {Dolag} K.,  {Saro} A.,  {Bachmann} L.,  {Borgani} S.,
  {Burkert} A.,  2014, \mn@doi [\mnras] {10.1093/mnras/stu1023}, \href
  {https://ui.adsabs.harvard.edu/abs/2014MNRAS.442.2304H} {442, 2304}

\bibitem[\protect\citeauthoryear{{Ivezi{\'c}} et~al.,}{{Ivezi{\'c}}
  et~al.}{2019}]{Ivezic2019LSST}
{Ivezi{\'c}} {\v Z}.,  et~al., 2019, \mn@doi [\apj] {10.3847/1538-4357/ab042c},
  \href {http://adsabs.harvard.edu/abs/2019ApJ...873..111I} {873, 111}

\bibitem[\protect\citeauthoryear{{Kaiser}}{{Kaiser}}{1986}]{Kaiser:1986}
{Kaiser} N.,  1986, \mn@doi [\mnras] {10.1093/mnras/222.2.323}, \href
  {https://ui.adsabs.harvard.edu/abs/1986MNRAS.222..323K} {222, 323}

\bibitem[\protect\citeauthoryear{{Katz} \& {White}}{{Katz} \&
  {White}}{1993}]{KatzWhite1993}
{Katz} N.,  {White} S. D.~M.,  1993, \mn@doi [\apj] {10.1086/172935}, \href
  {https://ui.adsabs.harvard.edu/abs/1993ApJ...412..455K} {412, 455}

\bibitem[\protect\citeauthoryear{{Kaviraj} et~al.,}{{Kaviraj}
  et~al.}{2017}]{Kaviraj2017HorizonAGN}
{Kaviraj} S.,  et~al., 2017, \mn@doi [\mnras] {10.1093/mnras/stx126}, \href
  {https://ui.adsabs.harvard.edu/abs/2017MNRAS.467.4739K} {467, 4739}

\bibitem[\protect\citeauthoryear{{Khandai}, {Di Matteo}, {Croft}, {Wilkins},
  {Feng}, {Tucker}, {DeGraf}  \& {Liu}}{{Khandai}
  et~al.}{2015}]{Khandai2015MassiveBlack}
{Khandai} N.,  {Di Matteo} T.,  {Croft} R.,  {Wilkins} S.,  {Feng} Y.,
  {Tucker} E.,  {DeGraf} C.,   {Liu} M.-S.,  2015, \mn@doi [\mnras]
  {10.1093/mnras/stv627}, \href
  {https://ui.adsabs.harvard.edu/abs/2015MNRAS.450.1349K} {450, 1349}

\bibitem[\protect\citeauthoryear{{Koulouridis} et~al.,}{{Koulouridis}
  et~al.}{2018}]{Koulouridis2018XXLSurvey}
{Koulouridis} E.,  et~al., 2018, \mn@doi [\aap] {10.1051/0004-6361/201730789},
  \href {https://ui.adsabs.harvard.edu/abs/2018A&A...620A...4K} {620, A4}

\bibitem[\protect\citeauthoryear{{Kravtsov} \& {Borgani}}{{Kravtsov} \&
  {Borgani}}{2012}]{KravtsovBorgani2012}
{Kravtsov} A.~V.,  {Borgani} S.,  2012, \mn@doi [\araa]
  {10.1146/annurev-astro-081811-125502}, \href
  {https://ui.adsabs.harvard.edu/abs/2012ARA&A..50..353K} {50, 353}

\bibitem[\protect\citeauthoryear{{Kravtsov}, {Vikhlinin}  \&
  {Meshcheryakov}}{{Kravtsov} et~al.}{2018}]{Kravtsov:2018}
{Kravtsov} A.~V.,  {Vikhlinin} A.~A.,   {Meshcheryakov} A.~V.,  2018, \mn@doi
  [Astronomy Letters] {10.1134/S1063773717120015}, \href
  {https://ui.adsabs.harvard.edu/abs/2018AstL...44....8K} {44, 8}

\bibitem[\protect\citeauthoryear{{Languignon}, {Le Petit}, {Rodrigo}, {Lemson},
  {Molinaro}  \& {Wozniak}}{{Languignon} et~al.}{2017}]{Languignon2017IVOA}
{Languignon} D.,  {Le Petit} F.,  {Rodrigo} C.,  {Lemson} G.,  {Molinaro} M.,
  {Wozniak} H.,  2017, Technical report, {Simulation Data Access Layer Version
  1.0}, \mn@doi{10.5479/ADS/bib/2017ivoa.spec.0320L.
}

\bibitem[\protect\citeauthoryear{{Laureijs} et~al.,}{{Laureijs}
  et~al.}{2011}]{Laureijs2011EUCLID}
{Laureijs} R.,  et~al., 2011, arXiv e-prints, \href
  {https://ui.adsabs.harvard.edu/abs/2011arXiv1110.3193L} {p. arXiv:1110.3193}

\bibitem[\protect\citeauthoryear{{Le Brun}, {McCarthy}, {Schaye}  \&
  {Ponman}}{{Le Brun} et~al.}{2014}]{Lebrun2014}
{Le Brun} A. M.~C.,  {McCarthy} I.~G.,  {Schaye} J.,   {Ponman} T.~J.,  2014,
  \mn@doi [\mnras] {10.1093/mnras/stu608}, \href
  {https://ui.adsabs.harvard.edu/abs/2014MNRAS.441.1270L} {441, 1270}

\bibitem[\protect\citeauthoryear{{Mantz}, {Allen}, {Rapetti}  \&
  {Ebeling}}{{Mantz} et~al.}{2010}]{Mantz2010XrayCountsCosmo}
{Mantz} A.,  {Allen} S.~W.,  {Rapetti} D.,   {Ebeling} H.,  2010, \mn@doi
  [\mnras] {10.1111/j.1365-2966.2010.16992.x}, \href
  {https://ui.adsabs.harvard.edu/abs/2010MNRAS.406.1759M} {406, 1759}

\bibitem[\protect\citeauthoryear{{Mantz} et~al.,}{{Mantz}
  et~al.}{2016}]{Mantz2016WtGScaling}
{Mantz} A.~B.,  et~al., 2016, \mn@doi [\mnras] {10.1093/mnras/stw2250}, \href
  {https://ui.adsabs.harvard.edu/abs/2016MNRAS.463.3582M} {463, 3582}

\bibitem[\protect\citeauthoryear{{Mantz}, {Allen}, {Morris}  \& {von der
  Linden}}{{Mantz} et~al.}{2018}]{Mantz2018Center-excised}
{Mantz} A.~B.,  {Allen} S.~W.,  {Morris} R.~G.,   {von der Linden} A.,  2018,
  \mn@doi [\mnras] {10.1093/mnras/stx2554}, \href
  {https://ui.adsabs.harvard.edu/abs/2018MNRAS.473.3072M} {473, 3072}

\bibitem[\protect\citeauthoryear{{Marinacci} et~al.,}{{Marinacci}
  et~al.}{2018}]{Marinacci2018FirstFields}
{Marinacci} F.,  et~al., 2018, \mn@doi [\mnras] {10.1093/mnras/sty2206}, \href
  {https://ui.adsabs.harvard.edu/abs/2018MNRAS.480.5113M} {480, 5113}

\bibitem[\protect\citeauthoryear{{Matthee}, {Schaye}, {Crain}, {Schaller},
  {Bower}  \& {Theuns}}{{Matthee}
  et~al.}{2017}]{Matthee2017TheOriginaOfScatter}
{Matthee} J.,  {Schaye} J.,  {Crain} R.~A.,  {Schaller} M.,  {Bower} R.,
  {Theuns} T.,  2017, \mn@doi [\mnras] {10.1093/mnras/stw2884}, \href
  {https://ui.adsabs.harvard.edu/abs/2017MNRAS.465.2381M} {465, 2381}

\bibitem[\protect\citeauthoryear{{McCarthy}, {Schaye}, {Bird}  \& {Le
  Brun}}{{McCarthy} et~al.}{2017}]{McCarthy2017TheCosmology}
{McCarthy} I.~G.,  {Schaye} J.,  {Bird} S.,   {Le Brun} A. M.~C.,  2017,
  \mn@doi [\mnras] {10.1093/mnras/stw2792}, \href
  {https://ui.adsabs.harvard.edu/abs/2017MNRAS.465.2936M} {465, 2936}

\bibitem[\protect\citeauthoryear{{McClintock} et~al.,}{{McClintock}
  et~al.}{2019}]{McClintock2019DESRMlensingmass}
{McClintock} T.,  et~al., 2019, \mn@doi [\mnras] {10.1093/mnras/sty2711}, \href
  {https://ui.adsabs.harvard.edu/abs/2019MNRAS.482.1352M} {482, 1352}

\bibitem[\protect\citeauthoryear{{McNamara} \& {Nulsen}}{{McNamara} \&
  {Nulsen}}{2012}]{McNamara2012AGNFeedback}
{McNamara} B.~R.,  {Nulsen} P.~E.~J.,  2012, \mn@doi [New Journal of Physics]
  {10.1088/1367-2630/14/5/055023}, \href
  {https://ui.adsabs.harvard.edu/abs/2012NJPh...14e5023M} {14, 055023}

\bibitem[\protect\citeauthoryear{{Merloni} et~al.,}{{Merloni}
  et~al.}{2012}]{Merloni2012eRositaScienceBook}
{Merloni} A.,  et~al., 2012, arXiv e-prints, \href
  {https://ui.adsabs.harvard.edu/abs/2012arXiv1209.3114M} {p. arXiv:1209.3114}

\bibitem[\protect\citeauthoryear{Moustakas et~al.,}{Moustakas
  et~al.}{2013}]{Moustakas:2013}
Moustakas J.,  et~al., 2013, The Astrophysical Journal, 767, 50

\bibitem[\protect\citeauthoryear{{Mulroy} et~al.,}{{Mulroy}
  et~al.}{2014}]{Mulroy2014LkMwl}
{Mulroy} S.~L.,  et~al., 2014, \mn@doi [\mnras] {10.1093/mnras/stu1387}, \href
  {https://ui.adsabs.harvard.edu/abs/2014MNRAS.443.3309M} {443, 3309}

\bibitem[\protect\citeauthoryear{{Mulroy} et~al.,}{{Mulroy}
  et~al.}{2019}]{Mulroy2019LoCuSS}
{Mulroy} S.~L.,  et~al., 2019, \mn@doi [\mnras] {10.1093/mnras/sty3484}, \href
  {https://ui.adsabs.harvard.edu/abs/2019MNRAS.484...60M} {484, 60}

\bibitem[\protect\citeauthoryear{{Naiman} et~al.,}{{Naiman}
  et~al.}{2018}]{Naiman2018FirstEuropium}
{Naiman} J.~P.,  et~al., 2018, \mn@doi [\mnras] {10.1093/mnras/sty618}, \href
  {https://ui.adsabs.harvard.edu/abs/2018MNRAS.477.1206N} {477, 1206}

\bibitem[\protect\citeauthoryear{{Nelson} et~al.,}{{Nelson}
  et~al.}{2018a}]{Nelson2018TNGcolors}
{Nelson} D.,  et~al., 2018a, \mn@doi [\mnras] {10.1093/mnras/stx3040}, \href
  {https://ui.adsabs.harvard.edu/abs/2018MNRAS.475..624N} {475, 624}

\bibitem[\protect\citeauthoryear{{Nelson} et~al.,}{{Nelson}
  et~al.}{2018b}]{Nelson2018FirstBimodality}
{Nelson} D.,  et~al., 2018b, \mn@doi [\mnras] {10.1093/mnras/stx3040}, \href
  {https://ui.adsabs.harvard.edu/abs/2018MNRAS.475..624N} {475, 624}

\bibitem[\protect\citeauthoryear{{Nelson} et~al.,}{{Nelson}
  et~al.}{2019}]{Nelson2019TNGPublicData}
{Nelson} D.,  et~al., 2019, \mn@doi [Computational Astrophysics and Cosmology]
  {10.1186/s40668-019-0028-x}, \href
  {https://ui.adsabs.harvard.edu/abs/2019ComAC...6....2N} {6, 2}

\bibitem[\protect\citeauthoryear{{Pillepich}, {Porciani}  \&
  {Reiprich}}{{Pillepich} et~al.}{2012}]{Pillepich2012eRositaForecasts}
{Pillepich} A.,  {Porciani} C.,   {Reiprich} T.~H.,  2012, \mn@doi [\mnras]
  {10.1111/j.1365-2966.2012.20443.x}, \href
  {https://ui.adsabs.harvard.edu/abs/2012MNRAS.422...44P} {422, 44}

\bibitem[\protect\citeauthoryear{{Pillepich} et~al.,}{{Pillepich}
  et~al.}{2018a}]{Pillepich2018SimulatingGalaxy}
{Pillepich} A.,  et~al., 2018a, \mn@doi [\mnras] {10.1093/mnras/stx2656}, \href
  {https://ui.adsabs.harvard.edu/abs/2018MNRAS.473.4077P} {473, 4077}

\bibitem[\protect\citeauthoryear{{Pillepich} et~al.,}{{Pillepich}
  et~al.}{2018b}]{Pillepich2018FirstGalaxies}
{Pillepich} A.,  et~al., 2018b, \mn@doi [\mnras] {10.1093/mnras/stx3112}, \href
  {https://ui.adsabs.harvard.edu/abs/2018MNRAS.475..648P} {475, 648}

\bibitem[\protect\citeauthoryear{{Pillepich}, {Reiprich}, {Porciani}, {Borm}
  \& {Merloni}}{{Pillepich} et~al.}{2018c}]{Pillepich2018DEForecastseRosita}
{Pillepich} A.,  {Reiprich} T.~H.,  {Porciani} C.,  {Borm} K.,   {Merloni} A.,
  2018c, \mn@doi [\mnras] {10.1093/mnras/sty2240}, \href
  {https://ui.adsabs.harvard.edu/abs/2018MNRAS.481..613P} {481, 613}

\bibitem[\protect\citeauthoryear{{Predehl} et~al.,}{{Predehl}
  et~al.}{2014}]{Predehl2014eROSITAonSRG}
{Predehl} P.,  et~al., 2014, {eROSITA on SRG}.
p. 91441T, \mn@doi{10.1117/12.2055426}

\bibitem[\protect\citeauthoryear{{Racca} et~al.,}{{Racca}
  et~al.}{2016}]{Racca2016EuclidDesign}
{Racca} G.~D.,  et~al., 2016, {The Euclid mission design}.
p. 99040O, \mn@doi{10.1117/12.2230762}

\bibitem[\protect\citeauthoryear{{Raddick}, {Kim}, {Lemson}, {Medvedev}  \&
  {Taghizadeh-Popp}}{{Raddick} et~al.}{2019}]{Raddick2019SciServerCompute}
{Raddick} M.~J.,  {Kim} J.~W.,  {Lemson} G.,  {Medvedev} D.,
  {Taghizadeh-Popp} M.,  2019, {SciServerCompute: Bring Analysis Close to the
  Data}.
p.~749

\bibitem[\protect\citeauthoryear{{Ragagnin}, {Dolag}, {Biffi}, {Cadolle Bel},
  {Hammer}, {Krukau}, {Petkova}  \& {Steinborn}}{{Ragagnin}
  et~al.}{2017}]{Ragagnin2017WebPortal}
{Ragagnin} A.,  {Dolag} K.,  {Biffi} V.,  {Cadolle Bel} M.,  {Hammer} N.~J.,
  {Krukau} A.,  {Petkova} M.,   {Steinborn} D.,  2017, \mn@doi [Astronomy and
  Computing] {10.1016/j.ascom.2017.05.001}, \href
  {https://ui.adsabs.harvard.edu/abs/2017A&C....20...52R} {20, 52}

\bibitem[\protect\citeauthoryear{{Ragone-Figueroa}, {Granato}, {Murante},
  {Borgani}  \& {Cui}}{{Ragone-Figueroa}
  et~al.}{2013}]{RagoneFigueroa2013BCGsims}
{Ragone-Figueroa} C.,  {Granato} G.~L.,  {Murante} G.,  {Borgani} S.,   {Cui}
  W.,  2013, \mn@doi [\mnras] {10.1093/mnras/stt1693}, \href
  {https://ui.adsabs.harvard.edu/abs/2013MNRAS.436.1750R} {436, 1750}

\bibitem[\protect\citeauthoryear{Rasia et~al.,}{Rasia
  et~al.}{2015}]{Rasia2015CoolCores}
Rasia E.,  et~al., 2015, The Astrophysical Journal Letters, 813, L17

\bibitem[\protect\citeauthoryear{{Rodriguez-Gomez} et~al.,}{{Rodriguez-Gomez}
  et~al.}{2016}]{Rodriguez2016StellarMassAssemblyIllustris}
{Rodriguez-Gomez} V.,  et~al., 2016, \mn@doi [\mnras] {10.1093/mnras/stw456},
  \href {https://ui.adsabs.harvard.edu/abs/2016MNRAS.458.2371R} {458, 2371}

\bibitem[\protect\citeauthoryear{{Rozo} et~al.,}{{Rozo}
  et~al.}{2010}]{Rozo2010SDSSclusterCosmo}
{Rozo} E.,  et~al., 2010, \mn@doi [\apj] {10.1088/0004-637X/708/1/645}, \href
  {https://ui.adsabs.harvard.edu/abs/2010ApJ...708..645R} {708, 645}

\bibitem[\protect\citeauthoryear{Salvadori}{Salvadori}{2019}]{Salvadori2019}
Salvadori S.,  2019, Uncertainty Quantification in CFD: The Matrix of
  Knowledge.
Springer International Publishing, Cham, pp 33--66,
  \mn@doi{10.1007/978-3-319-92943-9_2}, \url
  {https://doi.org/10.1007/978-3-319-92943-9_2}

\bibitem[\protect\citeauthoryear{{Scannapieco} et~al.,}{{Scannapieco}
  et~al.}{2012}]{Scannapieco2012AquilaCodeComp}
{Scannapieco} C.,  et~al., 2012, \mn@doi [\mnras]
  {10.1111/j.1365-2966.2012.20993.x}, \href
  {https://ui.adsabs.harvard.edu/abs/2012MNRAS.423.1726S} {423, 1726}

\bibitem[\protect\citeauthoryear{{Shaw}, {Holder}  \& {Dudley}}{{Shaw}
  et~al.}{2010}]{Shaw2010Non-GaussianRelations}
{Shaw} L.~D.,  {Holder} G.~P.,   {Dudley} J.,  2010, \mn@doi [\apj]
  {10.1088/0004-637X/716/1/281}, \href
  {https://ui.adsabs.harvard.edu/abs/2010ApJ...716..281S} {716, 281}

\bibitem[\protect\citeauthoryear{{Sohn}, {Geller}, {Rines}, {Hwang}, {Utsumi}
  \& {Diaferio}}{{Sohn} et~al.}{2018}]{Sohn2018HectospecRedMapper}
{Sohn} J.,  {Geller} M.~J.,  {Rines} K.~J.,  {Hwang} H.~S.,  {Utsumi} Y.,
  {Diaferio} A.,  2018, \mn@doi [\apj] {10.3847/1538-4357/aab20b}, \href
  {https://ui.adsabs.harvard.edu/abs/2018ApJ...856..172S} {856, 172}

\bibitem[\protect\citeauthoryear{{Spergel} et~al.,}{{Spergel}
  et~al.}{2015}]{Spergel2015WFIRST}
{Spergel} D.,  et~al., 2015, arXiv e-prints, \href
  {https://ui.adsabs.harvard.edu/abs/2015arXiv150303757S} {p. arXiv:1503.03757}

\bibitem[\protect\citeauthoryear{{Springel}}{{Springel}}{2005}]{Springel2005Gagdet2}
{Springel} V.,  2005, \mn@doi [\mnras] {10.1111/j.1365-2966.2005.09655.x},
  \href {https://ui.adsabs.harvard.edu/abs/2005MNRAS.364.1105S} {364, 1105}

\bibitem[\protect\citeauthoryear{{Springel}}{{Springel}}{2010}]{Springel2010EMesh}
{Springel} V.,  2010, \mn@doi [\mnras] {10.1111/j.1365-2966.2009.15715.x},
  \href {https://ui.adsabs.harvard.edu/abs/2010MNRAS.401..791S} {401, 791}

\bibitem[\protect\citeauthoryear{{Springel}, {White}, {Tormen}  \&
  {Kauffmann}}{{Springel} et~al.}{2001}]{Springel2001Subfind}
{Springel} V.,  {White} S. D.~M.,  {Tormen} G.,   {Kauffmann} G.,  2001,
  \mn@doi [\mnras] {10.1046/j.1365-8711.2001.04912.x}, \href
  {https://ui.adsabs.harvard.edu/abs/2001MNRAS.328..726S} {328, 726}

\bibitem[\protect\citeauthoryear{{Springel} et~al.,}{{Springel}
  et~al.}{2018}]{Springel2018FirstClustering}
{Springel} V.,  et~al., 2018, \mn@doi [\mnras] {10.1093/mnras/stx3304}, \href
  {https://ui.adsabs.harvard.edu/abs/2018MNRAS.475..676S} {475, 676}

\bibitem[\protect\citeauthoryear{{Stodden} et~al.,}{{Stodden}
  et~al.}{2016}]{Stodden:2016}
{Stodden} V.,  et~al., 2016, \mn@doi [Science] {10.1126/science.aah6168}, \href
  {https://ui.adsabs.harvard.edu/abs/2016Sci...354.1240S} {354, 1240}

\bibitem[\protect\citeauthoryear{{Tang}, {Lin}, {Cui}, {Kang}, {Wang},
  {Contini}  \& {Yu}}{{Tang} et~al.}{2018}]{Tang2018diffuselightsims}
{Tang} L.,  {Lin} W.,  {Cui} W.,  {Kang} X.,  {Wang} Y.,  {Contini} E.,   {Yu}
  Y.,  2018, \mn@doi [\apj] {10.3847/1538-4357/aabd78}, \href
  {https://ui.adsabs.harvard.edu/abs/2018ApJ...859...85T} {859, 85}

\bibitem[\protect\citeauthoryear{{The Dark Energy Survey Collaboration}}{{The
  Dark Energy Survey Collaboration}}{2005}]{DES2005}
{The Dark Energy Survey Collaboration} 2005, arXiv e-prints, \href
  {https://ui.adsabs.harvard.edu/abs/2005astro.ph.10346T} {pp
  astro--ph/0510346}

\bibitem[\protect\citeauthoryear{{Tremaine} \& {Richstone}}{{Tremaine} \&
  {Richstone}}{1977}]{TremaineRichstone1977}
{Tremaine} S.~D.,  {Richstone} D.~O.,  1977, \mn@doi [\apj] {10.1086/155049},
  \href {https://ui.adsabs.harvard.edu/abs/1977ApJ...212..311T} {212, 311}

\bibitem[\protect\citeauthoryear{{Tremmel} et~al.,}{{Tremmel}
  et~al.}{2019a}]{Tremmel2019Romulusc}
{Tremmel} M.,  et~al., 2019a, \mn@doi [\mnras] {10.1093/mnras/sty3336}, \href
  {https://ui.adsabs.harvard.edu/abs/2019MNRAS.483.3336T} {483, 3336}

\bibitem[\protect\citeauthoryear{{Tremmel} et~al.,}{{Tremmel}
  et~al.}{2019b}]{ROMULUSC2019}
{Tremmel} M.,  et~al., 2019b, \mn@doi [\mnras] {10.1093/mnras/sty3336}, \href
  {https://ui.adsabs.harvard.edu/abs/2019MNRAS.483.3336T} {483, 3336}

\bibitem[\protect\citeauthoryear{{Vikhlinin} et~al.,}{{Vikhlinin}
  et~al.}{2009}]{Vikhlinin2009}
{Vikhlinin} A.,  et~al., 2009, \mn@doi [\apj] {10.1088/0004-637X/692/2/1060},
  \href {https://ui.adsabs.harvard.edu/abs/2009ApJ...692.1060V} {692, 1060}

\bibitem[\protect\citeauthoryear{{Vogelsberger} et~al.,}{{Vogelsberger}
  et~al.}{2014}]{Vogelsberger2014Illustris}
{Vogelsberger} M.,  et~al., 2014, \mn@doi [\nat] {10.1038/nature13316}, \href
  {https://ui.adsabs.harvard.edu/abs/2014Natur.509..177V} {509, 177}

\bibitem[\protect\citeauthoryear{{Vogelsberger}, {Marinacci}, {Torrey}  \&
  {Puchwein}}{{Vogelsberger} et~al.}{2019}]{Vogelsberger2019SimReview}
{Vogelsberger} M.,  {Marinacci} F.,  {Torrey} P.,   {Puchwein} E.,  2019, arXiv
  e-prints, \href {https://ui.adsabs.harvard.edu/abs/2019arXiv190907976V} {p.
  arXiv:1909.07976}

\bibitem[\protect\citeauthoryear{Voit, Donahue, Bryan  \& McDonald}{Voit
  et~al.}{2015}]{Voit2015NaturePrecip}
Voit G.~M.,  Donahue M.,  Bryan G.~L.,   McDonald M.,  2015, Nature, 519, 203

\bibitem[\protect\citeauthoryear{{Wechsler} \& {Tinker}}{{Wechsler} \&
  {Tinker}}{2018}]{WechslerTinker2018GalaxyHaloConnection}
{Wechsler} R.~H.,  {Tinker} J.~L.,  2018, \mn@doi [\araa]
  {10.1146/annurev-astro-081817-051756}, \href
  {https://ui.adsabs.harvard.edu/abs/2018ARA&A..56..435W} {56, 435}

\bibitem[\protect\citeauthoryear{{Wu}, {Evrard}, {Hahn}, {Martizzi}, {Teyssier}
   \& {Wechsler}}{{Wu} et~al.}{2015}]{Wu:2015}
{Wu} H.-Y.,  {Evrard} A.~E.,  {Hahn} O.,  {Martizzi} D.,  {Teyssier} R.,
  {Wechsler} R.~H.,  2015, \mn@doi [\mnras] {10.1093/mnras/stv1434}, \href
  {https://ui.adsabs.harvard.edu/abs/2015MNRAS.452.1982W} {452, 1982}

\bibitem[\protect\citeauthoryear{{York} et~al.,}{{York}
  et~al.}{2000}]{York2000SDSS}
{York} D.~G.,  et~al., 2000, \mn@doi [\aj] {10.1086/301513}, \href
  {https://ui.adsabs.harvard.edu/abs/2000AJ....120.1579Y} {120, 1579}

\bibitem[\protect\citeauthoryear{{Zehavi}, {Contreras}, {Padilla}, {Smith},
  {Baugh}  \& {Norberg}}{{Zehavi} et~al.}{2018}]{Zehavi2018AssemblyBias}
{Zehavi} I.,  {Contreras} S.,  {Padilla} N.,  {Smith} N.~J.,  {Baugh} C.~M.,
  {Norberg} P.,  2018, \mn@doi [\apj] {10.3847/1538-4357/aaa54a}, \href
  {https://ui.adsabs.harvard.edu/abs/2018ApJ...853...84Z} {853, 84}

\bibitem[\protect\citeauthoryear{{Zentner}, {Hearin}  \& {van den
  Bosch}}{{Zentner} et~al.}{2014}]{Zentner2014GalaxyAssemblyBias}
{Zentner} A.~R.,  {Hearin} A.~P.,   {van den Bosch} F.~C.,  2014, \mn@doi
  [\mnras] {10.1093/mnras/stu1383}, \href
  {https://ui.adsabs.harvard.edu/abs/2014MNRAS.443.3044Z} {443, 3044}

\bibitem[\protect\citeauthoryear{{Zhang}, {Andernach}, {Caretta}, {Reiprich},
  {B{\"o}hringer}, {Puchwein}, {Sijacki}  \& {Girardi}}{{Zhang}
  et~al.}{2011}]{Zhang2011HIFLUGCS}
{Zhang} Y.~Y.,  {Andernach} H.,  {Caretta} C.~A.,  {Reiprich} T.~H.,
  {B{\"o}hringer} H.,  {Puchwein} E.,  {Sijacki} D.,   {Girardi} M.,  2011,
  \mn@doi [\aap] {10.1051/0004-6361/201015830}, \href
  {https://ui.adsabs.harvard.edu/abs/2011A&A...526A.105Z} {526, A105}

\bibitem[\protect\citeauthoryear{{Zhang} et~al.,}{{Zhang}
  et~al.}{2016}]{Zhang2016}
{Zhang} Y.,  et~al., 2016, \mn@doi [\apj] {10.3847/0004-637X/816/2/98}, \href
  {https://ui.adsabs.harvard.edu/abs/2016ApJ...816...98Z} {816, 98}

\bibitem[\protect\citeauthoryear{{Zhang} et~al.,}{{Zhang}
  et~al.}{2019}]{Zhang2019ICL}
{Zhang} Y.,  et~al., 2019, \mn@doi [\apj] {10.3847/1538-4357/ab0dfd}, \href
  {https://ui.adsabs.harvard.edu/abs/2019ApJ...874..165Z} {874, 165}

\bibitem[\protect\citeauthoryear{{ZuHone}, {Kowalik}, {{\"O}hman}, {Lau}  \&
  {Nagai}}{{ZuHone} et~al.}{2018}]{ZuHone2018ClusterMergerCatalog}
{ZuHone} J.~A.,  {Kowalik} K.,  {{\"O}hman} E.,  {Lau} E.,   {Nagai} D.,  2018,
  \mn@doi [\apjs] {10.3847/1538-4365/aa99db}, \href
  {https://ui.adsabs.harvard.edu/abs/2018ApJS..234....4Z} {234, 4}

\bibitem[\protect\citeauthoryear{{de Haan} et~al.,}{{de Haan}
  et~al.}{2016}]{deHaan2016SPTCosmo}
{de Haan} T.,  et~al., 2016, \mn@doi [\apj] {10.3847/0004-637X/832/1/95}, \href
  {https://ui.adsabs.harvard.edu/abs/2016ApJ...832...95D} {832, 95}

\bibitem[\protect\citeauthoryear{{van den Bosch} \& {Ogiya}}{{van den Bosch} \&
  {Ogiya}}{2018}]{vandenBosch2018}
{van den Bosch} F.~C.,  {Ogiya} G.,  2018, \mn@doi [\mnras]
  {10.1093/mnras/sty084}, \href
  {https://ui.adsabs.harvard.edu/abs/2018MNRAS.475.4066V} {475, 4066}

\makeatother
\end{thebibliography}

\appendix

\section{LLR Fits to BCG and Total stellar mass}
\label{Appendix:StellarLLRfits}

\begin{figure*}
    \includegraphics[width = 0.89\columnwidth]{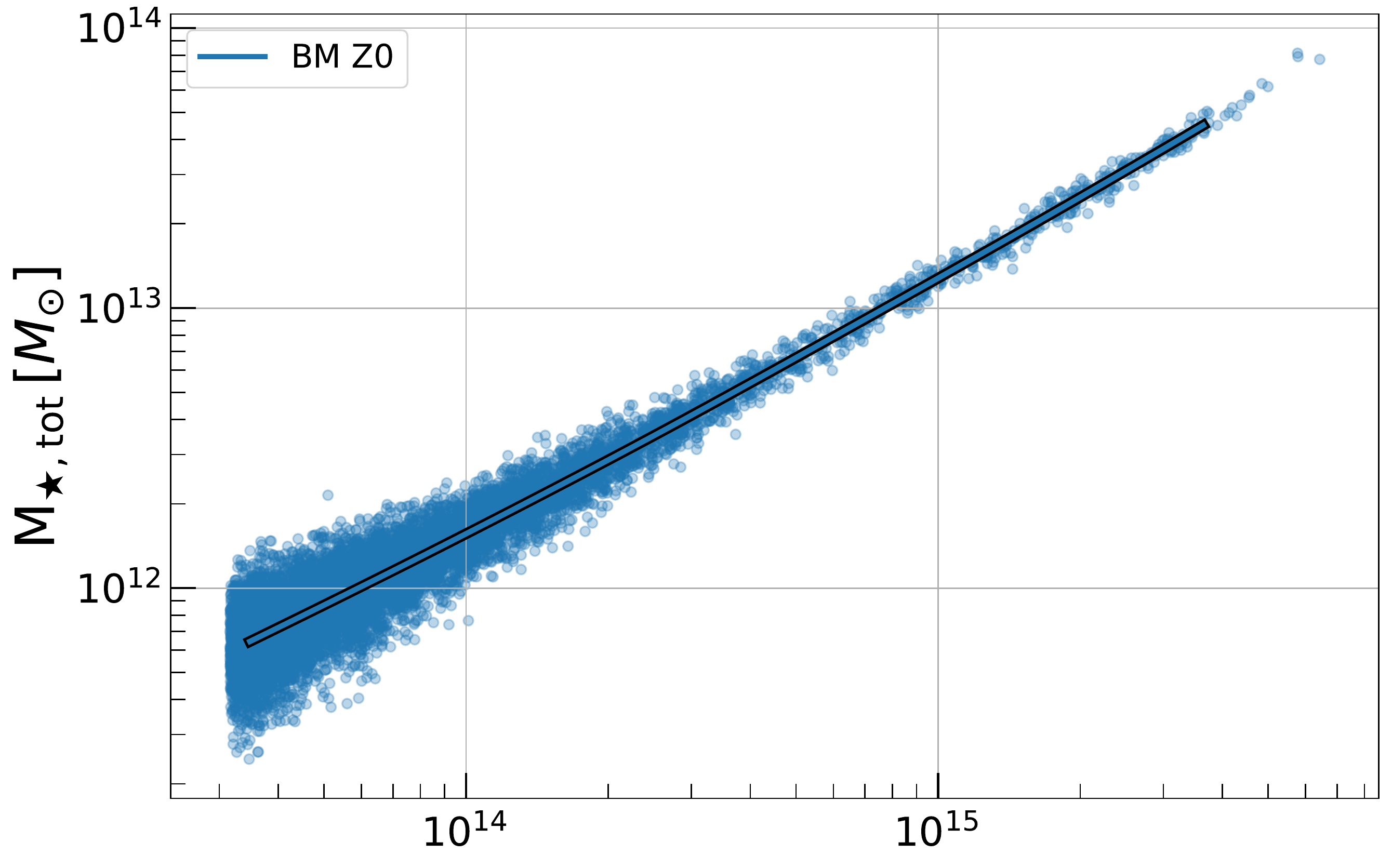}
    \includegraphics[width = 0.85\columnwidth]{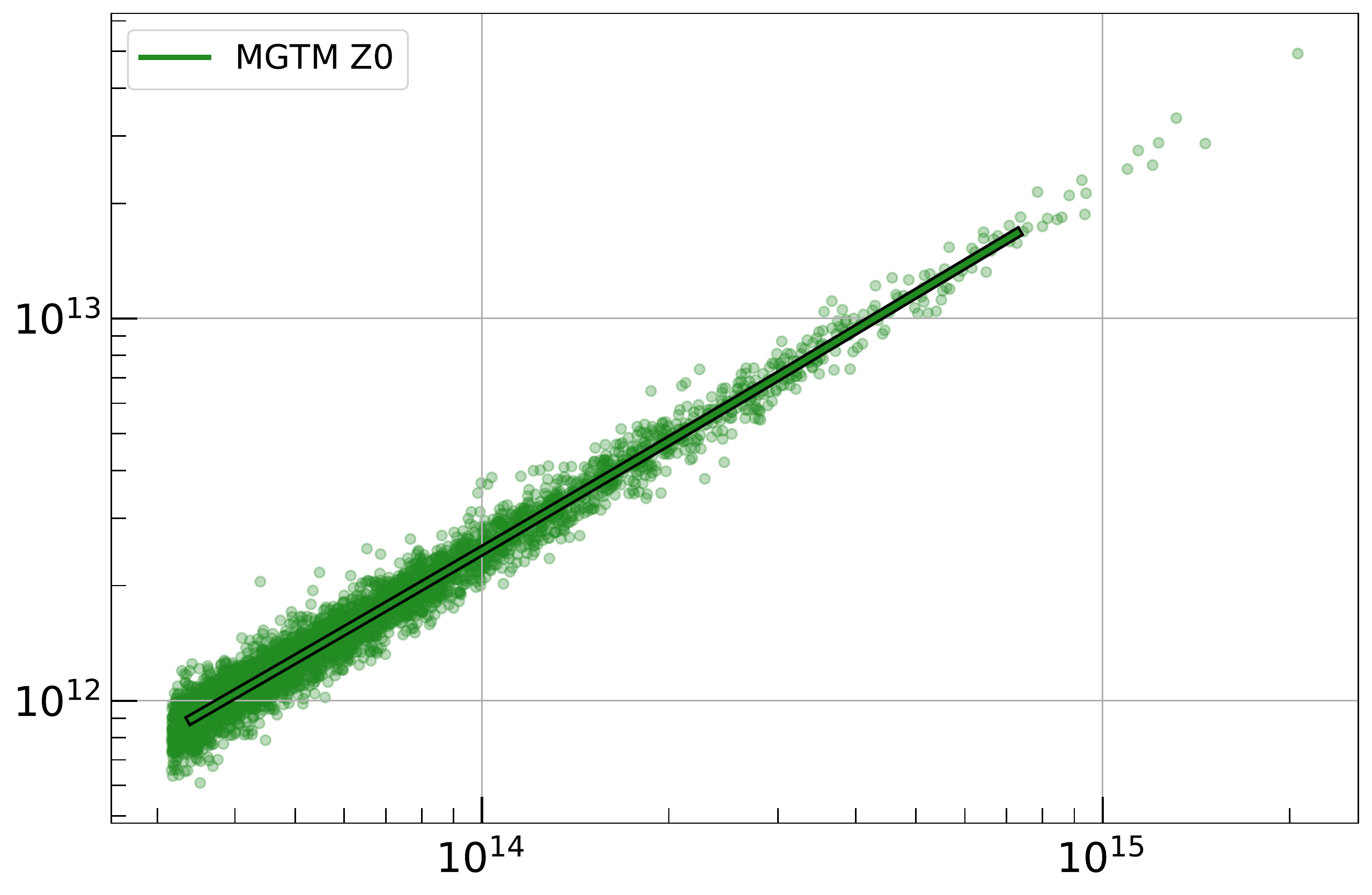}
    \includegraphics[width = 0.89\columnwidth]{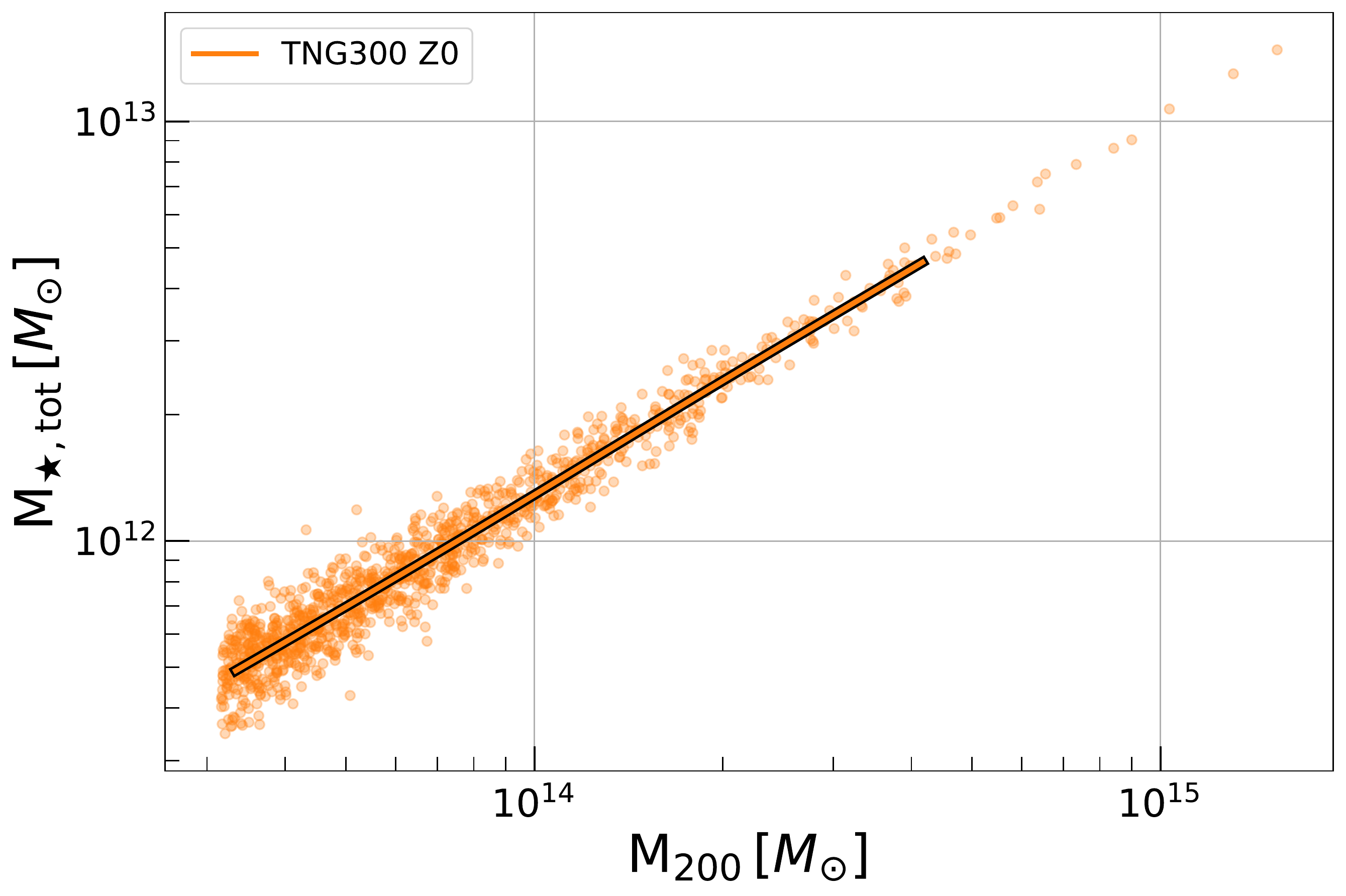}
    \includegraphics[width = 0.85\columnwidth]{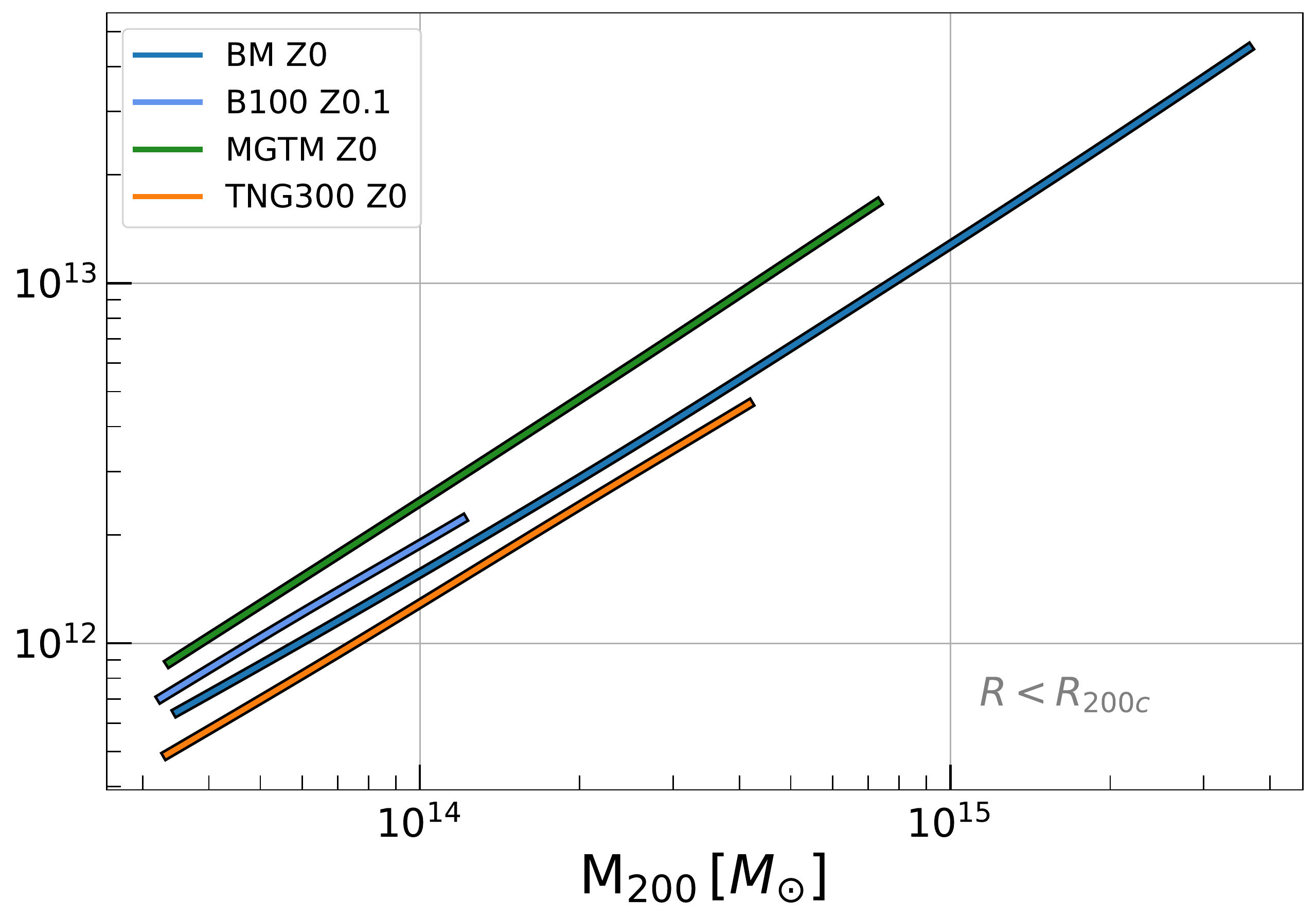}
    \caption{Scaling of total stellar mass within $\rtwoh$ at $z=0$ in the three simulations.  Format is identical to Figure~\ref{fig:LLR Fits}.}
    \label{fig:MStar_Z0_allsims}
\end{figure*}

Figure~\ref{fig:MStar_Z0_allsims} shows the scaling of total stellar mass within $\rtwoh$ with halo mass at $z=0$, including a panel for comparisons of the normalizations.  Note that we do not account for differences in the cosmic baryon fraction, $f_b = \Omega_b/\Omega_m$, of the simulations here.  Accounting for these differences brings the normalizations into slightly better agreement.

\begin{figure*}
    \includegraphics[width = 0.89\columnwidth]{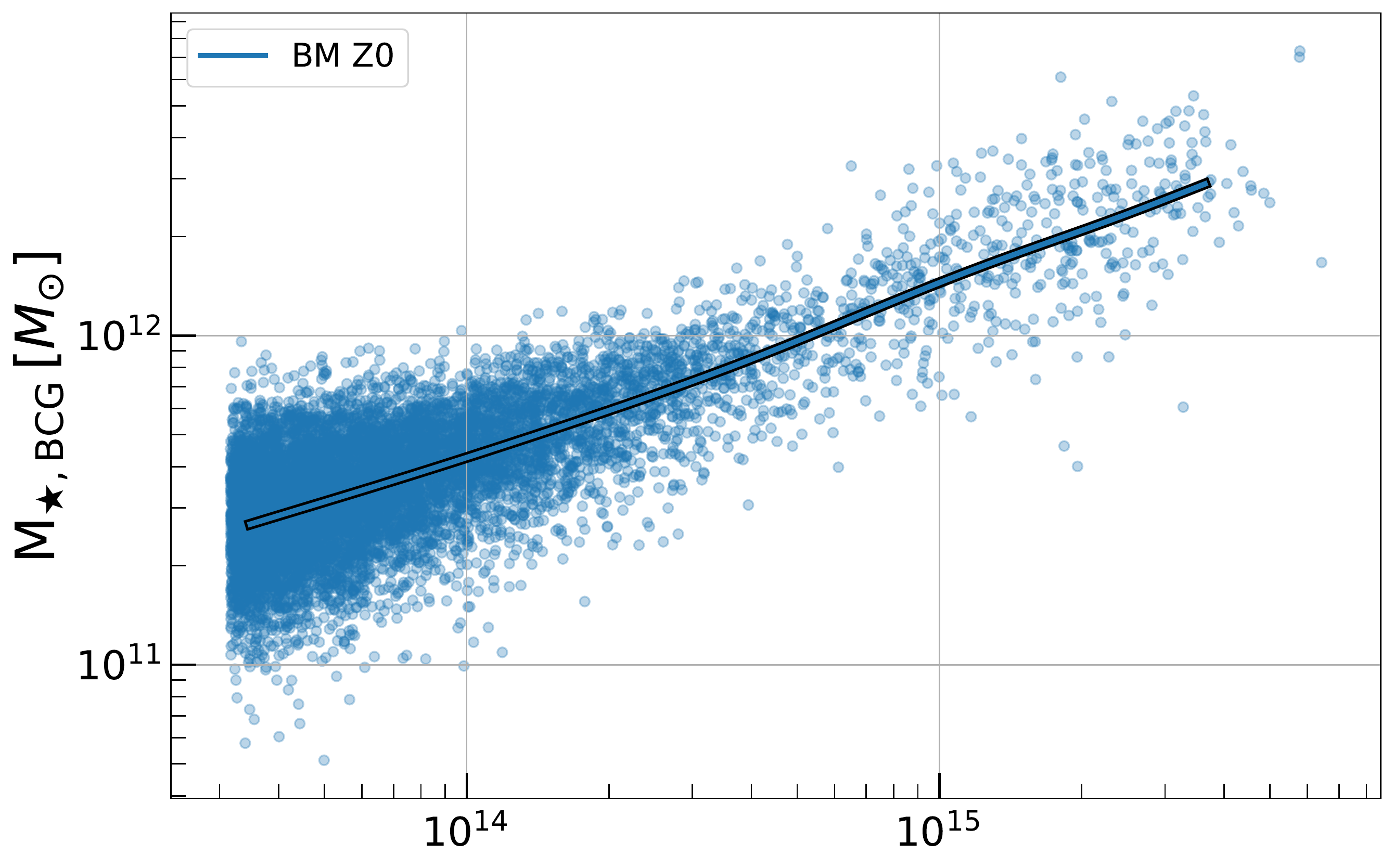}
    \includegraphics[width = 0.85\columnwidth]{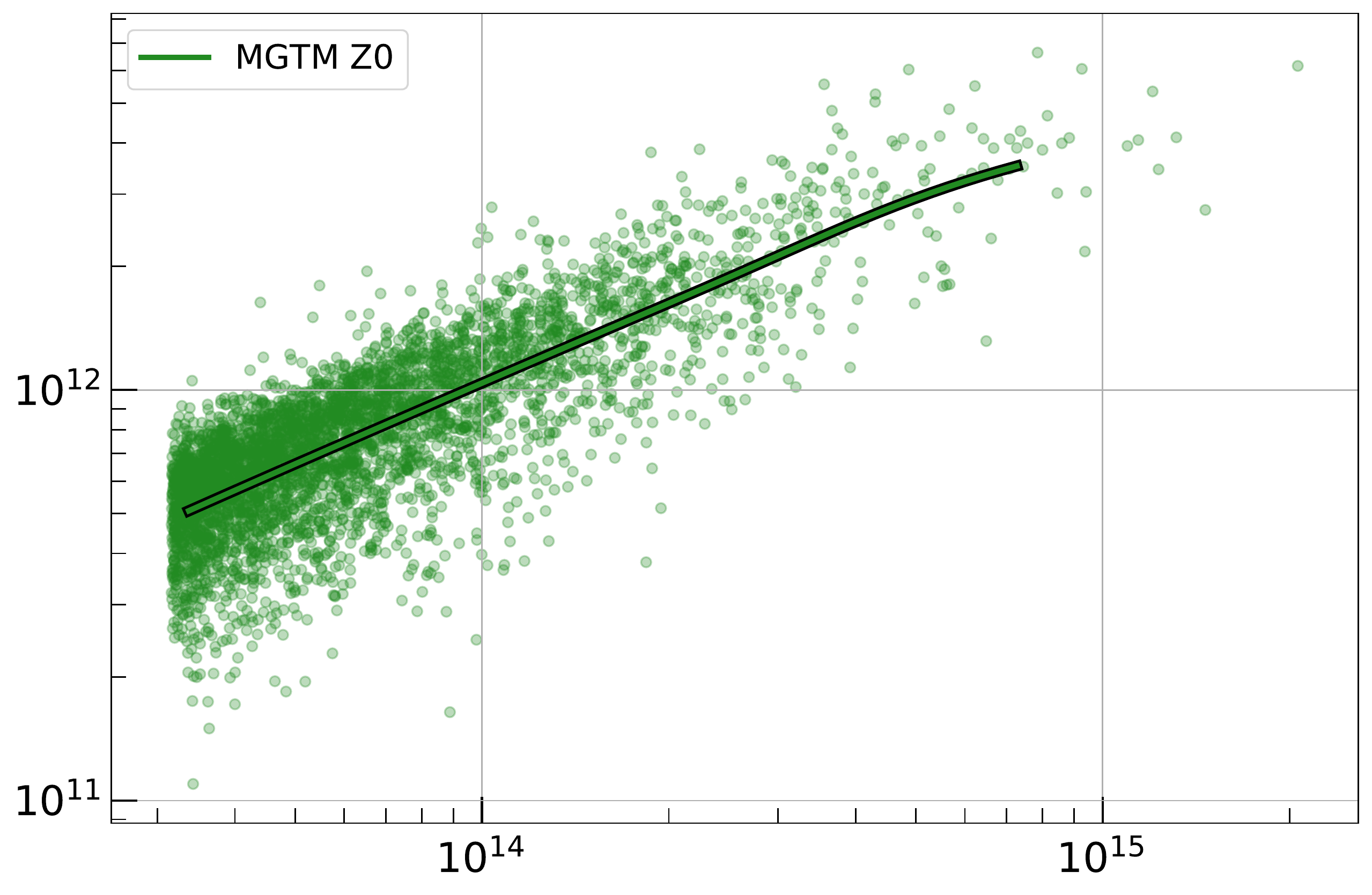}
    \includegraphics[width = 0.89\columnwidth]{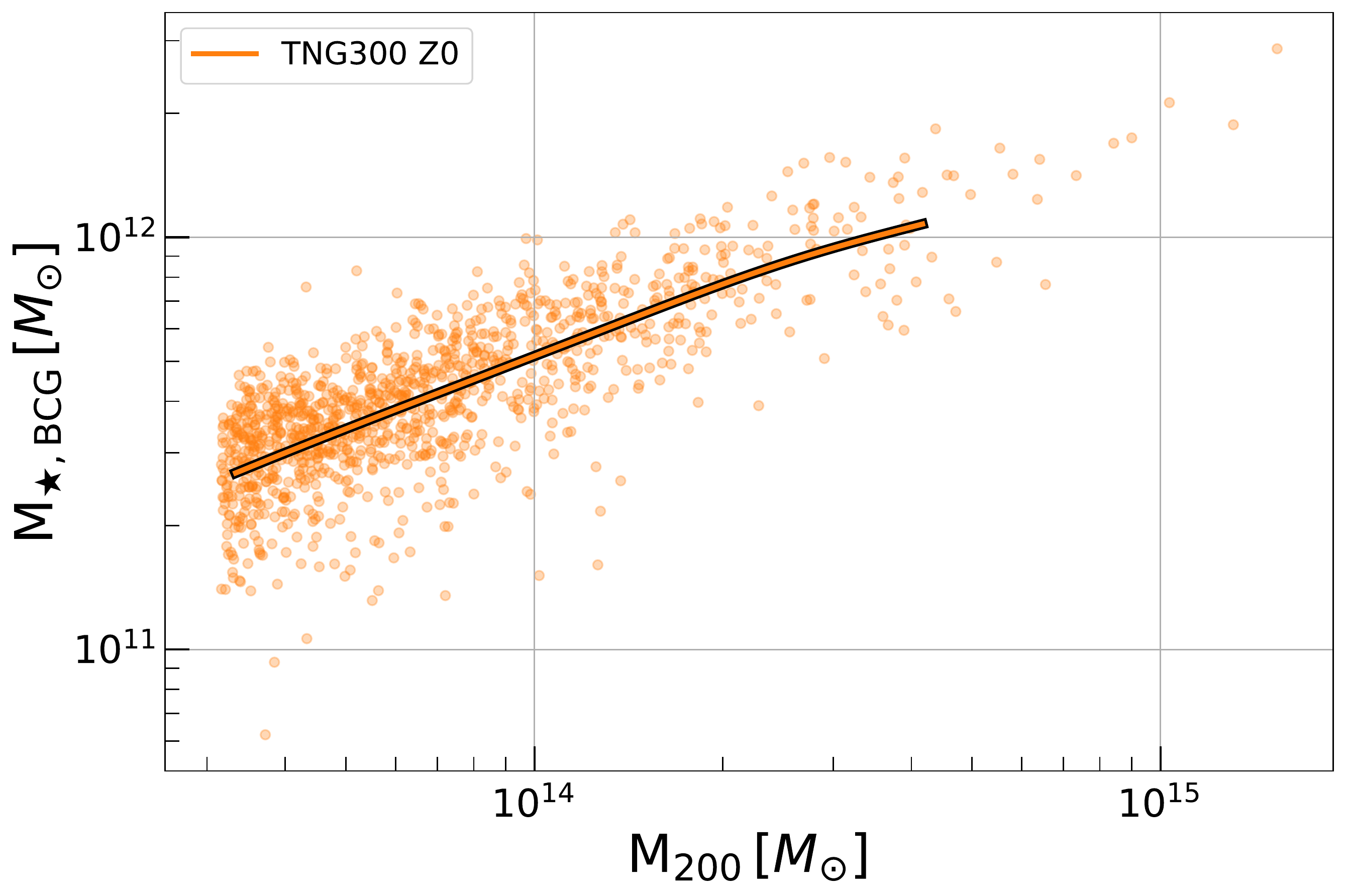}
    \includegraphics[width = 0.85\columnwidth]{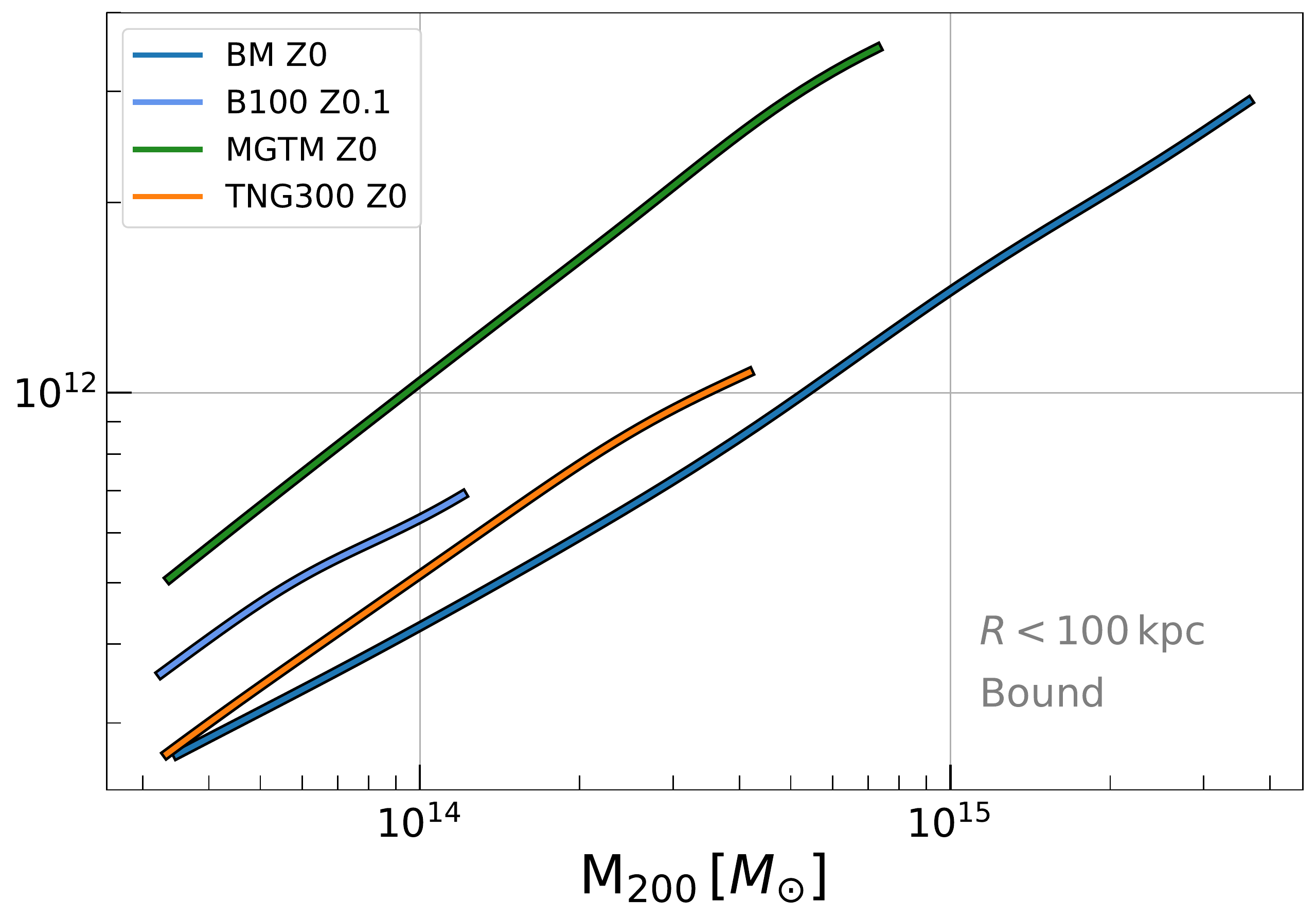}
    \caption{Scaling of central galaxy stellar mass within $\rtwoh$ at $z=0$ in the three simulations.  Format is identical to Figure~\ref{fig:LLR Fits}. 
    } 
    \label{fig:LLR_MStar_BCG100}
\end{figure*}

The scaling relations for $\MstarBCG-\Mhalo$, shown in Figure~\ref{fig:LLR_MStar_BCG100}, reveal that the central galaxies in MGTM are a factor $\sim 2-3$ more massive than those in TNG300 and BM.

\section{GMM, Skewness, and LLR fit parameters}
\label{Appendix:All_Params}
 
We provide individual simulation values of the $\Nsat$ skewness and GMM fit parameters in  Table~\ref{table:Skew_and_GMM_Params}. 

The remaining tables provide $z=0$ LLR fit parameters as a function of total halo mass, sampled in 0.1 dex intervals in the log of halo mass.  
 
 \begin{table*}
 \begin{center} 
 	\caption{Parameters of the $\Nsat$ Gaussian mixture model (Figure~\ref{fig:GMM_Params}), and the kernel skewness (Figure~\ref{fig:Sample_Skewness}) for all simulations having samples of $> 300$ halos with $\Mhalo > 10^{13.5} \msol$ at the available redshifts.}
    \begin{tabular}{ccccccccccc}
        	\label{table:Skew_and_GMM_Params}
        	\textbf{z} & & $\mu_1$ & $\mu_2$ & $\sigma_1$ & $\sigma_2$ & $f_1$ & & $\gamma(\Nsat)$ & $\gamma(\Mstar)$ & $\gamma(\MstarBCG)$\\
        	\hline
            \multicolumn{11}{c}{\textbf{BM}}\\
            \hline
         	 $0^a$ &\,\,\,& $0.25 \pm 0.03$    &$-0.91 \pm 0.12$   & $0.67 \pm 0.02$   & $1.05 \pm 0.04$ & $0.79 \pm 0.03$ & & $-0.85\pm 0.05$ & $-0.26\pm 0.02$ & $-0.31\pm 0.03$\\
             $0.5$ & & $0.31 \pm 0.02$    & $-0.90 \pm 0.08$  & $0.64 \pm 0.01$   & $1.05 \pm 0.03$ & $0.74 \pm 0.03$ & & $-0.93 \pm 0.04$ & $-0.19 \pm 0.03$ &  $-0.39 \pm 0.03$\\
         	 $1$ & & $0.29 \pm 0.02$      & $-0.99 \pm 0.1$   & $0.63 \pm 0.02$   & $1.06 \pm 0.04$ & $0.77 \pm 0.027$ & & $-0.99 \pm 0.05$ & $-0.05 \pm 0.04$ & $-0.34 \pm 0.04$\\
             $1.5$ & & $0.28 \pm 0.03$    & $-1.01 \pm 0.18$  & $0.67 \pm 0.02$   & $1.06 \pm 0.06$ & $0.78 \pm 0.04$ & & $-0.87 \pm 0.06$ & $-0.20 \pm 0.10$ & $-0.39 \pm 0.09$\\
             $2$ & & $0.24 \pm 0.06$      & $-1.03 \pm 0.35$  & $0.70 \pm 0.39$   & $1.06 \pm 0.11$ & $0.81 \pm 0.07$ & & $-0.73 \pm 0.09$ &  $-0.23 \pm 0.21$ & $-0.56\pm 0.14$\\ 
              \hline
              \multicolumn{11}{c}{\textbf{MGTM}}\\
            \hline
             $0$ & & $0.25 \pm 0.02$ & $-1.08 \pm 0.14$ & $0.71 \pm 0.02$ & $1.19 \pm 0.05$ & $0.82 \pm 0.03$ & & $-0.96 \pm 0.05$ & $-0.020 \pm 0.05$ & $-0.80 \pm 0.05$\\
             $0.5$ & & $0.23 \pm 0.02$ & $-1.02 \pm 0.15$ & $0.65 \pm 0.02$ & $1.05 \pm 0.05$ & $0.83 \pm 0.03$ & & $-0.89 \pm 0.07$ & $0.31 \pm 0.06$ & N/A$^b$\\
             $1$ & & $0.23 \pm 0.03$ & $-0.99 \pm 0.18$ & $0.65 \pm 0.02$ & $1.08 \pm 0.07$ & $0.82 \pm 0.04$ & & $-0.98 \pm 0.10$ & $0.33 \pm 0.06$ & N/A$^b$\\
              \hline
              \multicolumn{11}{c}{\textbf{TNG300}}\\
            \hline
             $0$ & & $0.25 \pm 0.05$ & $-1.09 \pm 0.25$ & $0.67 \pm 0.03$ & $1.04 \pm 0.08$ & $0.81 \pm 0.05$ & & $-0.90 \pm 0.07$ & $0.034 \pm 0.06$ & $-0.73 \pm 0.09$\\
             $0.5$ & & $0.23 \pm 0.05$ & $-1.05 \pm 0.3$ & $0.66 \pm 0.03$ & $1.04 \pm 0.09$ & $0.81 \pm 0.06$ & & $-0.92 \pm 0.08$ & $0.016 \pm 0.13$ & $-0.57 \pm 0.08$\\
             $1$ & & $0.28 \pm 0.11$ & $-0.72 \pm 0.35$ & $0.67 \pm 0.07$ & $0.98 \pm 0.11$ & $0.74 \pm 0.14$ & & $-0.70 \pm 0.18$ & $-0.06 \pm 0.14$ & $-0.59 \pm 0.15$\\
              
  		\end{tabular}
    \end{center}
    $^a$ For GMM parameters and $\Nsat$ kernel skewness, the sample uses $\Mhalo > 10^{13.8} \msol$. For the other skewness computations, it is $\Mhalo > 10^{13.5} \msol$\\
    $^b$ We do not have $\MstarBCG$ values for MGTM at $z > 0$.
 \end{table*}


\begin{table*}
\caption{LLR Fits for BM at $z = 0$ for $\Nsat$, $\Mstar$, and $\MstarBCG$. We show the decimal normalization ($\pi_{10} = \log_{10}e^\pi$), slope ($\alpha$), and scatter ($\sigma$, in $\ln$ terms) for each property, along with the correlation coefficients of property pairs ($r$).}
\label{Table:LLR_Fits_BM_Z0}
\begin{center}
    $\rm BM, z = 0$
\end{center}
\begin{tabular}{@{}ccccccccccccc@{}} 
    \toprule
     & \multicolumn{3}{c}{\textbf{$\Nsat$}} & \multicolumn{3}{c}{\textbf{$\Mstar$}} & \multicolumn{3}{c}{\textbf{$\MstarBCG$}} & \multicolumn{3}{c}{\textbf{$r$}}\\
    \midrule
$\log_{10}M_{\rm Halo}$ & $\pi_{10}$ & $\alpha$ & $\sigma$ & $\pi_{10}$ & $\alpha$ & $\sigma$ & $\pi_{10}$ & $\alpha$ & $\sigma$ & $\rm N_{Sat}-\rm M_{\star, tot}$ & $\rm N_{Sat}-\rm M_{\star, BCG}$ & $\rm M_{\star, tot}-\rm M_{\star, BCG}$ \\
\midrule
13.5 & 0.509 & 0.888 & 0.552 & 11.774 & 0.826 & 0.292 & 11.409 & 0.436 & 0.400 & 0.410 & -0.120 & 0.630 \\
13.6 & 0.600 & 0.911 & 0.543 & 11.857 & 0.828 & 0.279 & 11.452 & 0.434 & 0.390 & 0.439 & -0.129 & 0.582 \\
13.7 & 0.693 & 0.931 & 0.526 & 11.940 & 0.831 & 0.264 & 11.495 & 0.433 & 0.379 & 0.468 & -0.139 & 0.530 \\
13.8 & 0.788 & 0.947 & 0.502 & 12.023 & 0.836 & 0.247 & 11.539 & 0.436 & 0.366 & 0.493 & -0.152 & 0.476 \\
13.9 & 0.884 & 0.955 & 0.470 & 12.108 & 0.843 & 0.228 & 11.584 & 0.443 & 0.353 & 0.513 & -0.168 & 0.421 \\
14.0 & 0.980 & 0.957 & 0.435 & 12.194 & 0.852 & 0.210 & 11.629 & 0.451 & 0.340 & 0.529 & -0.189 & 0.368 \\
14.1 & 1.075 & 0.953 & 0.398 & 12.280 & 0.860 & 0.193 & 11.676 & 0.461 & 0.330 & 0.540 & -0.214 & 0.319 \\
14.2 & 1.169 & 0.946 & 0.364 & 12.368 & 0.869 & 0.177 & 11.723 & 0.469 & 0.322 & 0.548 & -0.239 & 0.277 \\
14.3 & 1.262 & 0.937 & 0.333 & 12.456 & 0.878 & 0.164 & 11.772 & 0.478 & 0.316 & 0.554 & -0.266 & 0.240 \\
14.4 & 1.354 & 0.928 & 0.307 & 12.546 & 0.890 & 0.151 & 11.821 & 0.489 & 0.311 & 0.559 & -0.295 & 0.203 \\
14.5 & 1.446 & 0.917 & 0.284 & 12.637 & 0.903 & 0.138 & 11.873 & 0.505 & 0.308 & 0.560 & -0.331 & 0.163 \\
14.6 & 1.536 & 0.908 & 0.263 & 12.730 & 0.916 & 0.124 & 11.927 & 0.526 & 0.308 & 0.555 & -0.371 & 0.123 \\
14.7 & 1.626 & 0.903 & 0.241 & 12.823 & 0.927 & 0.110 & 11.984 & 0.553 & 0.314 & 0.544 & -0.406 & 0.096 \\
14.8 & 1.716 & 0.902 & 0.218 & 12.917 & 0.936 & 0.098 & 12.043 & 0.579 & 0.327 & 0.522 & -0.429 & 0.091 \\
14.9 & 1.806 & 0.904 & 0.196 & 13.012 & 0.941 & 0.087 & 12.103 & 0.590 & 0.345 & 0.493 & -0.438 & 0.107 \\
15.0 & 1.897 & 0.909 & 0.179 & 13.106 & 0.946 & 0.079 & 12.161 & 0.576 & 0.366 & 0.469 & -0.437 & 0.127 \\
15.1 & 1.989 & 0.919 & 0.167 & 13.201 & 0.952 & 0.074 & 12.216 & 0.544 & 0.385 & 0.456 & -0.427 & 0.138 \\
15.2 & 2.082 & 0.936 & 0.158 & 13.297 & 0.960 & 0.072 & 12.268 & 0.516 & 0.398 & 0.449 & -0.412 & 0.141 \\
15.3 & 2.178 & 0.954 & 0.151 & 13.394 & 0.969 & 0.069 & 12.318 & 0.509 & 0.404 & 0.439 & -0.395 & 0.141 \\
15.4 & 2.275 & 0.967 & 0.145 & 13.492 & 0.977 & 0.067 & 12.371 & 0.522 & 0.404 & 0.424 & -0.379 & 0.142 \\
15.5 & 2.373 & 0.979 & 0.140 & 13.591 & 0.983 & 0.065 & 12.426 & 0.542 & 0.400 & 0.406 & -0.365 & 0.148 \\
\bottomrule
\end{tabular}
\end{table*}


\begin{table*}
\caption{LLR Fits for B100 at $z = 0.12$ for $\Nsat$, $\Mstar$, and $\MstarBCG$. We show the decimal normalization ($\pi_{10} = \log_{10}e^\pi$), slope ($\alpha$), and scatter ($\sigma$, in $\ln$ terms) for each property, along with the correlation coefficients of property pairs ($r$).}
\label{Table:LLR_Fits_B100_Z0}
\begin{center}
    $\rm B100, z = 0.12$
\end{center}
\begin{tabular}{@{}ccccccccccccc@{}} 
    \toprule
     & \multicolumn{3}{c}{\textbf{$\Nsat$}} & \multicolumn{3}{c}{\textbf{$\Mstar$}} & \multicolumn{3}{c}{\textbf{$\MstarBCG$}} & \multicolumn{3}{c}{\textbf{$r$}}\\
    \midrule
$\log_{10}M_{\rm Halo}$ & $\pi_{10}$ & $\alpha$ & $\sigma$ & $\pi_{10}$ & $\alpha$ & $\sigma$ & $\pi_{10}$ & $\alpha$ & $\sigma$ & $\rm N_{Sat}-\rm M_{\star, tot}$ & $\rm N_{Sat}-\rm M_{\star, BCG}$ & $\rm M_{\star, tot}-\rm M_{\star, BCG}$ \\
\midrule
13.5 & 0.646 & 1.105 & 0.516 & 11.836 & 0.899 & 0.180 & 11.545 & 0.628 & 0.314 & 0.269 & -0.367 & 0.507 \\
13.6 & 0.758 & 1.128 & 0.483 & 11.926 & 0.903 & 0.178 & 11.608 & 0.625 & 0.308 & 0.281 & -0.349 & 0.499 \\
13.7 & 0.871 & 1.127 & 0.439 & 12.016 & 0.897 & 0.172 & 11.667 & 0.592 & 0.301 & 0.300 & -0.325 & 0.485 \\
13.8 & 0.980 & 1.100 & 0.389 & 12.104 & 0.883 & 0.163 & 11.718 & 0.526 & 0.292 & 0.334 & -0.298 & 0.455 \\
13.9 & 1.084 & 1.059 & 0.341 & 12.190 & 0.868 & 0.151 & 11.759 & 0.455 & 0.285 & 0.385 & -0.279 & 0.400 \\
14.0 & 1.184 & 1.018 & 0.302 & 12.275 & 0.861 & 0.139 & 11.800 & 0.428 & 0.282 & 0.452 & -0.281 & 0.320 \\
\bottomrule
\end{tabular}
\end{table*}


\begin{table*}
\caption{LLR Fits for MGTM at $z = 0.03$ for $\Nsat$, $\Mstar$, and $\MstarBCG$. We show the decimal normalization ($\pi_{10} = \log_{10}e^\pi$), slope ($\alpha$), and scatter ($\sigma$, in $\ln$ terms) for each property, along with the correlation coefficients of property pairs ($r$).}
\label{Table:LLR_Fits_MGTM_Z0}
\begin{center}
    $\rm MGTM, z = 0.03$
\end{center}
\begin{tabular}{@{}ccccccccccccc@{}} 
    \toprule
     & \multicolumn{3}{c}{\textbf{$\Nsat$}} & \multicolumn{3}{c}{\textbf{$\Mstar$}} & \multicolumn{3}{c}{\textbf{$\MstarBCG$}} & \multicolumn{3}{c}{\textbf{$r$}}\\
    \midrule
$\log_{10}M_{\rm Halo}$ & $\pi_{10}$ & $\alpha$ & $\sigma$ & $\pi_{10}$ & $\alpha$ & $\sigma$ & $\pi_{10}$ & $\alpha$ & $\sigma$ & $\rm N_{Sat}-\rm M_{\star, tot}$ & $\rm N_{Sat}-\rm M_{\star, BCG}$ & $\rm M_{\star, tot}-\rm M_{\star, BCG}$ \\
\midrule
13.5 & 0.652 & 1.038 & 0.445 & 11.920 & 0.943 & 0.102 & 11.687 & 0.668 & 0.294 & -0.068 & -0.479 & 0.618 \\
13.6 & 0.756 & 1.018 & 0.429 & 12.014 & 0.941 & 0.102 & 11.754 & 0.664 & 0.297 & -0.066 & -0.473 & 0.618 \\
13.7 & 0.858 & 1.000 & 0.410 & 12.109 & 0.941 & 0.102 & 11.820 & 0.660 & 0.300 & -0.065 & -0.467 & 0.619 \\
13.8 & 0.957 & 0.987 & 0.387 & 12.203 & 0.941 & 0.101 & 11.886 & 0.657 & 0.305 & -0.067 & -0.461 & 0.623 \\
13.9 & 1.054 & 0.979 & 0.360 & 12.297 & 0.943 & 0.100 & 11.951 & 0.654 & 0.311 & -0.075 & -0.458 & 0.630 \\
14.0 & 1.152 & 0.977 & 0.332 & 12.392 & 0.946 & 0.099 & 12.016 & 0.652 & 0.318 & -0.090 & -0.459 & 0.640 \\
14.1 & 1.250 & 0.976 & 0.304 & 12.487 & 0.948 & 0.099 & 12.081 & 0.650 & 0.323 & -0.109 & -0.465 & 0.654 \\
14.2 & 1.347 & 0.970 & 0.277 & 12.582 & 0.951 & 0.098 & 12.146 & 0.647 & 0.327 & -0.130 & -0.471 & 0.669 \\
14.3 & 1.441 & 0.956 & 0.253 & 12.678 & 0.953 & 0.098 & 12.210 & 0.645 & 0.329 & -0.148 & -0.473 & 0.684 \\
14.4 & 1.533 & 0.935 & 0.230 & 12.774 & 0.956 & 0.098 & 12.275 & 0.649 & 0.331 & -0.161 & -0.471 & 0.698 \\
14.5 & 1.623 & 0.919 & 0.209 & 12.870 & 0.961 & 0.097 & 12.342 & 0.657 & 0.334 & -0.167 & -0.464 & 0.707 \\
14.6 & 1.715 & 0.919 & 0.189 & 12.968 & 0.967 & 0.096 & 12.408 & 0.654 & 0.339 & -0.169 & -0.459 & 0.710 \\
14.7 & 1.811 & 0.939 & 0.170 & 13.065 & 0.971 & 0.094 & 12.467 & 0.625 & 0.342 & -0.170 & -0.462 & 0.707 \\
14.8 & 1.912 & 0.970 & 0.151 & 13.162 & 0.971 & 0.091 & 12.518 & 0.572 & 0.343 & -0.175 & -0.473 & 0.700 \\
\bottomrule
\end{tabular}
\end{table*}


\begin{table*}
\caption{LLR Fits for TNG300 at $z = 0$ for $\Nsat$, $\Mstar$, and $\MstarBCG$. We show the decimal normalization ($\pi_{10} = \log_{10}e^\pi$), slope ($\alpha$), and scatter ($\sigma$, in $\ln$ terms) for each property, along with the correlation coefficients of property pairs ($r$).}
\label{Table:LLR_Fits_TNG300_Z0}
\begin{center}
    $\rm TNG300, z = 0$
\end{center}
\begin{tabular}{@{}ccccccccccccc@{}} 
    \toprule
     & \multicolumn{3}{c}{\textbf{$\Nsat$}} & \multicolumn{3}{c}{\textbf{$\Mstar$}} & \multicolumn{3}{c}{\textbf{$\MstarBCG$}} & \multicolumn{3}{c}{\textbf{$r$}}\\
    \midrule
$\log_{10}M_{\rm Halo}$ & $\pi_{10}$ & $\alpha$ & $\sigma$ & $\pi_{10}$ & $\alpha$ & $\sigma$ & $\pi_{10}$ & $\alpha$ & $\sigma$ & $\rm N_{Sat}-\rm M_{\star, tot}$ & $\rm N_{Sat}-\rm M_{\star, BCG}$ & $\rm M_{\star, tot}-\rm M_{\star, BCG}$ \\
\midrule
13.5 & 0.520 & 1.025 & 0.493 & 11.670 & 0.863 & 0.163 & 11.414 & 0.622 & 0.300 & 0.222 & -0.348 & 0.526 \\
13.6 & 0.623 & 1.025 & 0.477 & 11.757 & 0.865 & 0.155 & 11.476 & 0.609 & 0.297 & 0.234 & -0.355 & 0.492 \\
13.7 & 0.725 & 1.023 & 0.455 & 11.844 & 0.869 & 0.147 & 11.536 & 0.596 & 0.294 & 0.245 & -0.364 & 0.449 \\
13.8 & 0.827 & 1.021 & 0.429 & 11.931 & 0.874 & 0.138 & 11.594 & 0.588 & 0.293 & 0.255 & -0.374 & 0.404 \\
13.9 & 0.929 & 1.020 & 0.399 & 12.020 & 0.883 & 0.130 & 11.653 & 0.585 & 0.294 & 0.262 & -0.379 & 0.367 \\
14.0 & 1.031 & 1.018 & 0.367 & 12.110 & 0.892 & 0.122 & 11.711 & 0.585 & 0.295 & 0.269 & -0.381 & 0.339 \\
14.1 & 1.133 & 1.016 & 0.336 & 12.200 & 0.899 & 0.114 & 11.770 & 0.586 & 0.294 & 0.278 & -0.378 & 0.316 \\
14.2 & 1.234 & 1.015 & 0.305 & 12.291 & 0.902 & 0.108 & 11.829 & 0.585 & 0.290 & 0.291 & -0.373 & 0.296 \\
14.3 & 1.335 & 1.014 & 0.276 & 12.381 & 0.899 & 0.102 & 11.885 & 0.575 & 0.285 & 0.305 & -0.369 & 0.283 \\
14.4 & 1.437 & 1.016 & 0.248 & 12.470 & 0.894 & 0.096 & 11.938 & 0.546 & 0.283 & 0.314 & -0.370 & 0.278 \\
14.5 & 1.539 & 1.019 & 0.221 & 12.558 & 0.889 & 0.088 & 11.983 & 0.499 & 0.285 & 0.317 & -0.378 & 0.273 \\
14.6 & 1.642 & 1.022 & 0.199 & 12.646 & 0.886 & 0.081 & 12.024 & 0.451 & 0.290 & 0.314 & -0.392 & 0.257 \\
\bottomrule
\end{tabular}
\end{table*}

\bsp	
\label{lastpage}
\end{document}